\documentclass[onecolumn,showpacs,showkeys,preprintnumbers]{revtex4}

\usepackage{amsmath}
\usepackage{amssymb}

\usepackage{graphicx}
\usepackage{epsfig}
\usepackage{rotating}

\usepackage{dcolumn}
\usepackage{bm}

\def \ms {{\overline{\text{MS}}}}

\begin{document}

\preprint{\textsf{IFUP--TH 2004/10}}
\preprint{\textsf{arXiv:hep-ph/0402173}}

\title{Small $x$ behavior of parton distributions.\\
       A study of higher twist effects.}

\author{Alexei~Yu.~Illarionov\footnote{E-mail: Alexei.Illarionov@pi.infn.it,
 on leave of absence from the Joint Institute for Nuclear Research,
 141980 Dubna, Moscow region, Russia}
}%
\affiliation{%
   Dipartimento di Fisica ``Enrico Fermi'', Universit\`a di Pisa,
   and INFN, Sezione di Pisa,
   I-56100 Pisa, Italy
%
}%

\author{Anatoly~V.~Kotikov\footnote{E-mail: kotikov@thsun1.jinr.ru,
 on leave of absence from the Joint Institute for Nuclear Research,
 141980 Dubna, Moscow region, Russia}
}%
\affiliation{%
  Institut f\"ur Theoretische  Teilchenphysik,
  Universit\"at Karlsruhe,
  D-76128 Karlsruhe, Germany
%
}%

\author{Gonzalo~Parente~Bermudez\footnote{E-mail: gonzalo@fpaxp1.usc.es}
}%
\affiliation{%
   Departamento de F\'\i sica de Part\'\i culas,
   Universidade de Santiago de Compostela,
   E-15706 Santiago de Compostela, Spain
}%

\date{\today}

\begin{abstract}
Higher twist corrections to $F_2$ at small $x$ are studied for the case of
a flat initial condition for the twist-two QCD evolution in the
next-to-leading order approximation.
We present an analytical parameterization of the contributions from the
twist-two and higher twist operators of the Wilson operator product expansion.
Higher twist terms are evaluated using two different approaches, one
motivated by BFKL and the other motivated by the renormalon formalism.
The results of the latter approach are in very good agreement with deep
inelastic scattering data from HERA.
\end{abstract}



\maketitle


\section{Introduction}
 \label{Intro}

For more than a decade various models on the behavior of quarks and gluons
at small $x$ has been confronted with a large amount of experimental data from
HERA on the deep-inelastic scattering (DIS) structure function $F_2$
\cite{Adloff:2001,Adloff:1999,Adloff:1997,Aid:1996,Ahmed:1995,Abt:1993,
Chekanov:2001,Breitweg:2000,Breitweg:1999,Breitweg:1997,Derrick:1996:C72,
Derrick:1996:C69,Derrick:1995,Derrick:1993}.
In the small $x$ regime, non-perturbative effects are expected to give
a substantial contribution to $F_2$. However, what is observed up to very
low $Q^2 \sim 1~\text{GeV}^2$ values, traditionally explained by soft
processes, is described reasonably well by perturbative QCD evolution
(see for example \cite{Cooper-Sarkar:1998}).
Thus, it is important to identify the kinematical region where the 
well-established perturbative QCD formalism can be safely applied.

At small $x$ the $Q^2$ dependence of quarks and gluons is usually obtained
from the numerical solution of the Dokshitzer-Gribov-Lipatov-Altarelli-Parisi
(DGLAP) equations
\cite{Gribov:1972:1000+,Gribov:1972:500+,Lipatov:1975,Altarelli:1977,
Dokshitzer:1977}
\footnote{ The $x$ dependence can also be obtained from the 
Balitsky-Fadin-Kuraev-Lipatov (BFKL) equation
\cite{Lipatov:1976,Kuraev:1976,Kuraev:1977,Balitsky:1978,Lipatov:1986},
which is out of the scope of this work. However, in Section~\ref{Sec:5},
we use the twist-four anomalous dimensions from
Refs.~\cite{Levin:1992,Bartels:1991,Bartels:1993:PLB,Bartels:1993:ZP}
obtained from BFKL results.}.
The $x$ profile of partons at some initial $Q_0^2$ and the QCD energy scale
$\Lambda$ are determined from a fit to experimental data
\cite{Lai:1995,Lai:1997,Huston:1998,Lai:2000,Pumplin:2002,
Martin:1994,Martin:1996,Martin:1998,Martin:2001,Martin:2002,
Gluck:1992,Gluck:1993,Gluck:1995,Gluck:1998}.

On the other hand, when analyzing exclusively the small $x$ region, a much
simpler analysis can be done by using some of the existing analytical
approaches of DGLAP equations in the small $x$ limit
\cite{Ball:1994,Forte:1995,Ball:1997,Frichter:1995,Kotikov:1996:MPL,
Kotikov:1996:YF,Lopez:1996,Adel:1997,Mankiewicz:1997,Kotikov:1999}.
In Refs.~\cite{Ball:1994,Forte:1995,Ball:1997,Mankiewicz:1997,Kotikov:1999}
it was pointed out that HERA small $x$ data can be interpreted
in terms of the so called doubled asymptotic scaling (DAS) phenomenon related
to the asymptotic behavior of the DGLAP evolution discovered many years ago
in \cite{Gribov:1972:1000+,Gribov:1972:500+,DeRujula:1974}. 

In the present work we incorporate the contribution from higher
twist (HT) operators of the Wilson operator product expansion
to our previous analysis \cite{Kotikov:1999}.
The semi-analytical solution of DGLAP equations obtained in
Ref.~\cite{Kotikov:1999} using a flat initial condition,
is the next-to-leading order (NLO) extension of previous studies
performed at the leading order
(LO) in perturbative QCD \cite{Ball:1994,Mankiewicz:1997}.
The flat initial conditions at some initial value $Q^2_0$ correspond to
the case of parton distributions tending to some constant when $x \to 0$.

In Ref.~\cite{Kotikov:1999}, both the gluon and quark singlet densities are
presented in terms of the diagonal $'+'$ and $'-'$ components obtained from
the DGLAP equations in the Mellin moment space.
The $'-'$ components are constants at small $x$ for any values of $Q^2$,
whereas the $'+'$ components grow for $Q^2 \geq Q^2_0$ as
\footnote{
Since we are only interested in the small $x$ behavior and the initial
conditions are given by the flat ($x$-independent) functions (see
Eq.~(\ref{flat})), we use permanently the variable $z=x/x_0$ with values
$0<z<1$ with some arbitrary $x_0 \leq 1$.
}

\begin{equation}
\sim \exp{\left(2\sqrt{\left[
 a_+ \ln \left( \dfrac{a_s(Q^2_0)}{a_s(Q^2)} \right) -
 \left( b_+ +  a_+ \dfrac{\beta_1}{\beta_0} \right)
 \Bigl( a_s(Q^2_0) - a_s(Q^2) \Bigr)
\right] \ln \left( \dfrac{1}{z} \right)} \right)} \ ,
\label{intro:1}
\end{equation}
where $a_+ = 4C_A/\beta_0$ and $b_+ =  8[23 C_A - 26 C_F]T_Rf/(9\beta_0)$.
In Eq. (\ref{intro:1}) and hereafter we use the notation 
$a_s=\alpha_s/(4\pi)$.

The first two coefficients of the QCD $\beta$-function in the $\ms$-scheme
are $\beta_0 = (11/3) C_A - (4/3) T_R f$ and
$\beta_1 = (2/3)[17 C_A^2 - 10 C_A T_R f - 6 C_F T_R f]$ 
where $f$ is the number of active flavors. This new presentation as a
function of the $SU(N)$ group Casimirs, with $f$ active flavors,
$C_A = N$, $T_R = 1/2$, $T_F = T_R f$ and  $C_F = (N^2 - 1)/(2N)$
permits one to apply our results to, for example, the popular $N = 1$
supersymmetric model.
Of course, for $N = 3$ one obtains the QCD result \cite{Kotikov:1999}.

The analysis performed in our previous work \cite{Kotikov:1999} has shown
very good agreement with H1 and ZEUS 1994 data \cite{Aid:1996,
Derrick:1996:C72,Derrick:1996:C69} at $Q^2 \geq 1.5~\text{GeV}^2$. 
Here, we add the contribution from
higher twist operators with the hope to describe also 
more modern data \cite{Adloff:1999,Adloff:1997,Chekanov:2001,Breitweg:2000,Breitweg:1999,Breitweg:1997} at lower $Q^2$

Moreover, in comparison with Ref.~\cite{Kotikov:1999}, in the present
work we have solved the technical problem of ``backward'' evolution that
leads us now to have the normalization scale $Q^2_0$ of DGLAP evolution
in the middle point of the $Q^2$ range.

\subsection{Basic formulae}
 \label{Int1:Bf}

At this point of the introduction, we find convenient to present the basic
results of our article: the twist-four and twist-six corrections to $F_2$ in
the DAS
approach. Thus, a reader who has interest only
in application of the formulas to the analysis of $F_2$ can skip the following
sections and start to read Section~\ref{Sec:10}, where the fits of $F_2$ are
performed. We note, however, that some of the sections that follows
contain also the contribution of power corrections to the derivatives
$\partial F_2/\partial\ln Q^2$ and $\partial\ln F_2/\partial\ln(1/x)$ and to
the parton distributions (PD) (see the Sections~\ref{Sec:6}, \ref{Sec:7},
\ref{Sec:8} and \ref{Sec:9}, respectively).

The basic results of the present article are the twist-four and twist-six
corrections to $F_2$ 
\begin{equation}
 F_2(x, Q^2)   \ = \ F^{\tau2}_2(x,Q^2) \ + \
 \dfrac{1}{Q^2} \, F^{\tau4}_2(z, Q^2) \ + \
 \dfrac{1}{Q^4} \, F^{\tau6}_2(z, Q^2) \ ,
\label{intro:t2+t4+t6} 
\end{equation}
where
for the higher twist parts $F^{\tau4,6}_2$ BFKL-motivated evaluations
\cite{Levin:1992,Bartels:1991,Bartels:1993:PLB,Bartels:1993:ZP}
(in this case only the twist-four correction has been estimated)
and the calculations \cite{Stein:1998}
in the framework of the renormalon model (hereafter
marked with superindex $R$)
have been used.

The latter case is essentially more complete and the predicted HT
corrections can be expressed through the twist-two ones as follows
\begin{equation}
F^{R\tau4}_2(z, Q^2) \ = \ e \ \sum_{a=q,G} 
 {\mathrm a}_a^{\tau4} \,
 \widetilde{\mu}^{\tau4}_a (z, Q^2) \otimes f^{\tau2}_a(z, Q^2) 
\ = \ \sum_{a=q,G} F^{R\tau4}_{2,a}(z, Q^2) \ ,
\label{Ren:tw4In}
\end{equation}
where the symbol $\otimes$ marks the Mellin convolution (see
Eq.~(\ref{def:Mellin}) below), the functions
$\widetilde{\mu}^{\tau4}_a(z, Q^2)$ are given in \cite{Stein:1998}
and 
$e = (\sum^f_1 e_i^2) / f$ is the average charge square for $f$
active quarks. We call $F^{R\tau4}_{2,q}$ and $F^{R\tau4}_{2,G}$,
the HT corrections proportional to the
twist-two quark and gluon densities,
respectively.

Note that the parton distributions $f^{\tau2}_a(z, Q^2)$ are multiplied
by $z$, i.e., $f^{\tau2}_q(z, Q^2) = z \, q(z, Q^2)$ and
$f^{\tau2}_G(z, Q^2) = z \, G(z, Q^2)$. 
Note also that we neglect the non-singlet quark density $f_{\Delta}(z, Q^2)$
and the valent part $f_{V}(z,Q^2)$ of the singlet quark distributions,
because they have the following small-$x$ asymptotics:
$f_{\Delta}(z, Q^2) \sim f_{V}(z, Q^2) \sim x^{\lambda_V}$, where
$\lambda_V \sim 0.3 \div 0.5$.
Thus, our quark density $f^{\tau2}_a(z, Q^2)$ contains only the sea part
$f_{S}(z,Q^2)$, i.e. $f^{\tau2}_a(z, Q^2) = f_{S}(z, Q^2)$.

For the leading twist part
we have \cite{Kotikov:1999} at the LO and NLO approximations, 
respectively, 
\begin{subequations}
\label{intro:F2tw2}
\begin{gather}
 F^{\tau2}_{2,\text{LO}}(z, Q^2) \ = \ e \, f^{\tau2}_{q,\text{LO}}(z, Q^2) \ ,
\label{intro:F2tw2LO} \\
 F^{\tau2}_2(z, Q^2) \ = \ e \, \left( f^{\tau2}_q(z, Q^2) \ + \
  \dfrac{4T_Rf}{3} \, a_s(Q^2) \, f^{\tau2}_G(z, Q^2) \right) \ .
\label{intro:F2tw2NLO}
\end{gather}
\end{subequations}

Let us  keep the NLO relation (\ref{intro:F2tw2NLO}) beyond the
leading twist approximation. Then for the total  $F_2$ 
(see Eq.~(\ref{intro:t2+t4+t6})) we obtain
\begin{equation}
 F_2(z, Q^2) \ = \ e \, \left( f_q(z, Q^2) \ + \
  \dfrac{4T_Rf}{3} \, a_s(Q^2) \, f_G(z, Q^2) \right) \ ,
\label{intro:F2}
\end{equation}
where
$f_a(z, Q^2)$ are the parton distributions containing both the
twist-two part \cite{Kotikov:1999} (see next Section) and 
the twist-four and twist-six contributions
\begin{equation}
 f_a(x, Q^2)   \ = \ f^{\tau2}_a(x,Q^2) \ + \
 \dfrac{1}{Q^2} \, f^{R\tau4}_a(z, Q^2) \ + \
 \dfrac{1}{Q^4} \, f^{R\tau6}_a(z, Q^2) \ .
\label{intro:t2+Rht4+Rht6} 
\end{equation}
For the HT
part $f^{R\tau4,6}_a(z, Q^2)$
calculations in the framework of the renormalon model have been used
\footnote{Note that twist-four corrections are studied below in two
approaches based on BFKL and DGLAP equations (see the Section~\ref{Sec:5}). 
However, we give here the results only for the DGLAP approach based
on the infrared renormalon model because it contains a more complete
calculation and the agreement with experimental data is much better.}.
 
We would like to note that each HT term $f^{R\tau4,6}_a(z, Q^2)$
can be chosen in a quite arbitrary form and only the combination 
\begin{equation}
f^{R\tau4,6}_q(z, Q^2) \ + \
  \dfrac{4T_Rf}{3} \, a_s(Q^2) \, f^{R\tau4,6}_G(z, Q^2)
\label{intro:Rht4+Rht6} 
\end{equation}
is unique, because we kept the original twist-two relation, Eq. 
(\ref{intro:F2tw2NLO}),
to be same when HT corrections are
incorporated (see Eq.~(\ref{intro:F2})).

Note that in our previous studies \cite{Kotikov:2000,Kotikov:2001,Kotikov:2003}
we did not use the Eq.~(\ref{intro:F2}) to parameterize the HT corrections to
$F_2$. Instead we consider the following representation
\begin{equation}
 F^{R\tau4,6}_2(z, Q^2) \ = \ e \,
  \hat{f}^{R\tau4,6}_q(z, Q^2) \ ,
\label{intro:HTF2Old}
\end{equation}
coming from the LO relation (\ref{intro:F2tw2LO})
between $F_2$ and parton distributions. The choice (\ref{intro:HTF2Old}) 
looks quite natural because HT corrections have been obtained in
\cite{Stein:1998} at the LO approximation.
However, this choice is only useful to fit $F_2$ data and it has no
interest to study the parton distributions themselves:
note that the HT corrections to the gluon density are absent
in Eq.~(\ref{intro:HTF2Old}).
Indeed, in the calculation of $F_2$ at NLO one has to take 
a gluon density as in the r.h.s. of the 
Eq.~(\ref{intro:F2tw2NLO}).
So, one should take the condition
\begin{equation}
  \hat{f}^{R\tau4,6}_G(z, Q^2) \ = \ 0 \ ,
\label{intro:HTGOld}
\end{equation}
which is not so natural. Moreover, the choice (\ref{intro:HTF2Old})
and (\ref{intro:HTGOld}) leads to a quite complicated form for
the HT corrections to the quark density: 
there are two independent contributions 
$\sim A_q^{\tau2}$ and $\sim A_G^{\tau2}$
(see Refs.~\cite{Kotikov:2000,Kotikov:2001,Kotikov:2003} and formulas
therein).

In the work we also study $x$ and $Q^2$ dependences
of $\partial F_2/\partial\ln Q^2$ and
$\partial \ln F_2/\partial\ln (1/x)$,
that force to define the parton densities in a proper way.  
So, we take another quite \emph{natural} choice
\begin{subequations}
\label{Ren:f_q+f_G}
\begin{align}
f^{R\tau4,6}_q(z, Q^2) \, = & \, \, {\mathrm a}_q^{\tau4,6} \,
 \widetilde{\mu}^{\tau4,6}_q (z, Q^2) \otimes f^{\tau2}_q(z, Q^2) 
\ \equiv \ \dfrac{1}{e} \, F^{R\tau4,6}_{2,q}(z, Q^2) \ ,
\label{Ren:f_q} \\
f^{R\tau4,6}_G(z, Q^2) \, = & \, \,  \dfrac{3/4T_Rf}{a_s(Q^2)} \,
{\mathrm a}_G^{\tau4,6} \,
 \widetilde{\mu}^{\tau4,6}_G (z, Q^2) \otimes f^{\tau2}_G(z, Q^2) 
\ \equiv \ \dfrac{3/4T_Rf}{e a_s(Q^2)} \, F^{R\tau4,6}_{2,G}(z, Q^2) \ ,
\label{Ren:f_G}
\end{align}
\end{subequations}
i.e.,
the HT quark (gluon) part of $F_2$ relates only to the corresponding
quark (gluon) twist-two density.

Note once again that the choice (\ref{Ren:f_q+f_G}) corresponds 
exactly to the Eq.~(\ref{intro:F2}), i.e. to the extension
of the standard twist-two relation (\ref{intro:F2tw2NLO}) between 
$F_2$ and parton densities at the NLO formulas with the
purpose to include the HT contributions.

Note also that for both of the above parton density choices
the DGLAP equation will be
violated by the HT corrections
(see Section~\ref{Sec:6} and discussions therein).

\subsection{Higher twist terms in the renormalon model}
 \label{Int1:Rf}

As it has been already noted above
it is useful to split the parton distributions in two parts
\begin{equation}
 f_a(z, Q^2) \ = \
  f^{+}_a(z,Q^2) \ + \ f^{-}_a(z, Q^2) \ ,
\label{intro:f_a}
\end{equation}
where the both $'+'$ and $'-'$ components contain twist-two and HT parts.

The two component representation follows directly form the exact
solution of DGLAP equation in the Mellin moment space at the leading
twist approximation (see \cite{Kotikov:1999}).

The twist-two contribution is presented below in the Section~\ref{Sec:2} and
the twist-four and twist-six parts can be expressed through the 
twist-two one as follows (here for simplicity we restrict our
consideration by LO approximation):

\begin{subequations}
\label{intro:Rht4}

for the (singlet) quark distribution
\begin{align}
\dfrac{f^{R\tau4,+}_q(z, Q^2)}{f^{\tau2,+}_{q,\text{LO}}(z, Q^2)} \, =& \, 
\dfrac{64C_FT_Rf}{15\beta_0^2} \,  {\mathrm a}_q^{\tau4} \left\{
 \dfrac{2}{\rho_{\text{LO}}^2}
+ \ln \left(\dfrac{Q^2}{\left|{\mathrm a}_q^{\tau4}\right|}\right)
\dfrac{\widetilde{I}_0(\sigma_{\text{LO}})}{\rho_{\text{LO}} \, 
\widetilde{I}_1(\sigma_{\text{LO}})} 
\right\}
 ~ + ~ {\mathcal{O}}\left( \rho_{\text{LO}} \right ) \ ,
\label{intro:fq4+}\\
\dfrac{f^{R\tau4,-}_q(z, Q^2)}{f^{\tau2,-}_{q,\text{LO}}(z, Q^2)} \, =& \, 
\dfrac{64C_FT_Rf}{15\beta_0^2} \,  {\mathrm a}_q^{\tau4} \left\{
 \ln \left(\dfrac{1}{z_q}\right) \,
 \ln \left(\dfrac{Q^2}
  {z_q \left|{\mathrm a}_q^{\tau4}\right|}\right) -
 p^\prime(\nu_q) \right\} 
 ~ + ~ {\mathcal{O}}\left( z \right ) \ .
\label{intro:fq4-}
\end{align}

for the gluon distribution
\begin{align}
\dfrac{f^{R\tau4,+}_G(z, Q^2)}{f^{\tau2,+}_{G,\text{LO}}(z, Q^2)} \, =& \, 
\dfrac{8}{5\beta_0^2} \, \dfrac{{\mathrm a}_G^{\tau4}}{a_s(Q^2)} \, \left\{
 \dfrac{2}{\rho_{\text{LO}}}
\dfrac{\widetilde{I}_1(\sigma_{\text{LO}})}{
\widetilde{I}_0(\sigma_{\text{LO}})} 
+ \ln \left(\dfrac{Q^2}{\left|{\mathrm a}_G^{\tau4}\right|}\right)
\right\}
 ~ + ~ {\mathcal{O}}\left( \rho_{\text{LO}} \right ) \ ,
\label{intro:fG4+}\\
\dfrac{f^{R\tau4,-}_G(z, Q^2)}{f^{\tau2,-}_{G,\text{LO}}(z, Q^2)} \, =& \,
\dfrac{8}{5\beta_0^2} \, \dfrac{{\mathrm a}_G^{\tau4}}{a_s(Q^2)} \,
 \ln \left(\dfrac{Q^2}
  {z_G^2 \left|{\mathrm a}_G^{\tau4}\right|}\right)
 ~ + ~ {\mathcal{O}}\left( z \right ) \ ,
\label{intro:fG4-}
\end{align}
\end{subequations}
where ${\mathrm a}_a^{\tau4}$ are the magnitudes which should be 
extracted from the fits of the experimental data.
The variables $z_a = z\exp[p(\nu_a)]$, where
$p(\nu_a) = [\Psi(1+\nu_a) - \Psi(\nu_a)]$ and $\nu_a$ are the powers of
the $x \to 1$ asymptotics of the parton distributions, i.~e.
$f_a \sim (1 - x)^{\nu_a}$ at $x \to 1$. From the quark counting rules
we know that $\nu_q \approx 3$ and $\nu_G \approx 4$. Then, we get
$p(\nu_q) \approx 11/6$ and $p(\nu_G) \approx 25/12$, and there derivatives
$p^\prime(\nu_q) \approx -49/36$ and $p^\prime(\nu_G) \approx -205/144$
(see Appendix~B for further details).

The functions $\widetilde I_{\nu}$ in Eqs.~(\ref{intro:fq4+}, \ref{intro:fG4+})
are related to the modified Bessel function $I_{\nu}$ and to the Bessel
function $J_{\nu}$ by:
\begin{equation}
\widetilde I_{\nu}(\sigma)  \ = \
 \left\{\begin{array}{rl}
  I_{\nu}(\bar\sigma), & \text{if}~ \sigma^2 = \bar\sigma^2 \geq 0 \ , \\
  i^\nu J_{\nu}(\bar\sigma), & \text{if}~ \sigma^2 = -\bar\sigma^2 < 0 \ .
 \end{array} \right.
\label{def:I_nu}
\end{equation}
and
the $\sigma $ and $\rho $ values are given in the Section~\ref{Sec:2} 
by Eqs.~(\ref{def:sigma_LO}) and (\ref{def:rho_LO}) 
at the LO and by Eqs.~(\ref{def:s-rho}) NLO, respectively.

Note that the upper (down) line in the r.h.s. of Eq.~(\ref{def:I_nu})
corresponds to the solution of the DGLAP equation for the ``direct''
(``backward'') evolution in the DAS
approximation.

The twist-six part can be easy obtained from the corresponding twist-four
one as
\begin{equation}
 f^{R\tau6}_a(z, Q^2) \ =  \ - \dfrac{8}{7} \times \left[
  f^{R\tau4}_a(z, Q^2) ~\text{ with }~ 
  {\mathrm a}_a^{\tau4} \to {\mathrm a}_a^{\tau6} \, , \
  \ln \left(\dfrac{Q^2}{\left|{\mathrm a}_a^{\tau4}\right|}\right) \to
  \ln \left(\dfrac{Q^2}{\sqrt{\left|{\mathrm a}_a^{\tau6}\right|}}\right)
 \right] \ .
\label{intro:ht4-ht6}
\end{equation}


The article is organized as follows.
   In Section~\ref{Sec:2}
we shortly review basic formulae of
the solution of DGLAP equation at small $x$ values with the flat initial
conditions, given in \cite{Kotikov:1999}. We show the possibility to add the
backward evolution to the formulae.
In Sections~\ref{Sec:3} and \ref{Sec:4} we present the set of formulae
for the derivation $\partial F_2/\partial\ln Q^2$ and for the effective slope 
$\partial\ln  F_2/\partial\ln (1/x)$.
   Section~\ref{Sec:5} contains our suggestions about the contributions of
twist-four and twist-six operators of the Wilson operator product expansion.
   In Sections~\ref{Sec:6}--\ref{Sec:9} we consider the contributions of
the HT
operators to parton distributions and to derivatives of
$F_2$ in the framework of the infrared renormalon model.
   Section~\ref{Sec:10} contains the fits of experimental data for $F_2$,
predictions for its derivatives and some discussions of the obtained results.
   In the Appendix~A we present Mellin moments of renormalon
contributions, calculated in \cite{Stein:1998}, and obtain their
contributions to the PD corresponding moments.
   In the Appendix~B we illustrate the method
\cite{Kotikov:1994:YF,Kotikov:1994:PRD} of replacing at small $x$ the
convolution of two functions by simple products. The method is used in the
present work for the correct incorporation 
of renormalon-type contributions of higher twists terms into our formulae.  
The conclusions contains summary of the results and suggestions about
other applications of the approach.

\section{The contribution of twist-two operators}
 \label{Sec:2}

As in \cite{Kotikov:1999}, we will work with the small $x$ asymptotic
form of parton distributions 
in the framework of the DGLAP evolution
equations starting at some $Q^2_0$ with the flat function:
\begin{equation}
 f_a (Q^2_0) \ = \ A_a \hspace{5mm} (\text{hereafter} ~ a = q, G) \ ,
\label{flat}
\end{equation}
where 
$A_a$ are
unknown parameters that have to be determined from data. 

We shortly compile below the main results found in \cite{Kotikov:1999} at
the LO
NLO approximations. 

\subsection{Leading order}
 \label{Sec2:LO}

The small $x$ asymptotic results for PD,
$f^{\tau2}_{a, \text{LO}}$ ($a = q, G$) and $F^{\tau2}_{2,\text{LO}}$
at LO of perturbation theory and at twist-two 
in the operator product expansion have been found
in Ref.~\cite{Kotikov:1999}:
\begin{subequations}
\label{F2:tau2:LO}
\begin{gather}
 F^{\tau2}_{2,\text{LO}}(z, Q^2) \ = \ e \ f^{\tau2}_{q,\text{LO}}(z, Q^2) \ ,
\label{F2:LO} \\
 f^{\tau2}_{a,\text{LO}}(z, Q^2) \ = \
  f^{\tau2,+}_{a,\text{LO}}(z, Q^2) \ + \
  f^{\tau2,-}_{a,\text{LO}}(z, Q^2) \ .
\label{f_a:LO}
\end{gather}

After Mellin inversion of the explicit moment solution to DGLAP
equations, the $'+'$ and $'-'$ PD components are given by:
\begin{align}
f^{\tau2,+}_{G,\text{LO}}(z, Q^2) &=
 \left(A_G^{\tau2} + \dfrac{C_F}{C_A} A_q^{\tau2} \right) \;
 \widetilde I_0(\sigma_\text{LO}) \ \text{e}^{-\bar d_{+}(1) s_\text{LO}}
 ~ + ~ {\mathcal{O}}\left( \rho_\text{LO}\right ) \ ,
\label{fG+:LO} \\
f^{\tau2,+}_{q,\text{LO}}(z, Q^2) &= \dfrac{2T_Rf}{3C_A} \;
 \left(A_G^{\tau2} + \dfrac{C_F}{C_A} A_q^{\tau2} \right) \;
 \rho_\text{LO} \; \widetilde I_1(\sigma_\text{LO}) \;
 \text{e}^{-\bar d_{+}(1) s_\text{LO}}
 ~ + ~ {\mathcal{O}}\left( \rho_\text{LO}\right ) \ ,
\label{fq+:LO} \\
f^{\tau2,-}_{G,\text{LO}}(z, Q^2) &= - \dfrac{C_F}{C_A} \; A_q^{\tau2} \;
  \text{e}^{- d_{-}(1) s_\text{LO}} ~ + ~ {\mathcal{O}}\left( z \right ) \ ,
\label{fG-:LO} \\
f^{\tau2,-}_{q,\text{LO}}(z, Q^2) &=  A_q^{\tau2} \; 
\text{e}^{- d_{-}(1) s_\text{LO}}
 ~ + ~ {\mathcal{O}}\left( z \right ) \ ,
\label{fq-:LO}
\end{align}
\end{subequations}
where
\begin{equation}
\bar d_{+}(1) = 1 + \dfrac{8T_Rf}{3\beta_0}
 \left(1 - \dfrac{C_F}{C_A}\right) ~,~~
d_{-}(1) = \dfrac{8C_FT_Rf}{3C_A\beta_0}
\label{bard_pm1}
\end{equation}
are the regular parts of $d_{+}$ and $d_{-}$ anomalous dimensions,
respectively, in the limit $n \to 1$.
\footnote{
From now on, for a quantity $k(n)$ we use the notation $\hat k(n)$ for the
singular part when $n\to1$ and $\overline k(n)$ for the corresponding
regular part.}

We define the variable
\begin{equation}
 s = \ln \left(\dfrac{a_s(Q^2_0)}{a_s(Q^2)}\right) \ .
\label{def:s}
\end{equation}

At LO, in terms of the QCD scale $\Lambda_\text{LO}$, it has the form:
\begin{equation}
 s_\text{LO} = \ln \left(\dfrac{\ln(Q^2/\Lambda^2_\text{LO})}
               {\ln(Q^2_0/\Lambda^2_\text{LO})}\right) \ .
\label{def:s_LO}
\end{equation}

The argument $\sigma_\text{LO}$ in the LO is given by
\footnote{
Hereafter we use the variables $\sigma_\text{LO}$ and $\rho_\text{LO}$,
introduced in Refs.~\cite{Ball:1994,Forte:1995,Ball:1997} for the case
$Q^2 \geq Q_0^2$. In our article, they are generalized to arbitrary
values of $Q^2$ and beyond the LO approximation (see below).
}
\begin{equation}
\sigma_\text{LO} = 2\sqrt{\hat d_{GG} s_\text{LO} \ln(z)} \ ,
\label{def:sigma_LO}
\end{equation}
where
\begin{equation}
\hat d_{GG} = -4 C_A / \beta_0 
\label{def:d_GG}
\end{equation}
is the singular part when $n \to 1$ of
$d_{GG}=\gamma^{(0)}_{GG}(n)/(2\beta_0)$ being $\gamma^{(0)}_{GG}(n)$ 
the LO coefficient of the gluon-gluon anomalous dimension.

The prescription for the backward evolution given by Eq.~(\ref{def:I_nu})
is the result, in the more general case, of the following representation
of the series which appear in the inverse Mellin transformation of the
exact solution for PD moments. (see for example Eq.~(6) in
Ref.~\cite{Kotikov:1999}),
\begin{equation}
\sum^{\infty}_{k=0} \dfrac{t^k}{k!\Gamma(k+\nu +1)} \ = \
 t^{-\nu/2} \widetilde I_{\nu}(2\sqrt{t})  \ \equiv \
 {\left|t\right|}^{-\nu/2}
 \left\{\begin{array}{rl}
  I_{\nu}(2\sqrt{\left|t\right|}), & \text{if} ~ t \geq 0 \ , \\
  J_{\nu}(2\sqrt{\left|t\right|}), & \text{if} ~ t  <   0 \ .
 \end{array} \right.
\label{ser:I_nu}
\end{equation}

And finally, in Eq.~(\ref{fq+:LO})
\begin{equation}
 \rho_\text{LO} = \sqrt{\dfrac{\hat d_{GG} s_\text{LO}}{\ln(z)}}
                = \dfrac{\sigma_\text{LO}}{2\ln(1/z)} \ ,
\label{def:rho_LO} 
\end{equation}

Let us note, that
\begin{equation}
 {\rho^{-\nu}} \; {\widetilde{I}_{\nu}(\sigma)} \, \to \,
 \dfrac{1}{\nu!} \; \ln^\nu \left(1/z\right)
 ~~~\text{at} ~~~ Q^2 \, \to \, Q_0^2 \ .
\label{lim:I_nu} 
\end{equation}

\subsection{
Next-to-leading order}
 \label{Sec2:NLO}

The small $x$ behavior of the twist-2 parton densities $f^{\tau2}_a$
($a = q, G$) and of $F^{\tau2
}_2$ at the NLO approximation
has been presented in our previous paper \cite{Kotikov:1999}.
Here we give the result that can also be used for $Q^2$ below the initial
condition point $Q^2_0$ (where partons have the flat form in $x$ as
Eq.~(\ref{flat}))
\begin{subequations}
\label{F2:tau2:NLO}
\begin{gather}
 F^{\tau2}_2(z, Q^2) \ = \ e \,
  \left( f^{\tau2}_q(z, Q^2) \ + \
  \dfrac{4T_Rf}{3} \, a_s(Q^2) \, f^{\tau2}_G(z, Q^2) \right) \ ,
\label{F2:NLO} \\
 f^{\tau2}_a(z, Q^2) \ = \
  f^{\tau2,+}_a(z,Q^2) \ + \ f^{\tau2,-}_a(z, Q^2) \ .
\label{f_a:NLO}
\end{gather}

The $'+'$ and $'-'$ PD components in the equations above are:
\begin{align}
 f^{\tau2,+}_G(z, Q^2) \ =& \ A_G^+(Q^2, Q^2_0) \, \widetilde I_0(\sigma) \,
  \exp(- \bar d_{+}(1) s - \bar D_{+}(1) p) ~ + ~ {\mathcal{O}}(\rho) \ ,
\label{fG+:NLO} \\
 f^{\tau2,+}_q(z, Q^2) \ =& \ A_q^+(Q^2, Q^2_0) \,
  \left[ \left(1 - \bar{d}_{+-}^q(1) a_s(Q^2) \right) \, \rho
  \widetilde I_1(\sigma) + \dfrac{20C_A}{3} \, a_s(Q^2) \,
  \widetilde I_0(\sigma) \right]
\nonumber \\
 \times&
 \exp(- \bar d_{+}(1) s - \bar D_{+}(1) p) ~ + ~ {\mathcal{O}}(\rho) \ ,
\label{fq+:NLO} \\
 f^{\tau2,-}_a(z, Q^2) \ =& \ A_a^-(Q^2, Q^2_0) \,
  \exp(- d_{-}(1) s - D_{-}(1) p) ~ + ~ {\mathcal{O}}(z) \ ,
\label{fa-:NLO}
\end{align}
\end{subequations}
where $D_\pm(n) = d_{\pm\pm}(n) - (\beta_1/\beta_0)d_\pm(n)$;
$p = a_s(Q^2_0) - a_s(Q^2)$ and
\begin{equation}
 \sigma = 2 \sqrt{(\hat d_{+} s + \hat D_{+} p) \ln(z)} ~~ , ~~~~
 \rho = \sqrt{\dfrac{(\hat d_{+} s + \hat D_{+} p)}{\ln(z)}} =
        \dfrac{\sigma}{2\ln(1/z)} \ .
\label{def:s-rho}
\end{equation}
\begin{subequations}
\label{A_a+-:NLO}
\begin{align}
A_G^+(Q^2, Q^2_0) &=
 \left[1 - \bar{d}_{+-}^G(1) a_s(Q^2) \right] A_G^{\tau2}
 + \dfrac{C_F}{C_A} \left[1 - d_{-+}^G(1) a_s(Q^2_0)
 - \bar{d}_{+-}^G(1) a_s(Q^2) \right] A_q^{\tau2} \ , 
\label{A_G+:NLO} \\
A_G^-(Q^2, Q^2_0) &= A_G^{\tau2} - A_G^+(Q^2, Q^2_0) \ ,
\label{A_G-:NLO} \\
A_q^+(Q^2, Q^2_0) &= \dfrac{2T_Rf}{3C_A}
 \left(A_G^{\tau2} + \dfrac{C_F}{C_A} A_q^{\tau2} \right) \ ,
\label{A_q+:NLO} \\
A_q^-(Q^2, Q^2_0) &= A_q^{\tau2} -
 \dfrac{20C_A}{3} \, a_s(Q^2_0) \, A_q^+(Q^2, Q^2_0) \ .
\label{A_q-:NLO}
\end{align}
\end{subequations}
The different singular and regular parts of anomalous dimensions appearing
in Eqs. (\ref{F2:tau2:NLO})--(\ref{def:s-rho}) have the form
\footnote{%
The original results of \cite{Kotikov:1999} contain
an error in the term $\bar d^q_{+-}(1)$, where the correct number
$23$  at $C_F=4/3$ and $C_A=3$ was mistakenly replaced by $134/3$. With the 
wrong number the value of  $\bar d^q_{+-}(1)$ was approximately $10$
times higher then in the Table~\ref{Tab:Const}.
However, the results of fits do not depend practically
on the mistake.%
}:
\begin{subequations}
\label{sr-part:NLO}
\begin{align}
\hat{d}_{++} &= \dfrac{8T_Rf}{9\beta_0} \left(23C_A - 26C_F\right) \ , \ \
\hat{d}_{+-}^q = - \dfrac{20C_A}{3} \ , \ \
\hat{d}_{+-}^G = 0 \ ,
\label{hatd++} \\
\bar{d}_{++}(1) &= \dfrac{8}{3\beta_0}\biggl[
 \dfrac{C_A^2}{3}\left(36\zeta(3) + 33\zeta(2) - \dfrac{1643}{12}\right)
\nonumber \\
 &-\left(4C_F\zeta(2) + \dfrac{86}{9}C_A
  -\dfrac{547}{18}C_F + 3\dfrac{C_F^2}{C_A}\right)T_R f
  -\dfrac{26C_F}{9C_A} \left(1 - 2\dfrac{C_F}{C_A}\right)T_R^2 f^2
 \biggr] \ ,
\label{bard++} \\
\bar d_{+-}^q(1) &= C_A\left(9 - 3\dfrac{C_F}{C_A} - 4\zeta(2)\right)
 - \dfrac{26}{9} \left(1 - 2\dfrac{C_F}{C_A}\right)T_R f \ , \ \
\bar d_{+-}^G(1) = \dfrac{40C_FT_Rf}{9C_A} \ ,
\label{bard+-} \\
d_{--}(1) &= \dfrac{4C_AC_F}{\beta_0} \left(1 - 2\dfrac{C_F}{C_A}\right)
 \left(2\zeta(3) - 3\zeta(2) + \dfrac{13}{4} + \dfrac{52T_R^2f^2}{27C_A^2}
 \right)
\nonumber \\
 &+\dfrac{8C_F}{3\beta_0}
 \left(4\zeta(2) - \dfrac{47}{18} + 3\dfrac{C_F}{C_A}\right) T_R f \ ,
\label{d--} \\
d_{-+}^q(1) &= 0 \ , \ \
d_{-+}^G(1) = - \left(C_A
 + \dfrac{2}{3}\left(1 - 2\dfrac{C_F}{C_A}\right)T_R f\right) \ .
\label{bard-+}
\end{align}
\end{subequations}
The corresponding numerical values are collected in
Table~\ref{Tab:Const} (see Ref.~\cite{Kotikov:1999} for details).

\begin{table}
\caption{\label{Tab:Const}\sffamily
The values of the parameters used in the calculation of the parton
distributions as a function of the number of flavors.}
\centering
\vspace{0.3cm}
\begin{tabular}{|l||l|l||l|c||l|c||c|l|l|} \hline\hline
$f$ & ~~~~$\hat d_+$ & \hspace{7mm}$\hat D_+$ & ~~$\bar d_+(1)$
 & $\bar D_+(1)$ & ~$d_-(1)$ & $ D_-(1)$
 & $\bar d_{+-}^q(1)$ & $\bar d_{+-}^G(1)$ & $d_{-+}^G(1)$  \\ \hline
3 & $-$4/3 & 1180/81 & 101/81 &  $-$43.370269 & 16/81 
  & 1.974431 & 2.779310 & 80/27  & $-$29/9  \\
4 & $-$36/25 & 91096/5625 & 61/45  & $-$45.485532 & 64/225 
  & 3.108220 & 2.618816 & 320/81 & $-$89/27 \\
5 & $-$36/23 & 84964/4761 & 307/207  & $-$47.729779 & 80/207
  & 4.674958 & 2.458322 & 400/81 & $-$91/27 \\
6 & $-$12/7 & 8576/441 & 103/63  & $-$50.057345 & 32/63 
  & 6.864360 & 2.297828 & 160/27 & $-$31/9  \\
\hline\hline
\end{tabular}
\end{table}

We would like to note that the exact value of the variable $\sigma$ and the 
small $x$ asymptotics of the modified Bessel function
$$
 I_{\nu}(\sigma) ~\sim~ \exp{(\sigma)} ~~~\text{at} ~~~ \sigma \to \infty
$$
are given in Introduction (see Eq.~(\ref{intro:1})) with
$|\hat d_+| = a_+$ and $\hat D_+ = b_+ + a_+ (\beta_1/\beta_0)$.
So, the most important part from the NLO corrections (i.e. the singlet part
at $x \to 0$) is taken in a proper way: it comes directly into the argument
of the Bessel functions and does not spoil the applicability of perturbation
theory at low $x$ values.

We stress that the LO and NLO results given above coincide with
the ones in Ref.~\cite{Kotikov:1999} for positive values of $s$ and
$s_{LO}$ (i.e. for the case $Q^2 \geq Q^2_0$). 

Let us remind that these analytical expressions which have been
obtained from the exact solution to the moment space DGLAP evolution
equations in the asymptotic limit $n \to 1$ have been already used
in Ref.~\cite{Kotikov:1999} to reproduce the small $x$ behavior of
parton distributions and lastly of DIS structure functions themselves.
The consideration of negative values for $s$ and $s_{LO}$
leads us to apply the backward evolution in the present analysis and,
thus, to have the possibility to choose any normalization point
$Q^2_0$ and not only the low end of the $Q^2$-evolution as it was
done in Ref.~\cite{Kotikov:1999}.

\section{%
The contribution of twist-two operators to the derivative
  $\partial F_2/\partial\ln Q^2$}
 \label{Sec:3}

In QCD the scaling violation of $F_2(z, Q^2)$ are caused by gluon bremsstahlung
from quarks and quark pair creation from gluons. In the low $x$ domain the
latter process dominates the scaling violations. $F_2$ is then largely determined
by the sea quarks, whereas the $\partial F_2/\partial\ln Q^2$ is dominated by the
convolution of the splitting function $P_{qG}$ and the gluon density.
At the leading twist approximation the derivative
$\partial F_2/\partial\ln Q^2$ relates strongly to the gluon distribution
$f_G^{\tau2}(z, Q^2)$. Moreover, the derivative is measured with a good
accuracy. Then, the $\partial F_2/\partial\ln Q^2$ experimental data can be
successfully used to determine the characteristic properties of gluon 
distribution.

The $\partial F_2/\partial\ln Q^2$ data becomes even more important, when we
add higher twist corrections into consideration. In the case of the twist-four
terms (of sum of the twist-four and twist-six terms) in the renormalon model
there are
the two (four) additional parameters (see below
Section~\ref{Sec:4}) which may lead to
problems to fit all them together only
with help of $F_2$ experimental data.

\subsection{Leading order}
 \label{Sec3:LO}

Note that at the LO approximation there are the following properties
\begin{subequations}
\label{Der:Bessel:LO}
\begin{align}
\dfrac{\partial }{\partial \ln Q^2} \biggl[
 \dfrac{1}{\rho^{k}_\text{LO}} \widetilde{I}_k(\sigma_\text{LO}) \biggr]
 \ &= \ 4C_A \, a_s(Q^2) \,
  \dfrac{1}{\rho^{k+1}_\text{LO}} \widetilde{I}_{k+1}(\sigma_\text{LO}) \ ,
\label{Der:Bessel1} \\
\dfrac{\partial }{\partial \ln Q^2} \biggl[
 \rho^{k}_\text{LO} \widetilde{I}_k(\sigma_\text{LO}) \biggr]
\ &= \  4C_A \, a_s(Q^2) \,
  \rho^{k-1}_\text{LO} \widetilde{I}_{|k-1|}(\sigma_\text{LO})
 ~~~~(k = 0,1,2, \ldots) \ ,
\label{Der:Bessel2}
\end{align}
\end{subequations}
which lead to the following results
\begin{subequations}
\label{Der:GQ_LO}
\begin{align}
\dfrac{\partial f^{\tau2,+}_{G,\text{LO}}(z, Q^2)}{\partial \ln Q^2}  
 &= a_s(Q^2) \, \biggl[ \dfrac{4C_A}{ \rho_\text{LO}} \;
 \dfrac{\widetilde{I}_1(\sigma_\text{LO})}{\widetilde{I}_0(\sigma_\text{LO})} \;
 - \, \beta_0 \, \bar d_{+}(1) \biggr] \;  f^{\tau2,+}_{G,\text{LO}}(z, Q^2)
 ~ + ~ {\mathcal{O}}\left( \rho_\text{LO}\right ) \ ,
\label{dfG+:LO} \\
\dfrac{\partial f^{\tau2,+}_{q,\text{LO}}(z, Q^2)}{\partial \ln Q^2}
 &= a_s(Q^2) \, \biggl[ \dfrac{8T_Rf}{3} \; f^{\tau2,+}_{G,\text{LO}}(z, Q^2) - 
 \beta_0 \, \bar d_{+}(1) f^{\tau2,+}_{q,\text{LO}}(z, Q^2) \biggr]
 ~ + ~ {\mathcal{O}}\left( \rho_\text{LO}\right ) \ ,
\label{dfq+:LO} \\
\dfrac{\partial f^{\tau2,-}_{G,\text{LO}}(z, Q^2)}{\partial \ln Q^2} 
&= - a_s(Q^2) \, \dfrac{8C_F T_Rf}{3C_A} \; f^{\tau2,-}_{G,\text{LO}}(z, Q^2)
 ~ + ~ {\mathcal{O}}\left( z \right ) \ ,
\label{dfG-:LO} \\
\dfrac{f^{\tau2,-}_{q,\text{LO}}(z, Q^2)}{\partial \ln Q^2} 
&= a_s(Q^2) \, \dfrac{8 T_Rf}{3} \; f^{\tau2,-}_{G,\text{LO}}(z, Q^2)
 ~ + ~ {\mathcal{O}}\left( z \right ) \ .
\label{dfq-:LO}
\end{align}
\end{subequations}


Thus, we have 
\begin{equation}
\dfrac{\partial F^{\tau2}_{2, \text{LO}}(z, Q^2)}{\partial \ln Q^2} \ = \  e \,
 \dfrac{\partial f_{q, \text{LO}}^{\tau2}(z, Q^2)}{\partial \ln Q^2} 
\ = \
 e \, a_s(Q^2) \, \biggl[
 \dfrac{8T_Rf}{3} \, f_{G, \text{LO}}^{\tau2} (z, Q^2)  - 
  \beta_0 \, \bar d_{+}(1) 
f^{\tau2,+}_{q,\text{LO}}(z, Q^2)
\biggr]
\ 
\label{dlnQ:LO:1}
\end{equation}

The LO $Q^2$ evolution of the derivative
$\partial F_2^{\tau2}/\partial\ln Q^2$ is defined mostly
by the corresponding evolution of the gluon distribution
$f_{G, \text{LO}}^{\tau2}(z, Q^2)$,
i.e. by the Eqs.~(\ref{f_a:LO}, \ref{fG+:LO}) and (\ref{fG-:LO}).

\subsection{
Next-to-leading order}
 \label{Sec3:NLO}

At the NLO approximation of perturbation theory  the Eqs. 
(\ref{Der:Bessel:LO}) are replaced by
\begin{subequations}
\label{Der:Bessel:NLO}
\begin{align}
\dfrac{\partial }{\partial \ln Q^2} \biggl[
 \dfrac{1}{\rho^{k}} \widetilde{I}_k(\sigma) \biggr]
 \ &= \  a_s(Q^2) \, \Bigl[ 4C_A -  a_s(Q^2) \beta_0 \hat d_{++}
 \Bigr] \,
  \dfrac{1}{\rho^{k+1}} \widetilde{I}_{k+1}(\sigma) \ ,
\label{Der:BesselNLO1} \\
\dfrac{\partial }{\partial \ln Q^2} \biggl[
 \rho^{k} \widetilde{I}_k(\sigma) \biggr]
\ &= \   a_s(Q^2)  \, \Bigl[ 4C_A -  a_s(Q^2) \beta_0 \hat d_{++}
 \Bigr] \,
  \rho^{k-1} \widetilde{I}_{|k-1|}(\sigma)
 ~~~~(k = 0,1,2, \ldots) \ ,
\label{Der:BesselNLO2}
\end{align}
\end{subequations}
which leads to the following results
\begin{align}
\dfrac{\partial f^{\tau2,+}_{q}(z, Q^2)}{\partial \ln Q^2}
 &=  a_s(Q^2) \,
\dfrac{2T_Rf}{3C_A} \;
 \left(A_G^{\tau2} + \dfrac{C_F}{C_A} A_q^{\tau2} \right) \;
\Biggl[ 4C_A \widetilde I_0(\sigma)\;
 - \, \beta_0 \, \bar d_{+}(1) \rho \; 
\widetilde I_1(\sigma) \ \nonumber \\
&+  a_s(Q^2) \, \biggl\{
\dfrac{80}{3} \;
\dfrac{C^2_A}{ \rho} \; \widetilde
 I_1(\sigma) \; - \,
\biggl( \beta_0 \, \left[ \hat d_{++} \, + \, \dfrac{20}{3} \, C_A
\Bigl(1+ \bar d_{+}(1) \Bigr) \right] \, - \, 4C_A \bar d^q_{+-}(1)
\biggr) \, \widetilde I_0(\sigma) \; \ \nonumber \\
&+ \; \beta_0 \, \biggl( \bar d^q_{+-}(1) \Bigl(1+ \bar d_{+}(1) \Bigr) -
\bar d_{++}(1) \biggr) \; \rho \; \widetilde I_1(\sigma) 
\biggr\}
\Biggr] \;
 \exp(- \bar d_{+}(1) s - \bar D_{+}(1) p)
 ~ + ~ {\mathcal{O}}\left( \rho\right ) \  ,
\label{dfq+:NLO} \\
\dfrac{f^{\tau2,-}_{q}(z, Q^2)}{\partial \ln Q^2} 
&= - \beta_0 \, a_s(Q^2) \; 
\Biggl[  A_q^{\tau2} \, \Bigl(d_{-}(1) \, + \, a_s(Q^2)  \, d_{--}(1) \Bigr)
 \ \nonumber \\
& - \, 
\dfrac{40T_Rf}{9} \, a_s(Q^2_0) \; d_{-}(1) \,
 \left(A_G^{\tau2} + \dfrac{C_F}{C_A} A_q^{\tau2} \right) 
\Biggr] \;
 \exp(- d_{-}(1) s - D_{-}(1) p)
 ~ + ~ {\mathcal{O}}\left( z \right )  \ .
\label{dfq-:NLO}
\end{align}


Taking together equations (\ref{F2:tau2:NLO}), (\ref{dfG+:LO}), 
(\ref{dfG-:LO}), (\ref{dfq+:NLO}) and (\ref{dfq-:NLO}), after some algebra
we have got the final result
\begin{align}
\dfrac{\partial F_2^{\tau2}(z, Q^2)}{\partial \ln Q^2} \ = \ &
 e \, a_s(Q^2) \, \biggl[  \dfrac{8T_Rf}{3} \,\Bigl( f_G^{\tau2}(z, Q^2) 
\, + \, \Phi(z, Q^2) \Bigr) \, -
 \,   \beta_0 \, \bar d_{+}(1) f^{\tau2,+}_{q}(z, Q^2) \,  \ \nonumber \\
& - \,  a_s(Q^2) \, \beta_0 \, d_{--}(1) f^{\tau2,-}_{q}(z, Q^2)
\biggr] \ ,
\label{dlnQ:NLO:1}
\end{align}
where 
\begin{subequations}
\label{Phi:NLO}
\begin{gather}
 \Phi(z, Q^2) \ = \  \Phi^{+}(z, Q^2) \ + \ \Phi^{-}(z, Q^2) \ ,
\label{Phi+-} \\
 \Phi^{+}(z, Q^2) \ = \ \phi^{+}(z, Q^2) \,
  \exp(- \bar d_{+}(1) s - \bar D_{+}(1) p) \ + \ \mathcal{O}(\rho) \ ,
\label{Phi+} \\
 \Phi^{-}(z, Q^2) \ = \ \phi^{-}(z, Q^2) \,
  \exp(- d_{-}(1) s - D_{-}(1) p) \ + \ {\mathcal{O}}(z) \ .
\label{Phi-}
\end{gather}
The $'+'$ and $'-'$ components in the equations above are:
\begin{align}
 \phi^{+}(z, Q^2) \ =& \
  a_s(Q^2) \, 
 \left(A_G^{\tau2} + \dfrac{C_F}{C_A} A_q^{\tau2} \right) \;
  \biggl\{
\dfrac{26}{3}  \;
\dfrac{C_A}{ \rho} \; \widetilde I_1(\sigma) \; - \,
\biggl( \dfrac{\beta_0}{4C_A} \, \left[ \hat d_{++} \, 
+ \, \dfrac{2}{3} \, C_A \Bigl(13+ 3\bar d_{+}(1) \Bigr) \right] 
\, + \, \bar d^q_{+-}(1)  \ \nonumber \\
&  - \, \bar d^G_{+-}(1)
\biggr) \, \widetilde I_0(\sigma) \; + \; \dfrac{\beta_0}{4C_A} \, 
\biggl( \bar d^q_{+-}(1) - \bar d_{++}(1) \biggr) \; 
\rho \; \widetilde I_1(\sigma) 
\biggr\}  
\, + \; 
 a_s(Q^2_0) \,  A_q^{\tau2} \;  d^G_{-+}(1) \, \widetilde I_0(\sigma) \ ,
\label{phi+} \\
 \phi^{-}(z, Q^2) \ =& \
 \Bigl( a_s(Q^2_0) \, - \,  a_s(Q^2)\Bigr) \; 
  \biggl\{
 \bar d^G_{+-}(1) \; 
 \left(A_G^{\tau2} + \dfrac{C_F}{C_A} A_q^{\tau2} \right) \;
- \,  d^G_{-+}(1) \; \dfrac{C_F}{C_A} A_q^{\tau2}
\biggr\}
+ \, \dfrac{17C_F}{6} \,
 a_s(Q^2) \,  A_q^{\tau2}  \ .
\label{phi-}
\end{align}
\end{subequations}
%
The values of the coefficients are
given in Eqs.~(\ref{sr-part:NLO}).


Thus, the NLO $Q^2$ evolution of the derivative
$\partial F_2^{\tau2}/\partial \ln Q^2$ is defined mostly by the corresponding
evolution of  the gluon distribution $f_G^{\tau2}(z, Q^2)$,
i.e. by the Eqs.~(\ref{f_a:NLO}, \ref{fG+:NLO}) and (\ref{fa-:NLO}).

\section{The contribution of twist-two operators to the 
slopes of $F_2$ and of parton distributions}
 \label{Sec:4}

The behavior of $F_2$ and parton distributions
can mimic a power law shape over a limited region of $z, Q^2$:
 \begin{equation}
f_a(z, Q^2) \sim z^{-\lambda^\text{eff}_a(z, Q^2)}
 ~~~ \text{and} ~~~
F_2(z, Q^2) \sim z^{-\lambda^\text{eff}_{F2}(z, Q^2)} \ .
\label{power}
 \end{equation}
The slopes are effective ones because the 
parton distributions and $F_2$ have mostly the Bessel-like form.

Note that there are the following properties
\begin{subequations}
\label{Dz:Bessel}
\begin{align}
\dfrac{\partial}{\partial\ln(1/z)} \biggl[
 \dfrac{1}{\rho^{k}} \widetilde{I}_k(\sigma) \biggr]
 \ &= \
  \dfrac{1}{\rho^{k-1}} \widetilde{I}_{k-1}(\sigma) \ ,
\label{Dz:Bessel1} \\
\dfrac{\partial}{\partial \ln(1/z)} \biggl[
 \rho^{k} \widetilde{I}_k(\sigma) \biggr]
\ &= \
  \rho^{k+1} \widetilde{I}_{k+1}(\sigma)
 ~~~~(k = 0,1,2, \ldots) \ ,
\label{Dz:Bessel2}
\end{align}
\end{subequations}
which we will use below.

\subsection{Leading order}
 \label{Sec4:Slopes:LO}

The effective slopes have the form at the LO approximation
\begin{subequations}
\label{Slopes:LO}
\begin{gather}
\lambda^{\text{eff},\tau2}_{G,\text{LO}}(z,Q^2) \ = \
 \dfrac{f^{\tau2,+}_{G,\text{LO}}(z, Q^2)}{f^{\tau2}_{G,\text{LO}}(z, Q^2)}
 \, \rho_\text{LO} \,
  \dfrac{\widetilde{I}_1(\sigma_\text{LO})}
        {\widetilde{I}_0(\sigma_\text{LO})} \ ,
\label{G-Slope:LO} \\
\lambda^{\text{eff},\tau2}_{F2,\text{LO}}(z, Q^2) \ = \
 \lambda^{\text{eff},\tau2}_{q,\text{LO}}(z,Q^2) \ = \
 \dfrac{f^{\tau2,+}_{q,\text{LO}}(z, Q^2)}{f^{\tau2}_{q,\text{LO}}(z, Q^2)}
 \, \rho_\text{LO} \,
 \dfrac{\widetilde{I}_2(\sigma_\text{LO})}
       {\widetilde{I}_1(\sigma_\text{LO})} \ .
\label{q-Slope:LO}
\end{gather}
\end{subequations}

The effective slopes $\lambda^\text{eff}_a$ and $\lambda^\text{eff}_{F2}$
depend on the magnitudes $A^{\tau2}_a$ of the initial PD and also on the
chosen input values of $Q^2_0$ and $\Lambda$.
At quite large values of $Q^2 >> Q_0^2$, where the '$-$' component is not
relevant, the dependence on the magnitudes of the initial PD disappear,
having in this case for the asymptotic values:
\begin{subequations}
\label{as-Slopes:LO}
\begin{align}
 \lambda^{\text{eff},\tau2}_{G,\text{LO},\text{as}}(z, Q^2) \ =& \
  \rho_\text{LO} \, \dfrac{\widetilde{I}_1(\sigma_\text{LO})}{
\widetilde{I}_0(\sigma_\text{LO})}
  \approx \rho_\text{LO} - \dfrac{1}{4\ln{(1/z)}}  \ ,
\label{Gas-Slope:LO} \\
 \lambda^{\text{eff},\tau2}_{F2,\text{LO},\text{as}}(z, Q^2) \ =& \
 \lambda^{\text{eff},\tau2}_{q,\text{LO},\text{as}}(z,Q^2) \ = \
  \rho_\text{LO} \, \dfrac{\widetilde{I}_2(\sigma_\text{LO})}{
\widetilde{I}_1(\sigma_\text{LO})}
  \approx \rho_\text{LO} - \dfrac{3}{4\ln{(1/z)}} \ ,
\label{qas-Slope:LO}
\end{align}
\end{subequations}
where symbol $\approx$ marks approximations obtained by expansions of
modified Bessel functions ${I}_n(\sigma)$.
These approximations should be correct only at very large $\sigma $ values
(i.e. at very large $Q^2$ and/or very small $x$).
It is the case (see Figs. 2 and 6).

We would like to note that the slope
$\lambda^{\text{eff},\tau2}_{F2,\text{LO},\text{as}}(z, Q^2) = 
 \lambda^{\text{eff},\tau2}_{q,\text{LO},\text{as}}(z, Q^2)$ 
coincides at very large $\sigma $ with one obtained in
\cite{Navelet:1997} (see also \cite{Cooper-Sarkar:1998}) in the case of
flat input. Note that the slope
$\lambda^{\text{eff},\tau2}_{G,\text{LO},\text{as}}(z, Q^2)$
is large then the slope
$\lambda^{\text{eff},\tau2}_{F2,\text{LO},\text{as}}(z, Q^2) = 
 \lambda^{\text{eff},\tau2}_{q,\text{LO},\text{as}}(z, Q^2)$:
\begin{equation}
 \lambda^{\text{eff},\tau2}_{G,\text{LO},\text{as}}(z,Q^2) - 
 \lambda^{\text{eff},\tau2}_{F2,\text{LO},\text{as}}(z, Q^2) \ = \
 \rho_\text{LO} \,
 \left(\dfrac{\widetilde{I}_1(\sigma_\text{LO})}{
\widetilde{I}_0(\sigma_\text{LO})} -
       \dfrac{\widetilde{I}_2(\sigma_\text{LO})}{
\widetilde{I}_1(\sigma_\text{LO})} \right)
 \approx \dfrac{1}{2\ln{(1/z)}} \ ,
\label{Slopes:LO:Diff}
\end{equation}
that coincides with results of fits in Refs. \cite{Martin:2002,Gluck:1998}.

\subsection{
Next-to-leading order}
 \label{Sec4:Slopes:NLO}

At the NLO approximation of perturbation theory we have the following
properties of the effective slopes:
the quark and gluon ones $\lambda^{\text{eff},\tau2}_a(z, Q^2) =
 \partial\ln{f^{\tau2}_a(z, Q^2)}/\partial\ln{(1/z)}$
are reduced by the NLO terms that leads to the decreasing of the gluon
distribution at small $x$. For the quark case it is not the case, because
the normalization factor $A_q^{\tau2,+}$ of the '$+$' component produces an
additional contribution undamped as $\sim (\ln{z})^{-1}$.

Indeed, the effective slopes have the form,
\begin{subequations}
\label{Slopes:NLO}
\begin{align}
 \lambda^{\text{eff},\tau2}_G(z, Q^2) =&
  \dfrac{f^{\tau2,+}_G(z, Q^2)}{f^{\tau2}_G(z, Q^2)} \, \rho \,
  \dfrac{\widetilde{I}_1(\sigma)}{\widetilde{I}_0(\sigma)} \ ,
\label{G-Slope:NLO} \\
 \lambda^{\text{eff},\tau2}_q(z, Q^2) =&
  \dfrac{f^{\tau2,+}_q(z, Q^2)}{f^{\tau2}_q(z,Q^2)} \, \rho \,
  \dfrac{\widetilde{I}_2(\sigma)
  \left(1 - \bar{d}^q_{+-}(1) a_s(Q^2)\right)
 + (20C_A/3) a_s(Q^2) \widetilde{I}_1(\sigma)/\rho}
        {\widetilde{I}_1(\sigma)
  \left(1 - \bar{d}^q_{+-}(1) a_s(Q^2)\right)
 + (20C_A/3) a_s(Q^2) \widetilde{I}_0(\sigma)/\rho}
 \ ,
\label{q-Slope:NLO} \\
 \lambda^{\text{eff},\tau2}_{F2}(z, Q^2) =&
  \dfrac{\lambda^\text{eff}_q(z,Q^2) \, f^{\tau2}_q(z,Q^2) +
  (4T_Rf/3) \, a_s(Q^2) \, \lambda^\text{eff}_G(z, Q^2) \,
  f^{\tau2}_G(z, Q^2)}
  {f^{\tau2}_q(z, Q^2) + (4T_Rf/3) \, a_s(Q^2) \, f^{\tau2}_G(z,Q^2)}
 \ .
\label{F2-Slope:NLO}
\end{align}
\end{subequations}

The gluon effective slope $\lambda^{\text{eff},\tau2}_G(z, Q^2)$ is larger
than the quark slope $\lambda^{\text{eff},\tau2}_q(z, Q^2)$, which is in
excellent agreement with a recent MRS and GRV analysis
\cite{Martin:2002,Gluck:1998}.

For the asymptotic values we have got
\begin{subequations}
\label{as-Slopes:NLO}
 \begin{align}
\lambda^{\text{eff},\tau2}_{G,\text{as}}(z, Q^2) \ =& \
 \rho \, \dfrac{\widetilde{I}_1(\sigma)}{\widetilde{I}_0(\sigma)} \ 
\approx \  \rho - \dfrac{1}{4\ln{(1/z)}} \ ,
\label{Gas-Slope:NLO} \\
\lambda^{\text{eff},\tau2}_{q,\text{as}}(z, Q^2) \ =& \
 \rho \, \dfrac{\widetilde{I}_2(\sigma)
  \left(1 - \bar{d}^q_{+-}(1) a_s(Q^2)\right)
 + (20C_A/3) a_s(Q^2) \widetilde{I}_1(\sigma)/\rho}
               {\widetilde{I}_1(\sigma)
  \left(1 - \bar{d}^q_{+-}(1) a_s(Q^2)\right)
 + (20C_A/3) a_s(Q^2) \widetilde{I}_0(\sigma)/\rho}
\nonumber \\
 =& \ \rho \, \dfrac{\widetilde{I}_2(\sigma)}{\widetilde{I}_1(\sigma)}
 + \dfrac{20C_A}{3} \alpha(Q^2) \left(1 -
   \dfrac{\widetilde{I}_0(\sigma) \widetilde{I}_2(\sigma)}
         {\widetilde{I}_1^2(\sigma)} \right) \ \approx \
 \rho - \dfrac{3}{4\ln{(1/z)}} 
 + \dfrac{10C_A}{3} \dfrac{a_s(Q^2)}{\rho \ln(1/z)} \ ,
\label{qas-Slope:NLO} \\
\lambda^{\text{eff},\tau2}_{F2,\text{as}}(z, Q^2) \ =& \
 \rho \, \dfrac{\widetilde{I}_2(\sigma)}{\widetilde{I}_1(\sigma)}
 + \dfrac{26C_A}{3} \alpha(Q^2) \left(1 -
   \dfrac{\widetilde{I}_0(\sigma) \widetilde{I}_2(\sigma)}
         {\widetilde{I}_1^2(\sigma)} \right)  \ = \
\lambda^{\text{eff},\tau2}_{q,\text{as}}(z, Q^2) + 2C_A \, a_s(Q^2)
\left(1-
   \dfrac{\widetilde{I}_0(\sigma) \widetilde{I}_2(\sigma)}
         {\widetilde{I}_1^2(\sigma)} \right)
\nonumber \\
 \ \approx & \ \rho - \dfrac{3}{4\ln{(1/z)}} 
 + \dfrac{13C_A}{3} \dfrac{a_s(Q^2)}{\rho \ln(1/z)} \ = \
 \lambda^{\text{eff},\tau2}_{q,\text{as}}(z, Q^2) + 
 \dfrac{C_A a_s(Q^2)}{\rho \ln(1/z)}.
\label{F2as-Slope:NLO}
\end{align}
\end{subequations}

We would like to note that at the NLO approximation the slope 
$\lambda^{\text{eff},\tau2}_{F2,\text{as}}(z,Q^2)$ lies between quark and
gluon ones but closely to quark slope 
$\lambda^{\text{eff},\tau2}_{q,\text{as}}(z,Q^2)$, that is in agreement
with Refs. \cite{Martin:2002,Gluck:1998}.

Indeed,
\begin{subequations}
\label{Slopes:NLO:F2-a}
\begin{align}
 \lambda^{\text{eff},\tau2}_{G,\text{as}}(z, Q^2) \ - \
 \lambda^{\text{eff},\tau2}_{F2,\text{as}}(z, Q^2) \ =& \
 \left(\rho \, \dfrac{\widetilde{I}_1(\sigma)}{\widetilde{I}_0(\sigma)}
 + \dfrac{26C_A}{3} a_s(Q^2) \right)
\left(1 -
   \dfrac{\widetilde{I}_0(\sigma) \widetilde{I}_2(\sigma)}
         {\widetilde{I}_1^2(\sigma)} \right)
\nonumber \\
 \ \approx& \ \left(\rho - \dfrac{1}{4\ln{(1/z)}}
 + \dfrac{26C_A}{3} a_s(Q^2)\right) \dfrac{1}{2 \rho \ln{(1/z)}} \ ,
\label{Slopes:NLO:G-F2} \\
 \lambda^{\text{eff},\tau2}_{F2,\text{as}}(z, Q^2) \ - \
 \lambda^{\text{eff},\tau2}_{q,\text{as}}(z, Q^2) \ =& \
 2C_A \, a_s(Q^2) \left(1 -
   \dfrac{\widetilde{I}_0(\sigma) \widetilde{I}_2(\sigma)}
         {\widetilde{I}_1^2(\sigma)} \right) \ \approx \
 \dfrac{C_A a_s(Q^2)}{\rho \ln(1/z)} \ .
\label{Slopes:NLO:F2-q}
\end{align}
\end{subequations}

Both slopes $\lambda^{\text{eff},\tau2}_a(z, Q^2)$ decrease with decreasing
$z$. A $z$ dependence of the slope should not appear for a PD within a
Regge type asymptotic ($x^{-\lambda}$) and precise measurement of the slope 
$\lambda^{\text{eff},\tau2}_a(z, Q^2)$ may lead to the possibility to
verify the
type of small $x$ asymptotics of parton distributions.
The present data, however, are not enough to distinguish this slow
$x$-dependence of $\lambda^{\text{eff},\tau2}_a(z, Q^2)$ (see Fig. 2).

In the following Sections we study the higher-twist contributions to
 $F_2(x,Q^2)$, its derivatives  and parton distributions. 

\section{The higher twist contributions for $F_2$}
 \label{Sec:5}

In the Section 
we consider two different representations for twist-four effects. 
The first one comes from Regge-like analysis
\cite{Levin:1992,Bartels:1991,Bartels:1993:PLB,Bartels:1993:ZP}. Thus, 
it should have right asymptotics at $x \to 0$ limit,
but, unfortunately, the knowledge of its form is very restricted.

The second one is based on the IR-renormalon model. The predictions can
not reproduce the exact form of $x \to 0$
asymptotics, calculated in
Ref.~\cite{Levin:1992,Bartels:1991,Bartels:1993:PLB,Bartels:1993:ZP}
but gives rather good agreement with modern experimental data from HERA
(see Section~\ref{Sec:10}). We think this agreement is similar to one 
(see Ref. \cite{Cooper-Sarkar:1998}) at larger $Q^2$ values between DGLAP
approach (even for its analytical simplification: the generalized DAS
approach \cite{Kotikov:1999}) and experiment.

We would like to note here that in the analysis of experimental data
performed below we consider both LO and NLO approximations in the 
twist-two case and for HT corrections in the renormalon case.
In the BFKL-motivated approach, for simplicity
\footnote{
This simplification is connected also with quite poor present knowledge
about HT
contributions in the BFKL-motivated approach.}
we restrict the calculation of the HT
contribution to the
consideration of LO $Q^2$ evolution alone.

\subsection{BFKL-motivated estimations for twist-four operators}
 \label{Sec5:BFKL}

Twist-four operators are known \cite{Bukhvostov:1985} to have their own
evolution equations but the diagonalization of the operator anomalous
dimensions matrix is a very complicate problem. For our purpose, however,
as the relevant limit is  $n \to 1$, one can apply the results of
Refs.~\cite{Levin:1992,Bartels:1991,Bartels:1993:PLB,Bartels:1993:ZP},
which have very simple form and are given in the classical DAS
asymptotics considered in Section 2 of
Ref.~\cite{Kotikov:1999}

Here we show that the contribution from twist-four operators can be
represented in the same form as the twist-two operators by using the
twist-four anomalous dimensions instead of the twist-two ones.

For the singular part of twist-four anomalous dimensions we consider from
Ref.~\cite{Levin:1992} the result:
\begin{equation}
 \gamma^{\tau4}_{GG}(n - 1) \ = \
  2 \, \gamma^{(0)}_{GG} \! \left((n - 1)/2 \right) \, (1 + \varepsilon) \ ,
\label{BFKL:gt4}
 \end{equation}
where $\varepsilon $ is very small: $\varepsilon = 1/1224$.

Eq.~(\ref{BFKL:gt4}) allows us to find the relation between the singular 
part of twist-four operators anomalous dimensions, $\gamma^{\tau4}_{ab}(n)$
and $\gamma^{\tau4}_{\pm}(n)$, with the twist-two ones,
$\gamma^{(0)}_{ab}(n)$ and $\gamma^{(0)}_{\pm}(n)$.
It leads to the following relations:
\begin{equation}
 \hat d^{\tau4}_{+} \ = \ \hat d^{\tau4}_{GG} \ = \
  a^2 \hat d_{+}  \ = \  a^2 \hat d_{GG} \ ,~~~~~
 \hat d^{\tau4}_{-} \ = \   a^2 \hat d_{-}  \ = \  0 \ ,
\label{BFKL:rel}
\end{equation}
where $a^2 = 4(1 + \varepsilon)$ and $\hat d_{+} = \hat d_{GG}$ is
given by Eq.~(\ref{def:d_GG}).

The prediction for the regular parts $\overline d^{\tau4}_{+}(n)$ and
$d^{\tau4}_{-}(n)$ can not
be obtained from Eq. (\ref{BFKL:gt4}), but it should be essentially less
important in the kinematical range studied below. 
Then, in the analysis
presented below, we proceed by fixing this non-singular part by means of a
relation similar to Eq.~(\ref{BFKL:rel}):
\begin{equation}
 \bar d^{\tau4}_{+}(1) \ = \ b \, \bar d_{+}(1), ~~~~~ 
 d^{\tau4}_{-}(1) \ = \ b \, d_{-}(1), 
\label{BFKL:b}
\end{equation}
and further we examine different ``natural'' choices of $b$:
$b = 0, 1$ and $a^2/2$.

Note that the non-singular (when $n \to 1$) parts $\bar d^{\tau4}_{+}(1)$,
$d^{\tau4}_{-}(1)$ and $\bar d_{+}(1)$, $d_{-}(1)$ determine the
behavior of parton distributions 
and DIS structure functions at non-small $x$ values. Usually fits to
experimental data at intermediate and large values of $x$ are performed
with the help of the following forms for the structure function $F_2$:
\begin{gather}
 F_2(x, Q^2) \ = \ F_2^{\tau2}(x, Q^2) \ + \
  \dfrac{1}{Q^2} \, F_2^{\tau4}(x) ~~~ \text{or}
\label{def:F2tw4} \\
 F_2(x, Q^2) \ = \ F_2^{\tau2}(x, Q^2) \left(
  1 \ + \ \dfrac{1}{Q^2} \, f_2^{\tau4}(x) \right)
\label{def:f2tw4}
\end{gather}
with $Q^2$-independent functions $F_2^{\tau4}(x)$ or $f_2^{\tau4}(x)$.

In fact Eq.~(\ref{def:F2tw4}) is closed to our choice $b = 0$, i.e.
the twist-4 contribution does not evolve logarithmically with $Q^2$.
Also Eq.~(\ref{def:f2tw4}) is analogous to the choice $b = 1$, i.e.
twist-two and twist-four operators have the same logarithmic
$Q^2$-dependence at large and intermediate $x$ values.
Lastly, the choice $b = a^2/2$ corresponds to the hypothese
about applicability of Eq.~(\ref{BFKL:gt4}), obtained in the classical DAS
limit, to a more wide generalized DAS
approximation considered here.

By analogy with Section~\ref{Sec:2} we represent the twist-four contribution
split in the $'+'$ and $'-'$ parts:
\begin{subequations}
\label{F2:BFKL}
\begin{gather}
 F^{\tau4}_2(z, Q^2)  \ = \  e \, f^{\tau4}_q(z, Q^2) \ ,
\label{F2:BFKL1} \\
 f^{\tau4}_a(z, Q^2) \ = \
  f^{\tau4,+}_a(z, Q^2) \ + \ f^{\tau4,-}_a(z, Q^2) \ .
\label{f_a:BFKL}
\end{gather}
The $'+'$ and $'-'$ PD components are:
\begin{align}
 f^{\tau4,+}_G(z, Q^2) &=
  \left(A^{\tau4}_G + \dfrac{C_F}{C_A} A^{\tau4}_q \right) \,
  \widetilde I_0(a \, \sigma_\text{LO}) \,
  \text{e}^{-b \bar d_{+}(1) s_\text{LO}}
  ~ + ~ {\mathcal{O}}\left( \rho_\text{LO}\right ) \ ,
\label{fG+:BFKL} \\
 f^{\tau4,+}_q(z, Q^2) &= \dfrac{2T_Rf}{3C_A} \,
  \left(A^{\tau4}_G + \dfrac{C_F}{C_A} A^{\tau4}_q \right) \,
  \dfrac{b}{a} \rho_\text{LO} \, \widetilde I_1(a \, \sigma_\text{LO}) \,
  \text{e}^{-b \bar d_{+}(1) s_\text{LO}}
 ~ + ~ {\mathcal{O}}\left( \rho_\text{LO}\right ) \ ,
\label{fq+:BFKL} \\
 f^{\tau4,-}_G(z, Q^2) &= - \dfrac{C_F}{C_A} \, A_q^{\tau4} \,
\text{e}^{- b d_{-}(1) s_\text{LO}} ~ + ~ {\mathcal{O}}\left( z \right ) \ ,
\label{fG-:BFKL} \\
f^{\tau4,-}_q(z, Q^2) &=  A_q^{\tau4} \ \text{e}^{- b d_{-}(1) s_\text{LO}}
 ~ + ~ {\mathcal{O}}\left( z \right ) \ ,
\label{fq-:BFKL}
\end{align}
\end{subequations}
because the corresponding twist-four projectors (see ~\cite{Kotikov:1993})
have the following form \footnote{
The projectors $\varepsilon^{\tau4,\pm}_{ab}$ can be obtained from
Eq.~(10) in Ref.~\cite{Kotikov:1999} with the replacement $d_{\pm}(n) \to
d^{\tau4}_{\pm}(n) = \hat d^{\tau4}/(n-1) + \bar d^{\tau4}(n)$.
}:
\begin{equation}
\begin{split}
 \varepsilon^{\tau4,+}_{qq} \ = \
 \varepsilon^{\tau4,-}_{GG} \ = \
 \varepsilon^{+}_{qq} \, \dfrac{b}{a^2} ~~ &, ~~~
 \varepsilon^{\tau4,-}_{aa} \ = \
 1 - \varepsilon^{\tau4,+}_{aa} \ ,
\\
 \varepsilon^{\tau4,\pm}_{qG} \ = \
 \varepsilon^{\pm}_{qG} \, \dfrac{b}{a^2} ~~ &, ~~~
 \varepsilon^{\tau4,\pm}_{Gq} \ = \
 \varepsilon^{\pm}_{Gq} \ .
\end{split}
\label{BFKL:proj}
\end{equation}

In Eqs. (\ref{fG+:BFKL}--\ref{fq-:BFKL}) the twist-four parameters
$A_a^{\tau4}$ ($a = q, G$) have to be determined from fits to experimental
data.

The full contribution (i.e. the sum of twist-two and twist-four parts)
is given by:
\begin{gather}
 f_a(z,Q^2) \ = \ f^{\tau2}_a(z, Q^2) \ + \
  \dfrac{1}{Q^2} \, f^{\tau4}_a(z, Q^2) ~~~ \text{and}
\\
 F_2(z,Q^2)  \ = \ F^{\tau2}_2(z,Q^2) \ + \
 \dfrac{1}{Q^2} \, F^{\tau4}_2(z, Q^2) \ ,
\label{tw2+BFKL}
\end{gather}
where the leading twist contributions $f^{\tau2}_a(z, Q^2)$ and 
$F^{\tau2}_2(z, Q^2)$ are given at LO by Eqs.~(\ref{F2:LO},\ref{f_a:LO})
and at NLO by Eqs.~(\ref{F2:NLO},\ref{f_a:NLO}).

\subsection{Renormalon model predictions for twist-four operators}
 \label{Sec5:tw4}

The full small $x$ asymptotic results for parton densities
and $F_2$ structure function 
in the framework of the infrared renormalon model, i.e. $F^{R}_2$,
at LO of perturbation theory in the twist-four part:
\begin{equation}
\begin{split}
 F^{R}_2(z, Q^2)   \ = \ F^{\tau2}_2(z, & \ Q^2) \, + \,
 \dfrac{1}{Q^2} \, F^{R\tau4}_2(z, Q^2) \ ,
\end{split}
\label{F2:LO:Ren}
\end{equation}
where 
$F^{\tau2}_2(z, Q^2)$ is given by Eqs.~(\ref{F2:LO}--\ref{fq-:LO}) at
the LO approximation and by Eqs.~(\ref{F2:NLO}--\ref{fa-:NLO}) at NLO
one, respectively. The twist-four term $F^{R\tau4}_2(z, Q^2)$ has the form
(\ref{Ren:tw4In}), i.e.
\begin{equation}
 \dfrac{1}{e} \, F^{R\tau4}_2(z, Q^2) \ = \ \sum_{a=q,G} 
 {\mathrm a}_a^{\tau4} \,
 \widetilde{\mu}^{\tau4}_a (z, Q^2) \otimes f^{\tau2}_a(z, Q^2) \ ,
\nonumber \\
\end{equation}
where the symbol $\otimes$ marks the Mellin convolution
\begin{equation}
A(z) \otimes B(z) \ = \
 \int_z^1 \dfrac{dy}{y} \ A(y) \ B\left(\dfrac{z}{y}\right) \ .
\label{def:Mellin}
\end{equation}

The corresponding Mellin transforms of 
$\widetilde{\mu}^{\tau 4,6}_a(z, Q^2)$
\begin{equation}
 \mu^{\tau 4,6}_a(n, Q^2) \ = \
  \int_0^1 dz \, z^{n-1} \, \widetilde{\mu}^{\tau 4,6}_a (z, Q^2)
\label{MTtw4}
\end{equation}
are presented in the Appendix~A (see Eqs.~(\ref{App:muq}--\ref{App:bq:4})
and (\ref{App:muG})).

Looking the $n$-space representations for renormalon power-like corrections
given in Appendix~A and applying the technique to transform the Mellin
convolutions to standard products at small $x$
(see \cite{Kotikov:1994:YF,Kotikov:1994:PRD} and Appendix~B)
we can represent the Eq.~(\ref{Ren:tw4In}) in the form
\begin{align}
 \dfrac{1}{e} \, F^{R\tau4}_2(z, Q^2) = \dfrac{64T_Rf}{15\beta_0^2}
 \Biggl[ &{\mathrm a}_G^{\tau4} \left\{
 \widehat{\delta}^{-1} + \dfrac{101}{120} + \dfrac{1}{2}
 \ln \left(\dfrac{Q^2}{\left|{\mathrm a}_G^{\tau4}\right|}\right)
 \right\} f^{\tau2}_G(z, Q^2)
\nonumber \\
 +\,  2C_F &{\mathrm a}_q^{\tau4} \left\{ \widehat{\delta}^{-2}
 + \dfrac{11}{120} \widehat{\delta}^{-1} - \dfrac{2291}{3600} 
 + \dfrac{1}{2}
 \ln \left(\dfrac{Q^2}{\left|{\mathrm a}_q^{\tau4}\right|}\right)
 \left( \widehat{\delta}^{-1} - \dfrac{139}{120} \right)
 \right\} f^{\tau2}_q(z, Q^2) \Biggr] \ ,
\label{fq4:Ren}
\end{align}
The operators $\widehat{\delta}^{-1}$ and
$\widehat{\delta}^{-2}$ are defined as follows (see Appendix~B for details)
\begin{subequations}
\label{d:operators}
\begin{align}
 \widehat{\delta}^{-1} \left[f^{\tau2,-}_a(z, Q^2)\right] &=
  \dfrac{1}{\delta_R} \, f^{\tau2,-}_a(z, Q^2) ~ , ~~
 \widehat{\delta}^{-2} \left[f^{\tau2,-}_a(z, Q^2)\right]  =
  \dfrac{1}{\delta^2_R} \, f^{\tau2,-}_a(z, Q^2) \,
\label{d-:operators} \\
 \widehat{\delta}^{-1}
  \left[\rho^{k} \widetilde{I}_k(\sigma)\right] &=
  \rho^{k-1} \, \widetilde{I}_{\left|k-1\right|}(\sigma) ~~~ , ~~
 \widehat{\delta}^{-2}
  \left[\rho^{k} \widetilde{I}_k(\sigma)\right]  =
  \rho^{k-2} \, \widetilde I_{\left|k-2\right|}(\sigma) \ .
\label{d+:operators}
\end{align}
\end{subequations}
 
Note that the Eqs.~(\ref{F2:tau2:LO}) and Eqs.~(\ref{F2:tau2:NLO}) have
been obtained in \cite{Kotikov:1999} with the accuracy
$\mathcal{O}(\rho)$ for the $'+'$
component and with one $\mathcal{O}(z)$ for the $'-'$ component,
respectively. It leads to the fact that we should use only the most
singular terms in the r.h.s. of Eq.~(\ref{fq4:Ren}): i.e. the terms
$\widehat{\delta}^{-1}$ and $\sim \ln(Q^2/|{\mathrm a}_G^{\tau4}|)$ for the
gluon part and the terms $\widehat{\delta}^{-2}$ and
$\ln(Q^2/|{\mathrm a}_q^{\tau4}|) \, \widehat{\delta}^{-1}$
for the quark part.

Then, the Eq.~(\ref{fq4:Ren}) should be replaced by 
\begin{align}
 \dfrac{1}{e} \, F^{R\tau4}_2(z, Q^2) = \dfrac{64T_Rf}{15\beta_0^2}
 \Biggl[ &{\mathrm a}_G^{\tau4} \left\{
 \widehat{\delta}^{-1} + \dfrac{1}{2}
 \ln \left(\dfrac{Q^2}{\left|{\mathrm a}_G^{\tau4}\right|}\right)
 \right\} f^{\tau2}_G(z, Q^2)
\nonumber \\
 + \, 2C_F &{\mathrm a}_q^{\tau4} \left\{ \widehat{\delta}^{-2}
 + \dfrac{1}{2}
 \ln \left(\dfrac{Q^2}{\left|{\mathrm a}_q^{\tau4}\right|}\right) \,
 \widehat{\delta}^{-1} \right\} f^{\tau2}_q(z, Q^2) \Biggr] \ ,
\label{fq4:Ren1}
\end{align}

Applying the operators $\widehat{\delta}^{-1}$ and $\widehat{\delta}^{-2}$
separately to the $'+'$ and $'-'$ components of $f^{\tau2}_a(z, Q^2)$,
we obtain the following results for $F^{R\tau4}_2(z, Q^2)$:
\begin{subequations}
\label{q:tw4-Ren}
\begin{equation}
 F^{R\tau4}_2(z, Q^2) \ = \
  F^{R\tau4,+}_2(z, Q^2) \ + \ F^{R\tau4,-}_2(z, Q^2) \ ,
\label{fq4:Ren2}
\end{equation}
where
\begin{align}
 \dfrac{1}{e} \, F^{R\tau4,+}_2(z, Q^2) \, =& \, 
 \dfrac{32T_Rf}{15\beta_0^2} \, f^{\tau2,+}_{G}(z, Q^2) \,
 \Biggl[ {\mathrm a}_G^{\tau4} \left\{
 \dfrac{2}{\rho}
 \dfrac{\widetilde{I}_1(\sigma)}{\widetilde{I}_0(\sigma)} 
 + \ln \left(\dfrac{Q^2}{\left|{\mathrm a}_G^{\tau4}\right|}\right)
 \right\}
\nonumber \\
 +& \; \dfrac{4C_FT_Rf}{3C_A} {\mathrm a}_q^{\tau4} \Biggl(
 \left(1 - \bar{d}^q_{+-}(1) a_s(Q^2)\right) \left\{ \dfrac{2}{\rho}
 \dfrac{\widetilde{I}_1(\sigma)}{\widetilde{I}_0(\sigma)}
 + \ln \left(\dfrac{Q^2}{\left|{\mathrm a}_q^{\tau4}\right|}\right)
 \right\}
\nonumber \\
 +& \; \dfrac{20 C_A}{3} a_s(Q^2) \left\{ \dfrac{2}{\rho^2}
 \dfrac{\widetilde{I}_2(\sigma)}{\widetilde{I}_0(\sigma)}
 + \ln \left(\dfrac{Q^2}{\left|{\mathrm a}_q^{\tau4}\right|}\right)
 \dfrac{\widetilde{I}_1(\sigma)}{\rho \widetilde{I}_0(\sigma)}
 \right\} \Biggr) \Biggr] \ ,
\label{fq4+:Ren}\\
 \dfrac{1}{e} \, F^{R\tau4,-}_2(z, Q^2) \, =& \,
 \dfrac{32T_Rf}{15\beta_0^2} \, f^{\tau2,-}_{G}(z, Q^2) \,
 \Biggl[ {\mathrm a}_G^{\tau4}
 \ln \left(\dfrac{Q^2}{z_G^2 \left|{\mathrm a}_G^{\tau4}\right|}\right)
\nonumber \\
 -& \; 2 C_A {\mathrm a}_q^{\tau4} \left\{
 \ln \left(\dfrac{1}{z_q}\right) \,
 \ln \left(\dfrac{Q^2}{z_q \left|{\mathrm a}_q^{\tau4}\right|}\right) -
 p^\prime(\nu_q) \right\} \Biggr] \ .
\label{fq4-:Ren}
\end{align}
\end{subequations}

\subsection{Incorporation of twist-six contributions in the framework of
the renormalon model}
 \label{Sec5:tw6}

We shortly demonstrate the twist-six contributions in the framework of
the renormalon model.

When we added the twist-six part, the full small $x$ asymptotic results 
for PD and $F^\text{ren}_2$ structure function at NLO of
perturbation theory:
\begin{equation}
\begin{split}
 F^{R}_2(x, Q^2)   \ = \ F^{\tau2}_2(x,Q^2) \ +& \
 \dfrac{1}{Q^2} \, F^{R\tau4}_2(z, Q^2) \ + \
 \dfrac{1}{Q^4} \, F^{R\tau6}_2(z, Q^2) \ ,
\end{split}
\label{F2:LO:Ren6} 
\end{equation}

By analogy with twist-four case the twist-six term
$f^{R\tau6}_q(z, Q^2)$ has the form:
\begin{equation}
\dfrac{1}{e} \, F^{R\tau6}_2(z, Q^2) \ = \ \sum_{a=q,G} 
 {\mathrm a}_a^{\tau6} \,
 \widetilde{\mu}^{\tau6}_a (z, Q^2) \otimes f^{\tau2}_a(z, Q^2) \ ,
\label{Ren:tw6}
\end{equation}
where $\widetilde{\mu}^{\tau6}_a(z, Q^2)$ are given in \cite{Stein:1998}.
The corresponding Mellin transform of $\mu^{\tau6}_a(n, Q^2)$
is presented in the Appendix~A (see Eqs.~(\ref{App:muq}),
(\ref{App:Bq:6},\ref{App:bq:6}) and (\ref{App:muG})).

By analogy with the previous subsection applying the technique to transform
the Mellin convolutions to the standard products at small $x$ (see
\cite{Kotikov:1994:YF,Kotikov:1994:PRD} and Appendix B),
we can represent the Eq.~(\ref{Ren:tw6}) in the form
\begin{align}
\dfrac{1}{e} \,
F^{R\tau6}_2(z, Q^2) =& - \dfrac{8}{7} \times \dfrac{64T_Rf}{15\beta_0^2}
 \Biggl[ {\mathrm a}_G^{\tau6} \left\{
 \widehat{\delta}^{-1} + \dfrac{2663}{3360}
 + \dfrac{1}{2}
 \ln \left(\dfrac{Q^2}{\sqrt{\left|{\mathrm a}_G^{\tau6}\right|}}\right)
 \right\} f^{\tau2}_G(z, Q^2)
\nonumber \\
 +&\,  2C_F {\mathrm a}_q^{\tau6} \left\{ \widehat{\delta}^{-2}
 + \dfrac{143}{3360} \widehat{\delta}^{-1} - \dfrac{870637}{1411200} 
 + \dfrac{1}{2}
 \ln \left(\dfrac{Q^2}{\sqrt{\left|{\mathrm a}_q^{\tau6}\right|}}\right)
 \left( \widehat{\delta}^{-1} - \dfrac{3217}{3360} \right)
 \right\} f^{\tau2}_q(z, Q^2) \Biggr] 
\label{fq6:Ren}
\end{align}

 Considering only the most singular terms in the r.h.s. of
(\ref{fq6:Ren}), i.e. the terms $\widehat{\delta}^{-1}$ and
$\sim \ln(Q^2/\sqrt{\left|{\mathrm a}_G^{\tau6}\right|})$ for the gluon
 part  and the terms
$\widehat{\delta}^{-2}$ and
$\ln(Q^2/\sqrt{\left|{\mathrm a}_q^{\tau6}\right|}) \, \widehat{\delta}^{-1}$
for the quark part, we obtain immediately the following results:
\begin{align}
\dfrac{1}{e} \,
F^{R\tau6}_2(z, Q^2) = - \dfrac{8}{7} \times \dfrac{64T_Rf}{15\beta_0^2}
 \Biggl[ &{\mathrm a}_G^{\tau6} \left\{
 \widehat{\delta}^{-1} + \dfrac{1}{2}
 \ln \left(\dfrac{Q^2}{\sqrt{\left|{\mathrm a}_G^{\tau6}\right|}}\right)
 \right\} f^{\tau2}_G(z, Q^2)
\nonumber \\
 + \, 2C_F &{\mathrm a}_q^{\tau6} \left\{ \widehat{\delta}^{-2}
 + \dfrac{1}{2}
 \ln \left(\dfrac{Q^2}{\sqrt{\left|{\mathrm a}_q^{\tau6}\right|}}\right) \,
 \widehat{\delta}^{-1} \right\} f^{\tau2}_q(z, Q^2) \Biggr] \ ,
\label{fq6:Ren1}
\end{align}
which is very close to the twist-four one, see Eq.~(\ref{fq4:Ren1}):
\begin{equation}
\dfrac{1}{e} \, F^{R\tau6}_2(z, Q^2) = - \dfrac{8}{7} \times \left[
  f^{R\tau4}_q(z, Q^2) ~\text{ with }~
  {\mathrm a}_a^{\tau4} \to {\mathrm a}_a^{\tau6} \, , \
  \ln \left(\dfrac{Q^2}{\left|{\mathrm a}_a^{\tau4}\right|}\right) \to
  \ln \left(\dfrac{Q^2}{\sqrt{\left|{\mathrm a}_a^{\tau6}\right|}}\right)
 \right] \ .
\label{Rht4-Rht6}
\end{equation}

Note that the representation (\ref{Rht4-Rht6}) of the twist-six terms 
in the terms of the 
twist-four ones is universal and has quite compact form and, thus, it
will be often used below. 

Because the forms of the twist-four and twist-six contributions are very
similar, it is possible to present quite compact form for the full
contribution of the higher-twist operators $ F^{Rh\tau}_2(z, Q^2)$
\begin{subequations}
\label{F2:Rht}
\begin{equation}
\begin{split}
 F^{R}_2(z, Q^2) \ = \
  F^{\tau2}_2&(z,Q^2) \ + \ F^{Rh\tau}_2(z, Q^2) \ ,
\end{split}
\label{F2R:Rht}
\end{equation}
where
\begin{equation}
 F^{Rh\tau}_2(z, Q^2) \ = \
  F^{Rh\tau,+}_2(z, Q^2) \ + \ F^{Rh\tau,-}_2(z, Q^2)
\label{fq:Rht}
\end{equation}
and
\begin{align}
\dfrac{1}{e} \, 
F^{Rh\tau,+}_2(z, Q^2) \, =& \, \dfrac{32T_Rf}{15\beta_0^2} \,
f^{\tau2,+}_{G}(z, Q^2) \, \sum_{m = 4,6} k_m
 \Biggl[ \dfrac{{\mathrm a}_G^{\tau m}}{Q^{(m-2)}} \left\{ \dfrac{2}{\rho} 
\dfrac{\widetilde{I}_1(\sigma)}{\widetilde{I}_0(\sigma)}
+ \ln \left(\dfrac{Q^2} {\left|{\mathrm a}_G^{\tau m}\right|^{p_m}}\right)
\right\}
\nonumber \\
 +& \; \dfrac{4C_FT_Rf}{3C_A} \dfrac{{\mathrm a}_q^{\tau m}}{Q^{(m-2)}} \Biggl(
 \left(1 - \bar{d}^q_{+-}(1) a_s(Q^2)\right) \left\{ \dfrac{2}{\rho}
 \dfrac{\widetilde{I}_1(\sigma)}{\widetilde{I}_0(\sigma)}
 + \ln \left(\dfrac{Q^2}
  {\left|{\mathrm a}_q^{\tau m}\right|^{p_m}}\right)
 \right\}
\nonumber \\
 +& \; \dfrac{20 C_A}{3} a_s(Q^2) \left\{ \dfrac{2}{\rho^2}
 \dfrac{\widetilde{I}_2(\sigma)}{\widetilde{I}_0(\sigma)}
 + \ln \left(\dfrac{Q^2}
  {\left|{\mathrm a}_q^{\tau m}\right|^{p_m}}\right)
 \dfrac{\widetilde{I}_1(\sigma)}{\rho \widetilde{I}_0(\sigma)}
 \right\} \Biggr) \Biggr] \ ,
\label{fq+:Rht}\\
\dfrac{1}{e} \,
F^{Rh\tau,-}_2(z, Q^2) \, =& \, \dfrac{32T_Rf}{15\beta_0^2} \,
f^{\tau2,-}_{G}(z, Q^2) \,
\sum_{m = 4,6} k_m
 \Biggl[ \dfrac{{\mathrm a}_G^{\tau m}}{Q^{(m-2)}}
 \ln \left(\dfrac{Q^2}
 {z_G^2 \left| {\mathrm a}_G^{\tau m}\right|^{p_m}}\right)
\nonumber \\
 -& \; 2 C_A
        \dfrac{{\mathrm a}_q^{\tau m}}{Q^{(m-2)}} \left\{
 \ln \left(\dfrac{1}{z_q}\right) \,
 \ln \left(\dfrac{Q^2}
 {z_q \left| {\mathrm a}_q^{\tau m}\right|^{p_m}}\right) -
 p^\prime(\nu_q) \right\} \Biggr] \ ,
\label{fq-:Rht}
\end{align}
\end{subequations}
where $k_4 = 1$, $k_6 = -8/7$ and $p_4 = 1$, $p_6 = 1/2$.

\section{The higher twist contributions for the derivative
 $\partial F_2/\partial \ln Q^2$}
\label{Sec:6}

By analogy with the previous Section we consider firstly only
the twist-four terms in the framework of the infrared renormalon model.
The contribution of the twist-six terms will be incorporated shortly
at the end of this Section.

\subsection{Renormalon model predictions for twist-four operators}
 \label{Sec6:tw4}

Note that there are the following properties
\begin{equation}
\dfrac{d}{d\ln Q^2}
 \dfrac{1}{Q^2} \ = \  - \dfrac{1}{Q^2} ~ , ~~
\dfrac{d}{d\ln Q^2} \biggl[
 \dfrac{1}{Q^2} \ln \left( \dfrac{\Lambda^2}{Q^2} \right) \biggr] \ = \
 - \dfrac{1}{Q^2}
 \left( \ln \left( \dfrac{\Lambda^2}{Q^2} \right) + 1 \right) \ \approx \
 - \dfrac{1}{Q^2} \ln \left( \dfrac{\Lambda^2}{Q^2} \right) \ ,
\label{Der:tw4}
\end{equation}
where we keep only most important terms
(see discussions in the previous Section and Eq.~(\ref{fq4:Ren1})). 

In this approximation we easily obtain that
\begin{subequations}
\label{dF2:tw4}
\begin{equation}
\dfrac{\partial F_2^{R}(z, Q^2)}{\partial \ln Q^2} \ = \
 \dfrac{\partial F_2^{\tau2}(z, Q^2)}{\partial \ln Q^2} + \dfrac{1}{Q^2} \left(
 \dfrac{\partial F_2^{R\tau4}(z, Q^2)}{\partial \ln Q^2} -
 F_2^{R\tau4}(z, Q^2) \right)
\label{dF2:Ren4}
\end{equation}
and
\begin{equation}
\dfrac{\partial F_2^{R\tau4}(z, Q^2)}{\partial \ln Q^2} \ = \
 e \, \dfrac{8T_Rf}{3} \, a_s(Q^2) \,  \Phi^{R\tau4}_G(z, Q^2) \ .
\label{dF2:Ren4.1}
\end{equation}
The value of $F^{R\tau4}_2(z, Q^2)$ is given by 
Eqs.~(\ref{fq4:Ren2})--(\ref{fq4-:Ren}) and
\begin{align}
\Phi^{R\tau4}_G(z, Q^2) \, =& \, \dfrac{16C_A}{5\beta_0^2}
f^{\tau2,+}_{G}(z, Q^2) \,
 \Biggl[ {\mathrm a}_G^{\tau4} \left\{
 \dfrac{2}{\rho^2} 
\dfrac{\widetilde{I}_2(\sigma)}{\widetilde{I}_0(\sigma)}
+ \ln \left(\dfrac{Q^2}{\left|{\mathrm a}_G^{\tau4}\right|}\right)
 \dfrac{1}{\rho}
\dfrac{\widetilde{I}_1(\sigma)}{\widetilde{I}_0(\sigma)}
\right\}
\nonumber \\
 +& \; \dfrac{4C_FT_Rf}{3C_A} {\mathrm a}_q^{\tau4} \left\{
 \dfrac{2}{\rho^2}
\dfrac{\widetilde{I}_2(\sigma)}{\widetilde{I}_0(\sigma)}
+ \ln \left(\dfrac{Q^2}{\left|{\mathrm a}_q^{\tau4}\right|}\right)
 \dfrac{1}{\rho}
\dfrac{\widetilde{I}_1(\sigma)}{\widetilde{I}_0(\sigma)}
\right\}
 \Biggr] \ .
\label{PhiG+:tw4}
\end{align}
\end{subequations}

Thus, we see that the twist-four corrections to $F_2$ and 
$dF_2/d\ln Q^2$ have opposite signs, because $dF_2^{R\tau4}/d\ln Q^2
\sim a_s(Q^2)$ and the most important twist-four contribution is given
by $F_2^{R\tau4}(z,Q^2)$.

\subsection{Incorporation of twist-six contributions in the framework of
the renormalon model}
 \label{Sec6:tw6}

Following to the subsection~\ref{Sec5:tw6} of the previous Section and
considering the properties
\begin{equation}
\dfrac{d}{d\ln Q^2}
 \dfrac{1}{Q^4} \ = \ - \dfrac{2}{Q^4} ~,~~
\dfrac{d}{d\ln Q^2} \biggl[
 \dfrac{1}{Q^4} \ln \left( \dfrac{\Lambda^2}{Q^2} \right) \biggr] \ = \
 - \dfrac{1}{Q^2} 
 \left(2 \ln \left( \dfrac{\Lambda^2}{Q^2} \right) + 1 \right) \ \approx \
 - \dfrac{2}{Q^2} \ln \left( \dfrac{\Lambda^2}{Q^2} \right) \ ,
\label{Der:tw6}
\end{equation}
together with the one (\ref{Der:Bessel:LO}), we immediately obtain that
\begin{subequations}
\label{dF2:tw6}
\begin{equation}
\dfrac{\partial F_2^{R}(z, Q^2)}{\partial \ln Q^2} \ = \
 \dfrac{\partial F_2^{\tau2}(z, Q^2)}{\partial \ln Q^2} + \dfrac{1}{Q^2} \left(
 \dfrac{\partial F_2^{R\tau4}(z, Q^2)}{\partial \ln Q^2} -
 F_2^{R\tau4}(z, Q^2) \right) + \dfrac{1}{Q^4} \left(
 \dfrac{\partial F_2^{R\tau6}(z, Q^2)}{\partial \ln Q^2} -
 2 F_2^{R\tau6}(z, Q^2) \right)
\label{dF2:Ren6}
\end{equation}
and
\begin{equation}
 \dfrac{\partial F_2^{R\tau6}(z, Q^2)}{\partial \ln Q^2} \ = \
 e \, \dfrac{8T_Rf}{3} \, a_s(Q^2) \, \Phi^{R\tau6}_G(z, Q^2) \ .
\label{dF2:Ren6.1}
\end{equation}
The value of $f^{R\tau6}_q(z, Q^2)$ is given by Eq.~(\ref{Rht4-Rht6}) and
\begin{align}
\Phi^{R\tau6}_G(z, Q^2) 
 \ = \ & -  \dfrac{8}{7} \times \left[
  \Phi^{R\tau4}_G(z, Q^2) ~\text{ with }~ 
  {\mathrm a}_a^{\tau4} \to {\mathrm a}_a^{\tau6} \, , \
  \ln \left(\dfrac{Q^2}{\left|{\mathrm a}_a^{\tau4}\right|}\right) \to
  \ln \left(\dfrac{Q^2}{\sqrt{\left|{\mathrm a}_a^{\tau6}\right|}}\right)
 \right] \ .
\label{PhiG4-PhiG6}
\end{align}
\end{subequations}

Thus, we see that by analogy with the case of $F_2(z, Q^2)$ itself, 
for the derivation (\ref{dF2:tw6}) the twist-six terms partially 
compensate the contributions of the twist-four terms.

\section{Parton distribution functions in the renormalon model approach}
 \label{Sec:7}

It is clearly to see that the standard parton distributions
$f_q(x, Q^2)$ and
$f_G(x, Q^2)$ fitted with help of experimental data do not coincide with
the above twist-two ones $f^{\tau2}_q(z, Q^2)$ and $f^{\tau2}_G(z, Q^2)$.
These PD $f_q(z, Q^2)$ and $f_G(z, Q^2)$ are usually defined keeping their
twist-two relations (\ref{F2:LO}) or (\ref{F2:NLO}) with the structure
function $F_2(z, Q^2)$, i.e.

At LO
\begin{equation}
F_2(z, Q^2) \ = \ e \, f_q(z, Q^2) \ ,
\label{F2PD:LO}
\end{equation}

At NLO
\begin{equation}
F_2(z, Q^2) \ = \ e \, \left( f_q(z, Q^2) \, + \,
 \dfrac{8T_Rf}{3} \, a_s(Q^2) \, f_G(z, Q^2) \right) \ .
\label{F2PD:NLO}
\end{equation}

Thus, the parton distributions $f_q(z, Q^2)$ and $f_G(z, Q^2)$ can be strongly
deviated for the corresponding the twist-two densities $f^{\tau2}_q(z, Q^2)$
and $f^{\tau2}_G(z, Q^2)$ at quite low $Q^2$ values, because there are the
HT corrections to $F_2^{\tau2}(z, Q^2)$. 

The HT correction to the parton dstribution at the LO was presented
in the Introduction already. Here we present the results at the NLO.
As it was in the previous Section, we consider firstly the twist-four 
corrections.

\subsection{Twist-four corrections to (singlet) quark distribution}
 \label{Sec7:f_q}

Consider firstly the (singlet) quark parton distribution $f_q(z, Q^2)$.
From the Eq.~(\ref{F2:LO}) and the analysis of the Section~\ref{Sec:5}
we can obtain that
\begin{subequations}
\label{fq:total}
\begin{equation}
f^R_q(z, Q^2) \ = \  f^{\tau2}_q(z, Q^2) \ + \
    \dfrac{1}{Q^2} \, f^{R\tau4}_q(z, Q^2) \ ,
\label{fq:tot0}
\end{equation}
where
$f^{R\tau4}_q(z, Q^2)$ is given at the LO by Eqs.~(\ref{intro:fq4+}) and
 (\ref{intro:fq4-}).

It is useful to represent also the complete expressions directly for
$f^R_q(z, Q^2)$:
\begin{equation}
f^R_q(z, Q^2) \ = \ f^{R,+}_q(z, Q^2) \ + \ f^{R,-}_q(z, Q^2) \ ,
\label{fq:+-}
\end{equation}
where at the NLO
\begin{align}
\dfrac{f^{R,+}_q(z, Q^2)}{f^{\tau2,+}_{q}(z, Q^2)} \, = \, 1 \, +& \,
 \dfrac{64C_FT_Rf}{15\beta_0^2} \,  
 \dfrac{{\mathrm a}_q^{\tau4}}{Q^2} \Biggl\{ \dfrac{2}{\rho^2}
 \dfrac{\widetilde{I}_1(\sigma) \left(1 - \bar{d}^q_{+-}(1) a_s(Q^2)\right)
 + (20C_A/3) a_s(Q^2) \widetilde{I}_2(\sigma)/\rho}
       {\widetilde{I}_1(\sigma) \left(1 - \bar{d}^q_{+-}(1) a_s(Q^2)\right)
 + (20C_A/3) a_s(Q^2) \widetilde{I}_0(\sigma)/\rho}
\nonumber \\
 &\hspace{4mm}
 + \ln \left(\dfrac{Q^2}{\left|{\mathrm a}_q^{\tau4}\right|}\right)
 \dfrac{1}{\rho}
 \dfrac{\widetilde{I}_0(\sigma) \left(1 - \bar{d}^q_{+-}(1) a_s(Q^2)\right)
 + (20C_A/3) a_s(Q^2) \widetilde{I}_1(\sigma)/\rho}
       {\widetilde{I}_1(\sigma) \left(1 - \bar{d}^q_{+-}(1) a_s(Q^2)\right)
 + (20C_A/3) a_s(Q^2) \widetilde{I}_0(\sigma)/\rho}
\Biggr\}
 ~ + ~ {\mathcal{O}}\left( \rho \right ) \ ,
\label{fq+:tot}\\
\dfrac{f^{R,-}_q(z, Q^2)}{f^{\tau2,-}_{q}(z, Q^2)} \, = \, 1 \, +& \,
 \dfrac{64C_FT_Rf}{15\beta_0^2} \,  
 \dfrac{{\mathrm a}_q^{\tau4}}{Q^2} \left\{
 \ln \left(\dfrac{1}{z_q}\right) \,
 \ln \left(\dfrac{Q^2}
  {z_q \left|{\mathrm a}_q^{\tau4}\right|}\right) -
 p^\prime(\nu_q) \right\} 
 ~ + ~ {\mathcal{O}}\left( z \right ) \ .
\label{fq-:tot}
\end{align}
\end{subequations}

We clearly see that the twist-four terms are responsible for the
additional positive contributions to the quark distribution,
which are very important at low $Q^2$ values.

So, the experimentally extracted quark distribution $f_q(z, Q^2)$,
which have the leading twist relations (\ref{F2PD:LO}) and (\ref{F2PD:NLO})
with $F_2(z, Q^2)$, strongly deviates form the leading twist quark
distribution $f_q^{\tau2}(z, Q^2)$. At quite low $Q^2$ values, where
$f^{\tau2}_q(z, Q^2)$ had the quite flat behavior closed to (\ref{flat}),
the full quark distribution  $f^R_q(z, Q^2)$ will rise at $z \to 0$
(see Eqs.~(\ref{fq+:tot}) and (\ref{fq-:tot})), because 
${\mathrm a}_q^{\tau4}>0$ (see Tables~\ref{Tab:H1+ZEUS:HT} ,\ref{Tab:Rht}).
This rise is in full agreement with the corresponding experimental data
(see Tables~\ref{Tab:H1+ZEUS:HT} ,\ref{Tab:Rht}, Figure~\ref{Fig:PDFs},
Section~\ref{Sec:10} and discussions therein).

\subsection{Twist-four corrections to gluon distribution}
 \label{Sec7:f_G}

Consider now the gluon parton distribution $f_G(z, Q^2)$.
From the Eq.~(\ref{F2:LO}) and the analysis of the Section~\ref{Sec:5}
we can obtain that
\begin{subequations}
\label{fG:total}
\begin{equation}
f^R_G(z, Q^2) \ = \  f^{\tau2}_G(z, Q^2) \ + \
    \dfrac{1}{Q^2} \, f^{R\tau4}_G(z, Q^2) \ .
\label{fG:tot0}
\end{equation}
where $f^{R\tau4}_q(z, Q^2)$ is given at the LO by Eqs.~(\ref{intro:fG4+}) and
(\ref{intro:fG4-}).

For the gluon distribution in the NLO we have the similar relations
\begin{equation}
f^R_G(z, Q^2) \ = \ f^{R,+}_G(z, Q^2) \ + \ f^{R,-}_G(z, Q^2) \ ,
\label{fG:+-}
\end{equation}
\begin{align}
\dfrac{f^{R,+}_G(z, Q^2)}{f^{\tau2,+}_{G}(z, Q^2)} \, = \, 1 \, +& \, 
 \dfrac{8}{5\beta_0^2} \, \dfrac{{\mathrm a}_G^{\tau4}}{a_s(Q^2) \, Q^2} \,
 \left\{ \dfrac{2}{\rho}
\dfrac{\widetilde{I}_1(\sigma)}{\widetilde{I}_0(\sigma)} 
+ \ln \left(\dfrac{Q^2}{\left|{\mathrm a}_G^{\tau4}\right|}\right)
\right\}
 ~ + ~ {\mathcal{O}}\left( \rho \right ) \ ,
\label{fG+:tot}\\
\dfrac{f^{R,-}_G(z, Q^2)}{f^{\tau2,-}_{G}(z, Q^2)} \, = \, 1 \, +&
\dfrac{8}{5\beta_0^2} \, \dfrac{{\mathrm a}_G^{\tau4}}{a_s(Q^2) \, Q^2} \,
 \ln \left(\dfrac{Q^2}
  {z_G^2 \left|{\mathrm a}_G^{\tau4}\right|}\right)
 ~ + ~ {\mathcal{O}}\left( z \right ) \ .
\label{fG-:tot}
\end{align}
\end{subequations}

So, as in the case of the quark distribution, the experimentally
extracted gluon density $f_G(z, Q^2)$, which has the leading twist relation
with $F_2(z, Q^2)$ and $dF_2/d\ln Q^2$, strongly deviates form the
leading twist gluon distribution $f_G^{\tau2}(z, Q^2)$.
At quite low $Q^2$ values: $Q^2 \sim Q_0^2$ , where $f^{\tau2}_G(z,
Q^2)$ had the quite flat behavior closed to (\ref{flat}), the full gluon 
distribution $f^R_q(z,Q^2)$ 
falls at $x \to 0$, because 
${\mathrm a}_G^{\tau4}<0$ (see Tables ~\ref{Tab:H1+ZEUS:HT} ,\ref{Tab:Rht}). 
The behavior is in full agreement with the corresponding experimental data
(see Tables~\ref{Tab:H1+ZEUS:HT} ,\ref{Tab:Rht}, Figure~\ref{Fig:PDFs},
Section~\ref{Sec:10} and discussions therein).

\subsection{Twist-six corrections to parton distributions}
 \label{Sec7:tw6}

We shortly demonstrate the twist-six contributions to parton 
distribution in the framework of the renormalon model.
When we added the twist-six part, the full small $x$ asymptotic results 
for parton distributions
is
\begin{equation}
f_a(z, Q^2) \ = \  f^{\tau2}_a(z, Q^2) \ + \
 \dfrac{1}{Q^2} \, f^{R\tau4}_a(z, Q^2) \ + \
 \dfrac{1}{Q^4} \, f^{R\tau6}_a(z, Q^2) \ = \
 f^{\tau2}_a(z, Q^2) \ + \ f^{Rh\tau}_a(z, Q^2) \ ,
\label{fa:total}
\end{equation}
where
$f^{R\tau6}_a(z, Q^2)$
are given by Eqs. (\ref{intro:ht4-ht6}):
\begin{equation}
 f^{R\tau6}_a(z, Q^2) \ =  \ - \dfrac{8}{7} \times \left[
  f^{R\tau4}_a(z, Q^2) ~\text{ with }~ 
  {\mathrm a}_a^{\tau4} \to {\mathrm a}_a^{\tau6} \, , \
  \ln \left(\dfrac{Q^2}{\left|{\mathrm a}_a^{\tau4}\right|}\right) \to
  \ln \left(\dfrac{Q^2}{\sqrt{\left|{\mathrm a}_a^{\tau6}\right|}}\right)
 \right] \ ,
\nonumber 
\end{equation}

The twist-six corrections do not change the results for parton
distributions obtained in the previous subsection.

\section{The higher twist contributions to the
slopes of $F_2$ and of parton distributions}
 \label{Sec:8}

Consider the power-like corrections to the twist-two effective slopes 
$\lambda^{\text{eff},\tau2}_{F2}(z,Q^2)$ and
 $\lambda^{\text{eff},\tau2}_{a}(z,Q^2)$ $(a=q,G)$
introduced in the Section~\ref{Sec:4}.
The effective slopes have the following form
\begin{align}
\lambda^\text{eff}_{F2}(z, Q^2) \ =& \ \dfrac{\partial}{\partial\ln(1/z)} 
 \ln \left[ F_2^{\tau2}(z, Q^2) \, + \,
 \dfrac{1}{Q^2} \, F_2^{R\tau4}(z, Q^2) \, + \,
 \dfrac{1}{Q^4} \, F_2^{R\tau6}(z,Q^2) \right] \ ,
\label{F2-Slope:HT} \\
\lambda^\text{eff}_a(z, Q^2) \ =& \ \dfrac{\partial}{\partial\ln(1/z)} 
 \ln \left[ f_a^{\tau2}(z, Q^2) \, + \,
 \dfrac{1}{Q^2} \, f_a^{R\tau4}(z,Q^2) \, + \,
 \dfrac{1}{Q^4} \, f_a^{R\tau6}(z,Q^2) \right] \ .
\label{a-Slope:HT}
\end{align}

Using Eqs.~(\ref{Dz:Bessel}), the derivations
$\partial F_2^{\tau2}/\partial\ln(1/z)$,
$\partial f_a^{\tau2}/\partial\ln(1/z)$ and
$\partial f_a^{R\tau m}/\partial\ln(1/z)$,
$(m = 4,6)$, can be represented as the sum of two components
($'+'$ and $'-'$) which are obtained from the corresponding
($'+'$ and $'-'$) PD functions. One can show that
\begin{subequations}
\label{dlnz:Rht}
\begin{align}
\dfrac{\partial f^{R\tau4,+}_q(z, Q^2)}{\partial\ln(1/z)} \, &= \,
 \dfrac{64 C_F T_R f}{15 \beta_0^2} \, {\mathrm a}_q^{\tau4} \,
 \Biggl\{ \dfrac{2}{\rho}
 \dfrac{\widetilde{I}_0(\sigma) \left(1 - \bar{d}^q_{+-}(1) a_s(Q^2)\right)
 + (20C_A/3) a_s(Q^2) \widetilde{I}_1(\sigma)/\rho}
       {\widetilde{I}_1(\sigma) \left(1 - \bar{d}^q_{+-}(1) a_s(Q^2)\right)
 + (20C_A/3) a_s(Q^2) \widetilde{I}_0(\sigma)/\rho}
\nonumber \\
 &\hspace{2cm}
 + \ln \left(\dfrac{Q^2}{\left|{\mathrm a}_q^{\tau4}\right|}\right) \Biggr\} \,
 f^{\tau2,+}_q(z, Q^2) \ ,
\label{dlnz:q4+} \\
\dfrac{\partial f^{R\tau4,-}_q(z, Q^2)}{\partial\ln(1/z)} \, &= \, 
 \dfrac{64 C_F T_R f}{15 \beta_0^2} \, {\mathrm a}_q^{\tau4} 
 \ln \left(\dfrac{Q^2}{z_q^2 \left|{\mathrm a}_q^{\tau4}\right|}\right)
 f^{\tau2,-}_q(z, Q^2) \ ,
\label{dlnz:q4-} \\
\dfrac{\partial f^{R\tau4,+}_G(z, Q^2)}{\partial\ln(1/z)} \, &= \,
 \dfrac{8}{5 \beta_0^2} \, \dfrac{{\mathrm a}_G^{\tau4}}{a_s(Q^2)} \,
 \left\{ 2 + \ln \left(\dfrac{Q^2}{\left|{\mathrm a}_G^{\tau4}\right|}\right)
 \, \rho \, \dfrac{\widetilde{I}_1(\sigma)} {\widetilde{I}_0(\sigma)} \right\}
 f^{\tau2,+}_G(z, Q^2) \ ,
\label{dlnz:G4+} \\
\dfrac{\partial f^{R\tau4,-}_G(z, Q^2)}{\partial\ln(1/z)} \, &= \, 
 \dfrac{16}{5 \beta_0^2} \, \dfrac{{\mathrm a}_G^{\tau4}}{a_s(Q^2)} \,
 f^{\tau2,-}_G(z, Q^2) \ ,
\label{dlnz:G4-} \\
\dfrac{\partial f^{R\tau6,\pm}_a(z, Q^2)}{\partial\ln(1/z)} \, &= \,
 - \dfrac{8}{7} \times
 \Biggl[
 \dfrac{\partial f^{R\tau4,\pm}_a(z, Q^2)}{\partial\ln(1/z)}
 ~\mbox{ with }~ 
 {\mathrm a}_a^{\tau4} \to {\mathrm a}_a^{\tau6}~,~~
 \ln \left(\dfrac{Q^2}{\left|{\mathrm a}_a^{\tau4}\right|}\right) \to
 \ln \left(\dfrac{Q^2}{\sqrt{\left|{\mathrm a}_a^{\tau6}\right|}}\right)
 \Biggr] \ .
\label{dlnx6:Ren-}
\end{align}
\end{subequations}


The Eqs.~(\ref{F2-Slope:HT}) and (\ref{a-Slope:HT}) together with the
Eqs.~(\ref{Slopes:NLO}) and (\ref{dlnz:Rht}) give a complete information
about the full and asymptotical values of the slopes
$\lambda^\text{eff}_{F2}(z, Q^2)$ and $\lambda^\text{eff}_{a}(z, Q^2)$.
The results will be demonstrated on Figs.~2, 6 and 7.

It is possible, however, to give a simple demonstration of the effect
of the HT corrections.
Following the Section~\ref{Sec:4}, we can prepare also the results for the
 higher twist corrections to the asymptotical values of 
$\lambda^{\text{eff},\tau2}_{F2}(z, Q^2)$ and
$\lambda^{\text{eff},\tau2}_{a}(z, Q^2)$, which can be obtained
by neglecting the $'-'$ components.  Restricting ourselves by 
the twist-four case we can estimate the value of the slopes
$\lambda^\text{eff}_{F2,\text{as}}(z,Q^2)$ and 
$\lambda^\text{eff}_{a,\text{as}}(z,Q^2)$ in the form
\begin{align}
\lambda^\text{eff}_{F2, \text{as}}(z, Q^2) \, &= \,
 \lambda^{\text{eff}, \tau2}_{F2, \text{as}}(z, Q^2) +
 \dfrac{1}{Q^2} \, \lambda^{\text{eff}, R\tau4}_{F2, \text{as}}(z, Q^2) 
 + \mathcal{O} \left( \dfrac{1}{Q^4} \right) \ ,
\label{SlopF2:tw4} \\
\lambda^\text{eff}_{a, \text{as}}(z, Q^2) \, &= \,
 \lambda^{\text{eff}, \tau2}_{a, \text{as}}(z, Q^2) +
 \dfrac{1}{Q^2} \lambda^{\text{eff},R\tau4}_{a,\text{as}}(z,Q^2) 
 + \mathcal{O} \left( \dfrac{1}{Q^4} \right) \ ,
\label{Slopa:tw4}
\end{align}
where at the LO
\begin{subequations}
\label{dlnz:Rht:eff}
\begin{align}
\lambda^{\text{eff}, R\tau4}_{F2, \text{as}}(z, Q^2) \, &= \, 
 \dfrac{16 C_A}{5\beta_0^2} \Biggl[ {\mathrm a}_G^{\tau4} \Biggl\{ 2
 \dfrac{\widetilde{I}_0(\sigma_\text{LO}) - \widetilde{I}_2(\sigma_\text{LO})}
 {\rho_\text{LO} \widetilde{I}_1(\sigma_\text{LO})}
 + \ln \left(\dfrac{Q^2}{{\mathrm a}_G^{\tau4}}\right) \left(1 -
 \dfrac{\widetilde{I}_0(\sigma_\text{LO}) \widetilde{I}_2(\sigma_\text{LO})}
       {\widetilde{I}_1^2(\sigma_\text{LO})} \right) \Biggr\}
\nonumber \\
&+ \,  \dfrac{4C_FT_Rf}{3C_A} {\mathrm a}_q^{\tau4} \Biggl\{ 2
 \dfrac{\widetilde{I}_0(\sigma_\text{LO}) - \widetilde{I}_2(\sigma_\text{LO})}
 {\rho_\text{LO} \widetilde{I}_1(\sigma_\text{LO})}
 + \ln \left(\dfrac{Q^2}{{\mathrm a}_G^{\tau4}}\right) \left(1 -
 \dfrac{\widetilde{I}_0(\sigma_\text{LO})\widetilde{I}_2(\sigma_\text{LO})}
       {\widetilde{I}_1^2(\sigma_\text{LO})} \right) \Biggr\} \Biggr]
\label{SlopF2:Bessel} \\
&\approx \, \dfrac{16 C_A}{5\beta_0^2} \dfrac{1}{2\rho_\text{LO} \ln(1/z)}
 \Biggl[ {\mathrm a}_G^{\tau4} \Biggl\{ \dfrac{4}{\rho_\text{LO}} +
 \ln \left(\dfrac{Q^2}{{\mathrm a}_G^{\tau4}}\right) \Biggr\} 
 + \dfrac{4C_FT_Rf}{3C_A} {\mathrm a}_q^{\tau4} 
 \Biggl\{ \dfrac{4}{\rho_\text{LO}} +
 \ln \left(\dfrac{Q^2}{{\mathrm a}_q^{\tau4}}\right) \Biggr\} \Biggr]
\label{SlopF2:tw4+} \\
\lambda^{\text{eff}, R\tau4}_{q, \text{as}}(z, Q^2) \, &= \,
 \dfrac{64C_FT_Rf}{15\beta_0^2} \, {\mathrm a}_q^{\tau4} \Biggl\{
 2\dfrac{\widetilde{I}_0(\sigma_\text{LO}) - \widetilde{I}_2(\sigma_\text{LO})}
        {\rho \widetilde{I}_1(\sigma_\text{LO})} +
 \ln \left(\dfrac{Q^2}{{\mathrm a}_q^{\tau4}}\right) \left(1 -
 \dfrac{\widetilde{I}_0(\sigma_\text{LO})\widetilde{I}_2(\sigma_\text{LO})}
       {\widetilde{I}_1^2(\sigma_\text{LO})} \right) \Biggr\}
\label{Slopq:Bessel} \\
&\hspace{6.5mm}
 \approx \,
 \dfrac{64C_FT_Rf}{15\beta_0^2} \dfrac{{\mathrm a}_q^{\tau4}}
 {2\rho_\text{LO}\ln(1/z)} \Biggl\{ \dfrac{4}{\rho_\text{LO}} +
 \ln \left(\dfrac{Q^2}{{\mathrm a}_q^{\tau4}}\right) \Biggr\}
\label{Slopq:tw4+}\\
\lambda^{\text{eff}, R\tau4}_{G, \text{as}}(z, Q^2) \, &= \,
 \dfrac{16}{5\beta_0^2} \, \dfrac{{\mathrm a}_G^{\tau4}}{a_s(Q^2)} \, 
 \left(1 - \dfrac{\widetilde{I}_1^2(\sigma_\text{LO})}
                 {\widetilde{I}_0^2(\sigma_\text{LO})}\right)
 \ \approx \
 \dfrac{8 }{5\beta_0^2} \, \dfrac{{\mathrm a}_G^{\tau4}}{a_s(Q^2)} \, 
 \dfrac{1}{\rho_\text{LO}\ln(1/z)} \ .
\label{Slopg:tw4+}
\end{align}
\end{subequations}

From the equations (\ref{SlopF2:tw4+}) and (\ref{Slopq:tw4+}) it possible
to see that the slopes $ \lambda^\text{eff}_{F2, \text{as}}(z, Q^2)$ and
$\lambda^\text{eff}_{q, \text{as}}(z, Q^2)$ have got the positive 
twist-four corrections, that is in full agreement with the corresponding
experimental H1 and ZEUS data for the slope $ \lambda_{F2}$ at low $Q^2$
values (see Fig. 7). 
However, the difference between the twist-four corrections to these
slopes is negative , because 
${\mathrm a}_G^{\tau4}<0$ (see Tables 4, 6--8):
\begin{equation}
 \lambda^{\text{eff}, R\tau4}_{F2, \text{as}}(z, Q^2) \, - \,
 \lambda^{\text{eff}, R\tau4}_{q,  \text{as}}(z, Q^2) \ \approx \
 \dfrac{8C_A}{5\beta_0^2} \, \dfrac{{\mathrm a}_G^{\tau4}}
 {\rho_\text{LO}\ln(1/z)} \,
 \Biggl\{ \dfrac{4}{\rho_\text{LO}} +
 \ln \left(\dfrac{Q^2}{{\mathrm a}_q^{\tau4}}\right) \Biggr\} \ .
\label{Slope:Rht:F2-q}
\end{equation}

Thus, the inequality $  \lambda^\text{eff}_{F2, \text{as}}(z, Q^2) >
\lambda^\text{eff}_{q, \text{as}}(z, Q^2)$ coming form Eq. 
(\ref{Slopes:NLO:F2-q}) takes place
for not very small $Q^2$, because 
it is suppressed by power corrections.

We would like to note that the equations (\ref{dlnz:Rht:eff}) are valid only
at not very small $Q^2$ values, where we can neglect the terms $\sim 1/Q^4$
coming from expanding the denominator and from the twist-six terms.
The small $Q^2$ behavior of
$\lambda^{\text{eff}}
_{a, \text{as}}(z, Q^2)$ can be easy demonstrated
at the point $Q^2 = Q^2_0$ in the following Section.

\section{ Parton distributions in the renormalon model at $Q^2_0$}
 \label{Sec:9}

As it has been already shown in the previous Section the total PD functions
$f_q(z, Q^2)$ and $f_G(z, Q^2)$ fitted in experiments data do not coincide
with the above twist-two ones $f^{\tau2}_q(z, Q^2)$ and
$f^{\tau2}_G(z, Q^2)$. It is very useful to demonstrate
the difference at $Q^2_0$, at the starting point of the DGLAP evolution.

We begin the analysis with the consideration only the twist-four terms.
The results can be calculated from the final formulae of the previous
Section but it is simpler to 
repeat all calculations given in the Section V. At $Q^2=Q^2_0$ all
results simplify essentially because the leading-twist parton
distributions are constant $A_q$ and $A_G$ at the point.

\subsection{ 
Parton distributions at $Q^2_0$}
 \label{Sec9:f_q}

From the Eqs.~(\ref{flat}) and (\ref{fq:total}) we can easy obtain at 
$Q^2 = Q^2_0$,
that
\begin{subequations}
\label{fQ0:total}
\begin{equation}
f_a(z, Q^2_0) \ = \ A_a^{\tau2} \ + \
 \dfrac{1}{Q^2_0} \, f^{R\tau4}_a(z, Q^2_0)  \ + \
 \dfrac{1}{Q^4_0} \, f^{R\tau6}_a(z, Q^2_0) \ ,
\label{fq:Q_0}
\end{equation}
where at the LO
\begin{align}
f^{R\tau4}_q(z, Q^2_0) \, =& \,
 \dfrac{64C_FT_Rf}{15\beta_0^2} \, {\mathrm a}_q^{\tau4} \, \Biggl[ A_q^{\tau2}
 \left\{ \ln \left(\dfrac{1}{z_q}\right) \,
 \ln \left(\dfrac{Q_0^2}{z_q \left|{\mathrm a}_q^{\tau4}\right|}\right) -
 p^\prime(\nu_q) \right\}
\nonumber \\
 &\hspace{8mm}
 + \, \dfrac{2T_Rf}{3C_A}
 \left(A_G^{\tau2} + \dfrac{C_F}{C_A} A_q^{\tau2} \right)
 \ln \left(\dfrac{Q_0^2}{z^2 \left|{\mathrm a}_q^{\tau4}\right|}\right)
 \Biggr]
\label{fq4:Q_0}\\
f^{R\tau4}_G(z, Q^2_0) \, =& \,
 \dfrac{8}{5\beta_0^2} \, \dfrac{{\mathrm a}_G^{\tau4}}{a_s(Q_0^2)} \,
 \Biggl[ A_G^{\tau2}
 \ln \left(\dfrac{Q_0^2}{z^2 \left|{\mathrm a}_G^{\tau4}\right|}\right)
 + 2 \dfrac{C_F}{C_A} A_q^{\tau2} p(\nu_G) \Biggr] \ ,
\label{fG4:Q_0}\\
f^{R\tau6}_a(z, Q^2_0) \ =& \ -\dfrac{8}{7} \times \left[
 f^{R\tau4}_a(z, Q^2_0) ~\text{ with }~ 
  {\mathrm a}_a^{\tau4} \to {\mathrm a}_a^{\tau6} \, , \
  \ln \left(\dfrac{Q^2}{\left|{\mathrm a}_a^{\tau4}\right|}\right) \to
  \ln \left(\dfrac{Q^2}{\sqrt{\left|{\mathrm a}_a^{\tau6}\right|}}\right)
 \right] \ .
\label{fa6:Q_0}
\end{align}
\end{subequations}

Thus, the total parton distributions $f_q(z, Q^2_0)$ and $f_G(z, Q^2_0)$ 
are strongly deviated for the corresponding the twist-two densities
$f^{\tau2}_q(z, Q^2_0) = A^{\tau2}_q$ and
$f^{\tau2}_G(z, Q^2_0) = A^{\tau2}_G$.
Because usually the fitted values of ${\mathrm a}_q^{\tau4}$ 
(${\mathrm a}_G^{\tau4}$) are positive (negative), the twist-four terms
lead to positive and negative contributions in the case of quark and
gluon densities, respectively. The twist-six terms do not change the
results essentially.

\subsection{ The effective
slopes of $F_2$ and of parton distributions at $Q^2_0$}
 \label{Sec9:slopes}

To estimate the values of the effective slopes at low $Q^2$ values we
can look on their behavior at $Q^2_0$, where our formulae simplifies 
essentially. In the approximation, when the twist six contributions are
negligible, we can easy obtain from the Eqs.~(\ref{fQ0:total})
\begin{subequations}
\label{dlnz:Q04}
\begin{align}
 \lambda^{\text{eff}, R}_{q}(z, Q^2_0) \, &= \, 
 \dfrac{64C_FT_Rf}{15\beta_0^2} \, \dfrac{{\mathrm a}_q^{\tau4}}{Q^2_0} \,
 \Biggl\{
 \ln \left(\dfrac{Q_0^2}{z_q^2 \left|{\mathrm a}_G^{\tau4}\right|}\right) \,
 + \, \dfrac{4T_Rf}{3C_A} \left( \dfrac{A_G^{\tau2}}{A_q^{\tau2}} +
 \dfrac{C_F}{C_A} \right) \Biggr\} \ ,
\label{Slopq:Q0} \\
\lambda^{\text{eff}, R}_{G}(z, Q^2_0) \, &= \, 
 \dfrac{16}{5\beta_0^2} \, \dfrac{{\mathrm a}_G^{\tau4}}{a_s(Q^2_0) Q^2_0} \ ,
\label{SlopG:Q0} \\
\lambda^{\text{eff}, R}_{F2}(z,Q^2_0) \, &= \, 
 \dfrac{64C_FT_Rf}{15\beta_0^2} \, \dfrac{1}{Q^2_0} \, \left[
 {\mathrm a}_q^{\tau4} \Biggl\{
 \ln \left(\dfrac{Q_0^2}{z_q^2 \left|{\mathrm a}_G^{\tau4}\right|}\right) \,
 + \, \dfrac{4T_Rf}{3C_A} \left( \dfrac{A_G^{\tau2}}{A_q^{\tau2}} +
 \dfrac{C_F}{C_A} \right) \Biggr\} \, + \, \dfrac{{\mathrm a}_G^{\tau4}}{C_F}
 \dfrac{A_G^{\tau2}}{A_q^{\tau2}} \right] \ .
\label{SlopF2:Q0} 
\end{align}
\end{subequations}

Because ${\mathrm a}_G^{\tau4}<0$,
it is easy to see that $\lambda^{\text{eff}, R}_{F2}(z, Q^2_0) <
\lambda^{\text{eff}, R}_{q}(z, Q^2_0)$. This indicates that the inequality
$\lambda^{\text{eff}}_{F2}(z, Q^2) > \lambda^{\text{eff}}_{q}(z, Q^2)$
seems to be correct only at quite large 
$Q^2$ values (see also the previous section and discussions therein),
where the twist-two terms give basic contributions. 

Note also, that at $Q^2_0$ the slope $\lambda^{\text{eff}, R}_{q}(z, Q^2_0)$
rises at $x \to 0$, but the gluon slope $\lambda^{\text{eff}, R}_{G}(z, Q^2_0)$
is negative and $x$-independent.
Thus, $\lambda^{\text{eff}, R}_{q}(z,Q^2) >
\lambda^{\text{eff}, R}_{G}(z, Q^2_0)$ at low $Q^2$ that is in full agreement
with the recent experimental data from HERA (see, for example, the
review \cite{Cooper-Sarkar:1998}).
%
%
The twist-six terms do not change the above
results essentially.

\section{Results of the fits}
 \label{Sec:10}

With the help of the results obtained in the previous sections we have
analyzed $F_2(x, Q^2)$ HERA data at small $x$ from the H1
\cite{Adloff:2001,Adloff:1999,Adloff:1997,Aid:1996,Ahmed:1995,Abt:1993}
and ZEUS
\cite{Chekanov:2001,Breitweg:2000,Breitweg:1999,Breitweg:1997,Derrick:1996:C72,Derrick:1996:C69,Derrick:1995,Derrick:1993} collaborations as separately,
as well as together. 

Without higher-twist corrections our solution of the
DGLAP equations depends on five parameters, i.e. $Q_0^2$, $x_0$,
$A^{\tau2}_G$, $A^{\tau2}_q$ and $\Lambda$ (or, equally well, on
$\alpha_\text{s}(M_{\text{Z}})$). 
The incorporation of twist-four and twist-six corrections leads to
two and four additional parameters, respectively.

In order to keep the analysis as simple as possible we have fixed
$\Lambda_{\ms}$ to the values given in Eq.~(\ref{Lambda:LO+NLO}),
which corresponds to $\alpha_\text{s}(M_{\text{Z}}) = 0.1166$, obtained
recently by ZEUS \cite{Chekanov:2001}.
The analyzed data region was restricted to $x < 0.01$ to stay in the
kinematic region where our results are expected to be applicable.
The $\chi^2$ minimizations were done with MINUIT \cite{MINUIT}.
In the fits the errors 
are statistical and systematical added in quadrature. 
Finally, the number of active flavors was fixed to $f = 3$ and $4$ for
comparison.

\subsection{Leading twist approximation}
 \label{Sec10:Fit:LT}

Tables~\ref{Tab:H1+ZEUS:96/97} and \ref{Tab:H1+ZEUS} summarize the results
of the fits to H1 and ZEUS data using twist-two formulas at LO
(\ref{F2:tau2:LO}) and NLO (\ref{F2:tau2:NLO})
approximations.

\begin{figure}[t]
\begin{center}
  \setlength{\unitlength}{1.4mm}
\begin{picture}(125,125)   
 \put(5,5){
  \centering\epsfig{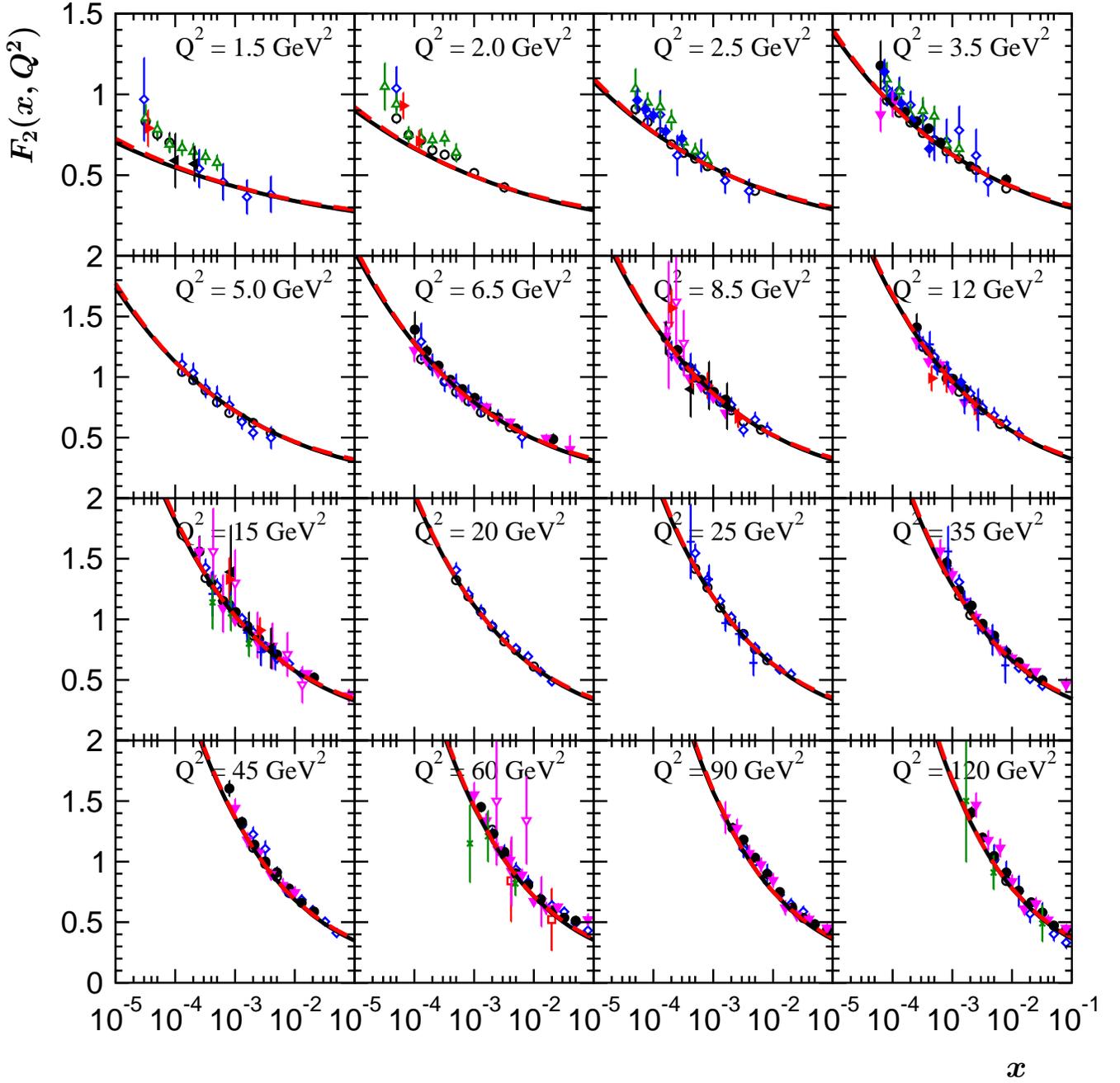}}
%
 \put(115,0){\bfseries\Large
  $\boldsymbol{x}$}
 \put(0,105){\bfseries\Large
  \begin{sideways}
   $\boldsymbol{F_2(x, Q^2)}$
  \end{sideways}}
\end{picture}
\end{center}
 \caption{\label{Fig:F2-LOt2}%
$F_2^{\tau2}(x, Q^2)$ as a function of $x$ for different $Q^2$ bins.
The experimental points are from H1\protect\cite{Adloff:1999,Adloff:1997,Aid:1996,Ahmed:1995,Abt:1993} (open points) and ZEUS\protect\cite{Chekanov:2001,Breitweg:2000,Breitweg:1999,Breitweg:1997,Derrick:1996:C72,Derrick:1996:C69,Derrick:1995,Derrick:1993} (solid points).
The solid, black line represents the NLO fit with
$\chi^2/\text{n.d.f.} = 798/611 = 1.31$
[$A_G^{\tau2} = -.020, ~A_q^{\tau2} = .903, ~Q_0^2 = .495~\text{GeV}^2$].
The long dashed, red line represents the LO fit with
$\chi^2/\text{n.d.f.} = 754/611 = 1.23$
[$A_G^{\tau2} = .784, ~A_q^{\tau2} = .801, ~Q_0^2 = .304~\text{GeV}^2$].
}%
\end{figure}

We can see in Tables ~\ref{Tab:H1+ZEUS:96/97} and \ref{Tab:H1+ZEUS} 
and in Fig.~\ref{Fig:F2-LOt2} that the qualities of the fits are very similar
for the LO and NLO approximations. 
This suggest that perturbation theory works well in the small $x$ regime.
This is in accord with Refs.~\cite{Shirkov:1995,Kotikov:1994:R=,Kim:1998}
(see also recent review \cite{Andersson:2002}), where it was shown, that the
argument of the strong coupling constant is effectively much larger as $Q^2$
in the small $x$ domain.

Hovewer the similarity of the results found at LO and NLO fits
does not agree with our previous analysis
\cite{Kotikov:1999}, where NLO corrections essentially improved the
comparison between QCD and experiment.
This disagreement relates mostly to the incorrect use of the same value of
the QCD parameter $\Lambda$ in \cite{Kotikov:1999} in both LO and NLO
cases. By contrast, $\Lambda$ should be different
(see \cite{Bethke:2002}). They are extracted from 
$\alpha_\text{s}(M_{\text{Z}})$ by using $b$- and $c$-quarks thresholds
following to \cite{Chetyrkin:1997}. 
The values of $\Lambda$ obtained by this procedure and
used hereafter in all the fits are:
\begin{equation}
\begin{array}{lll}
\Lambda_{\text{LO}}(f=5) \ = \ 80.80~\text{[MeV]} \ , &
\Lambda_{\text{LO}}(f=4) \ = \ 111.8~\text{[MeV]} \ , &
\Lambda_{\text{LO}}(f=3) \ = \ 136.8~\text{[MeV]} \ ,
 \\
\Lambda_{\ms}(f=5) \ = \ 195.7~\text{[MeV]} \ , &
\Lambda_{\ms}(f=4) \ = \ 284.0~\text{[MeV]} \ , &
\Lambda_{\ms}(f=3) \ = \ 347.2~\text{[MeV]} \ ,
\end{array}
\label{Lambda:LO+NLO}
\end{equation}
obtained from ZEUS result $\alpha_\text{s}(M_{\text{Z}})=0.1166$ (see 
\cite{Chekanov:2001}).

Table~\ref{Tab:H1+ZEUS:96/97} contains the results of separate fits to
H1 and ZEUS data with a low  $Q^2$ cut, $Q^2_{\text{cut}}$, that
increases step by step. We observe that the agreement between theory and 
experiment improves when increasing the value of
$Q^2_{\text{cut}}$. For $Q^2 \geq 2.5~\text{GeV}^2$ the agreement is good
(see Tables~\ref{Tab:H1+ZEUS:96/97} and \ref{Tab:H1+ZEUS}).

Note that the separated fits of H1 and ZEUS data lead to purely comparable
values of the parameters $Q_0^2$, $x_0$, $A^{\tau2}_G$, $A^{\tau2}_q$.
Thus we may fit to the combined data set.
The results of such combined fits can be found in the last rows of
Table~\ref{Tab:H1+ZEUS:96/97} and Table~\ref{Tab:H1+ZEUS}.

Looking carefully on that Tables, we arrive to the following conclusions:
\begin{itemize}
\item
In the leading twist approximation the
preferred number of flavors $f$ is four.
\item
The value of the quark distribution does not depend on the specific
$Q^2_\text{cut}$ values within the limits of experimental errors.
The magnitude of the gluon density and $Q^2_0$ decrease slowly with
decreasing $Q^2_\text{cut}$.
\item
A strong reduction of the magnitude of the gluon density is observed
when NLO corrections are included.

The suppression of the gluon density rise with $Q^2$ at NLO 
in comparison with the LO prediction is well-known effect
\cite{Tung:1989,Kotikov:1993} but in addition we also observe
a strong reduction of the gluon magnitude at $Q^2_0$. 

At least partialy,
this effect can be explained based on the GRV-like point of
view \cite{Gluck:1992,Gluck:1993,Gluck:1995,Gluck:1998}, where 
at low $Q^2$ values there are only valence quarks and all other types of
partons are generated in the $Q^2$-evolution. Thus, the slowe rise with 
$Q^2$ when NLO corrections are included directly implies a reduction of 
the magnitude at a given $Q_0^2$.

It should be mentioned that
a similar relative reduction of gluon normalization is obtained 
in the analyses \cite{Forshaw:1995,Ball:1994}, when the $\ln(1/x)$
resummation was included.
Thus, the correct incorporation of NLO terms has a similar tendency.
\item
The fitted $Q^2_0$ values are essentially higher at NLO:
$Q^2_0 \sim 0.5 \div 0.6~\text{GeV}^2$, in comparison with LO fits, 
where $Q^2_0 \sim 0.3 \div 0.4~\text{GeV}^2$, and
comparable to those obtained earlier in \cite{Kotikov:1999}.

Partialy, the effect can be explained by different $\Lambda$ values
at LO and NLO approximations. Note, however, that the ratio
$\Lambda_{\ms}^2/\Lambda_{\text{LO}}^2 \sim 6.4$ and, thus, the $Q^2$ 
dependence of $F_2$ data itself should be important in the 
definition of $Q^2_0$.
\end{itemize}

Considering
Tables~\ref{Tab:H1+ZEUS:96/97} and \ref{Tab:H1+ZEUS} and Fig. 1
we find good agreement with data only at $Q^2 \geq 2.5~\text{GeV}^2$.
The situation is little bit worse than it was before in
\cite{Kotikov:1999}, 
mainly due to the  strong improvement of experimental data. To expand the
range of applicability of our analysis to $Q^2 < 2.5~\text{GeV}^2$
we add to our fits HT
corrections presented in the previous sections.

Let's consider both types of estimations of the HT corrections separately.

\subsection{BFKL-motivated estimations for twist-four operators}
 \label{Sec10:Fit:BFKL}

Tables \ref{Tab:H196/97:HT}--\ref{Tab:H1+ZEUS:HT} and Fig. 4 contain the 
results of the fits to H1 and ZEUS data using Eqs.~(\ref{F2:tau2:LO}),
(\ref{F2:BFKL}) and (\ref{tw2+BFKL}) at LO and 
(\ref{F2:tau2:NLO}), (\ref{F2:BFKL}) and (\ref{tw2+BFKL}) at NLO.  

The results demonstrate good agreement between the theoretical predictions
having BFKL-like
twist-four term and experimental data of the H1 
\cite{Aid:1996,Adloff:1997} and ZEUS \cite{Derrick:1996:C72,Derrick:1996:C69} collaborations.

The fits of H1 \cite{Aid:1996,Adloff:1997} and ZEUS 
\cite{Derrick:1996:C72,Derrick:1996:C69} data
demonstrate a strong improvement of the agreement between theory and 
experiment (see  Fig. 4), essentially at LO and in the case $f=4$. 
%

The values of parameters in the twist-two terms do not change drastically.
$Q^2_0$  rises 100 MeV and 150 MeV at LO and NLO, respectively. 
The  gluon density in the twist-two term rises essentially
and the quark distribution decreases slowly. The changes are compensated by
a negative gluon and a positive quark twist-four magnitudes, respectively.

We found also a tiny
dependence on the real value of the parameter 
`$b$', that supports our hypothese (see Section~\ref{Sec:3}) about the
irrelevance of the exact form for the nonsingular (at $n\to1$) terms in the
twist-four anomalous dimensions.

An interesting fact is that the value of the sum
$A^{\tau4}_G + 4/9 A^{\tau4}_q$ is very close to zero. Hence, HERA
data do not seem to support a strong increase of the twist-four
terms at small $x$, contrary to the expectation from various
BFKL-motivated 
estimations \cite{Levin:1992,Bartels:1991,Bartels:1993:PLB,Bartels:1993:ZP}.
However a small value for the the twist-four terms has also been found
in  a model-dependent analysis \cite{Bartels:2000}.

\subsection{Renormalon model predictions for higher twist  operators}
 \label{Sec10:Fit:Renorm}

Tables~\ref{Tab:H196/97:HT}--\ref{Tab:Rht} and Figs.~\ref{Fig:F2-t2HT}
and \ref{Fig:F2sm-t2HT} contain the results of the fits to H1 and ZEUS data
using Eqs.~(\ref{F2:tau2:LO}) and  (\ref{F2:Rht}) at LO and 
(\ref{F2:tau2:NLO}) and (\ref{F2:Rht}) at NLO.
The results demonstrate excellent agreement between theoretical predictions
and experimental data.
The $\chi^2 $ decreases very strongly. 

Consider separately the fits of data for $Q^2 \geq 1.5~\text{GeV}^2$ and
$Q^2 \geq 0.5~\text{GeV}^2$,
presented in Tables~\ref{Tab:H196/97:HT}--\ref{Tab:H1+ZEUS:HT} 
(and on Fig. 4) and Table~\ref{Tab:Rht} (and on Fig. 5), respectively.

Looking carefully Tables~\ref{Tab:H196/97:HT}--\ref{Tab:H1+ZEUS:HT} and 
Fig.~\ref{Fig:F2-t2HT}, we arrive to the following conclusions:
\begin{itemize}
\item 
For the data usage of $f = 4$ is strongly preffered.
\item
The values of parameters in the twist-two terms do not change essentially.

We see, however, for H1 data in Table~\ref{Tab:H196/97:HT} and for combibed
data in Table~\ref{Tab:H1+ZEUS:HT} some rise of gluon terms when higher twist
terms are incorporated. The rise exists
for both the LO and NLO approximations and it is compensated by 
negative gluon twist-four magnitude.
The twist-six gluon magnitude has different signs (it is negative and
positive at LO and NLO approximations, respectively) but the
combination of the higher twist terms gives negative contribution
for the gluon case.

Note that the phenomenon is similar to one observed for BFKL-motivated
twist-four corrections (see previous subsection) and can be considered
as quite general property of the HT corrections.
\item
For the ZEUS data in Table~\ref{Tab:ZEUS96/97:HT} the influence of the higher
twist terms is not so important.
\item
In contrary to the gluon case, the higher twist corrections
for the quark density are mostly positive that leads to different
small-$x$ asymptotics of gluon and quark distributions at low $Q^2$
values, observed recently at HERA experiments \cite{Chekanov:2002}
(see a detailed discussions in the subsection F).
\item
The fitted value of $Q^2_0$ tends to be little higher (at LO $Q^2_0 \sim
0.5 $ GeV$^2$ and at NLO $Q^2_0 \sim
0.7 \div 0.8 $ GeV$^2$) when the twist-four corrections have been added.
It is in agreement with the results when
BFKL-motivated twist-four corrections have been considered (see the
previous sunbsection).
The incorporation of twist-six terms returns the  $Q^2_0$ values to the
ones, obtained in the twist-two approximation.

\end{itemize}

Looking carefully Table~\ref{Tab:Rht} and Fig.~\ref{Fig:F2sm-t2HT}, we see
full support of above results: the agreement with experimental data improves
drastically, essentially for $0.5~\text{GeV}^2 \leq Q2 \leq 2.5~\text{GeV}^2$.
We should note, however, about following  excepting features:
\begin{itemize}
\item 
Usage of $f = 3$ is 
preffered, that is natural choice at low $Q^2$ values.
\item
The twist-six corrections are important to stabilize the HT
contributions and,
thus, the results of Table~\ref{Tab:Rht}
are comparable with ones in Tables~\ref{Tab:H196/97:HT}--\ref{Tab:H1+ZEUS:HT}
only when the twist-six corrections taken into account.
\end{itemize}

\subsection{Leading and higher twist approximations
for the derivative $\partial F_2/\partial\ln Q^2$}
 \label{Sec10:Fit:deriv}

The results for the derivative $\partial F_2/\partial\ln Q^2$ are shown on
Fig.~\ref{Fig:dF2-LOt2} and Fig.~\ref{Fig:dF2-t2Rht} together with H1
experimental data \cite{Adloff:1999}.

Fig.~\ref{Fig:dF2-LOt2} contains only the leading twist theoretical 
predictions.
As in the case of $F_2$ data we have very good agreement between
our formulae and experimental data at $Q^2 \geq 3$ GeV$^2$.

When we added the HT corrections, the theoretical results begin to be in
agreement with experiment also at $Q^2 < 3~\text{GeV}^2$ (see
Fig.~\ref{Fig:dF2-t2Rht}), especially  when we used the results of
$F_2$ data fits  at $Q^2 \geq 0.5~\text{GeV}^2$. The corresponding
results for $\partial F_2/\partial\ln Q^2$ are shown as the dashed curve
for the NLO and as the dash-dotted curve for the LO fits.

Both curves are hardly distinguished from each other. It meens,
that in this kinematical region of small $x$ the order of perturbation theory
inside the leading twist does not matter. The importance has
the number of twists taking into account.

Note that the HT corrections to $F_2$ and $\partial F_2/\partial\ln
Q^2$ structure functions are opposite in sign that demonstrates the
importance, respectively, the quark density and gluon one for the
functions (see also the following subsection and discussions
therein). The fact is in full agreement with results of Section~\ref{Sec:6}.

%
%
%
%

Thus, our quite simple formulas obtained in the generalized DAS approach
are very convenient also to the
study the derivative $\partial F_2/\partial\ln Q^2$, which is very
important to extract gluon density and the longitudinal $F_L$ or the
ration $R=\sigma_L/\sigma_T$ (see 
\cite{Kotikov:1994vb,Kotikov:1996:PL}
 and \cite{Kotikov:1995,Kotikov:1997:MPL,Kotikov:1997:JETP},
respectively).

\subsection{Effective slope $\boldsymbol{\lambda_{F_2}^\text{eff}(x, Q^2)}$}
 \label{Sec10:Fit:slope}

\begin{figure}[t]
\begin{center}
  \setlength{\unitlength}{1.4mm}
\begin{picture}(125,125)   
 \put(5,5){
  \centering\epsfig{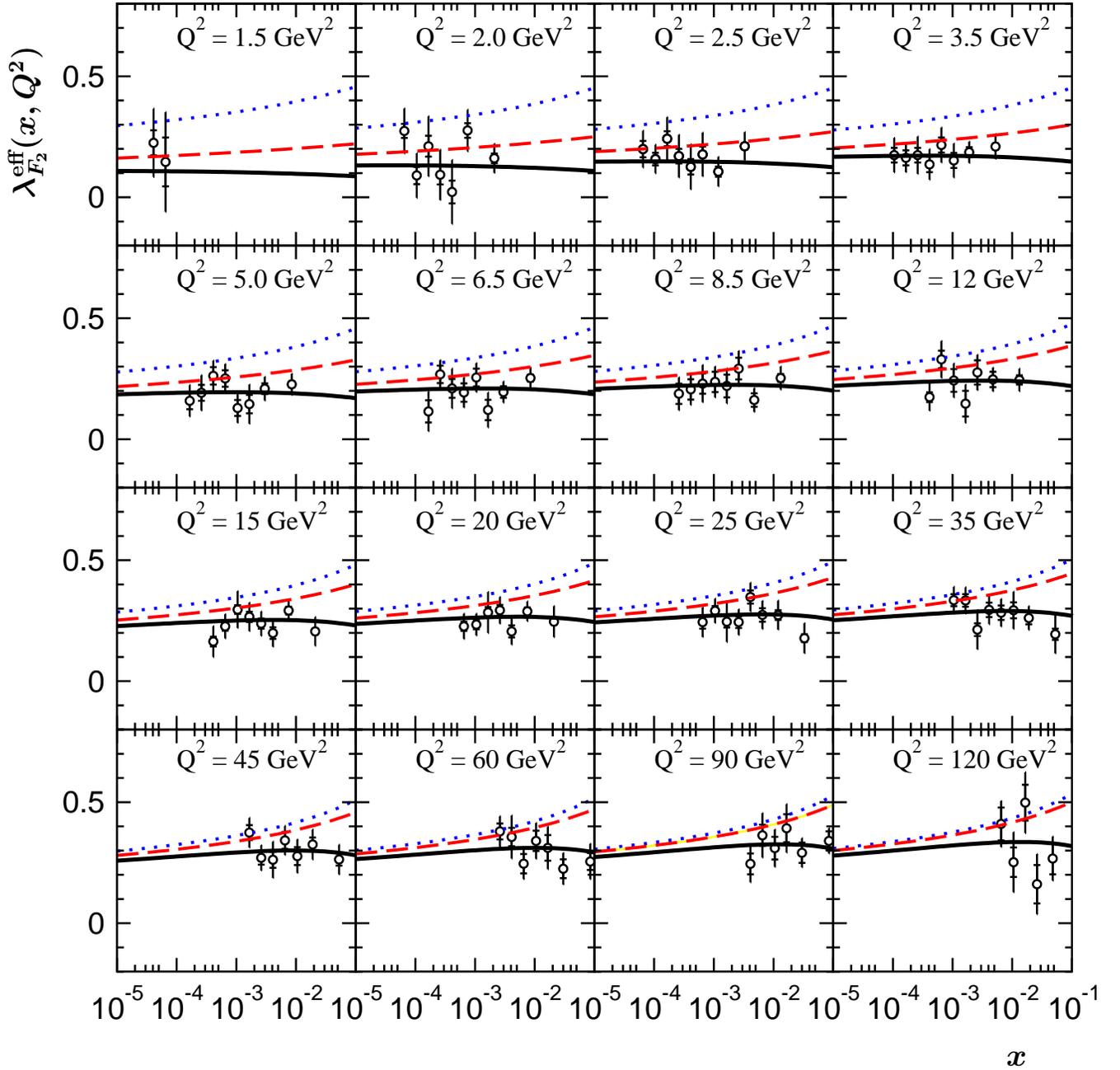}}
%
 \put(115,0){\bfseries\Large
  $\boldsymbol{x}$}
 \put(0,100){\bfseries\Large
  \begin{sideways}
   $\boldsymbol{\lambda_{F_2}^\text{eff}(x, Q^2)}$
  \end{sideways}}
\end{picture}
\end{center}
 \caption{\label{Fig:LH1-LOt2}%
The derivative function (effective slope)
$\lambda_{F_2}^{\text{eff}, \tau2}(x, Q^2) =
\partial\ln{F_2^{\tau2}(x, Q^2)}/\partial\ln{(1/x)}$
as a function of $x$ for different $Q^2$ bins.
The experimental points are from H1\protect\cite{Adloff:1999}. The outer error
bars include statistical and systematical errors added in quadrature, while the
inner error bars correspond to statistical errors only.
The solid, black line represents the NLO fit
with $\chi^2/\text{n.d.f.} = 798/611 = 1.31$
[$A_G^{\tau2} = -.020, ~A_q^{\tau2} = .903, ~Q_0^2 = .495~\text{GeV}^2$],
while the long dashed, red line (hardly distinguished from the solid one) is
the LO fit with $\chi^2/\text{n.d.f.} = 754/611 = 1.23$
[$A_G^{\tau2} = .784, ~A_q^{\tau2} = .801, ~Q_0^2 = .304~\text{GeV}^2$].
The dotted, blue line corresponds to the asymptotic expression
$\lambda_{F_2,\text{as}}^{\text{eff},\tau2}(x, Q^2)$ in
Eq.~(\ref{F2as-Slope:NLO}).
}%
\end{figure}

\begin{figure}[t]
\begin{center}
  \setlength{\unitlength}{1.4mm}
\begin{picture}(125,125)   
 \put(5,5){
  \centering\epsfig{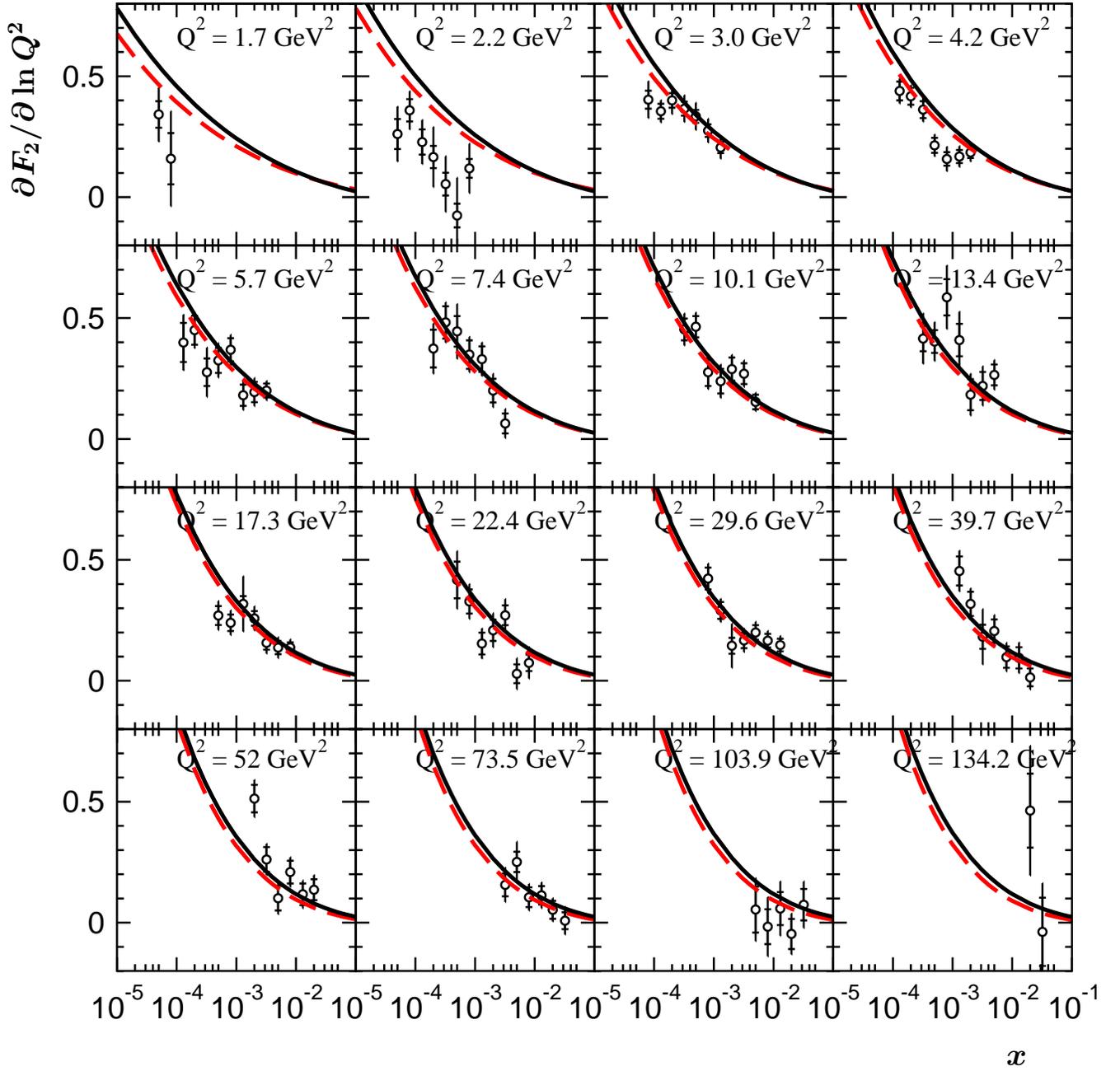}}
%
 \put(115,0){\bfseries\Large
  $\boldsymbol{x}$}
 \put(0,100){\bfseries\Large
  \begin{sideways}
   $\boldsymbol{\partial{F_2}/\partial\ln{Q^2}}$
  \end{sideways}}
\end{picture}
\end{center}
 \caption{\label{Fig:dF2-LOt2}%
The derivative function $\partial F_2^{\tau2}(x, Q^2)/\partial\ln{Q^2}$ taken
at fixed $Q^2$ and plotted as a function of $x$.
The experimental points are from H1\protect\cite{Adloff:1999}. The outer error
bars represent the quadratic sum of statistical and systematical errors. The
inner error bars show the statistical error only.
The solid, black line represents the NLO fit
with $\chi^2/\text{n.d.f.} = 798/611 = 1.31$
[$A_G^{\tau2} = -.020, ~A_q^{\tau2} = .903, ~Q_0^2 = .495~\text{GeV}^2$],
while the long dashed, red line is the LO fit
with $\chi^2/\text{n.d.f.} = 754/611 = 1.23$
[$A_G^{\tau2} = .784, ~A_q^{\tau2} = .801, ~Q_0^2 = .304~\text{GeV}^2$].
}%
\end{figure}

\begin{figure}[t]
\begin{center}
  \setlength{\unitlength}{1.4mm}
\begin{picture}(125,125)   
 \put(5,5){
  \centering\epsfig{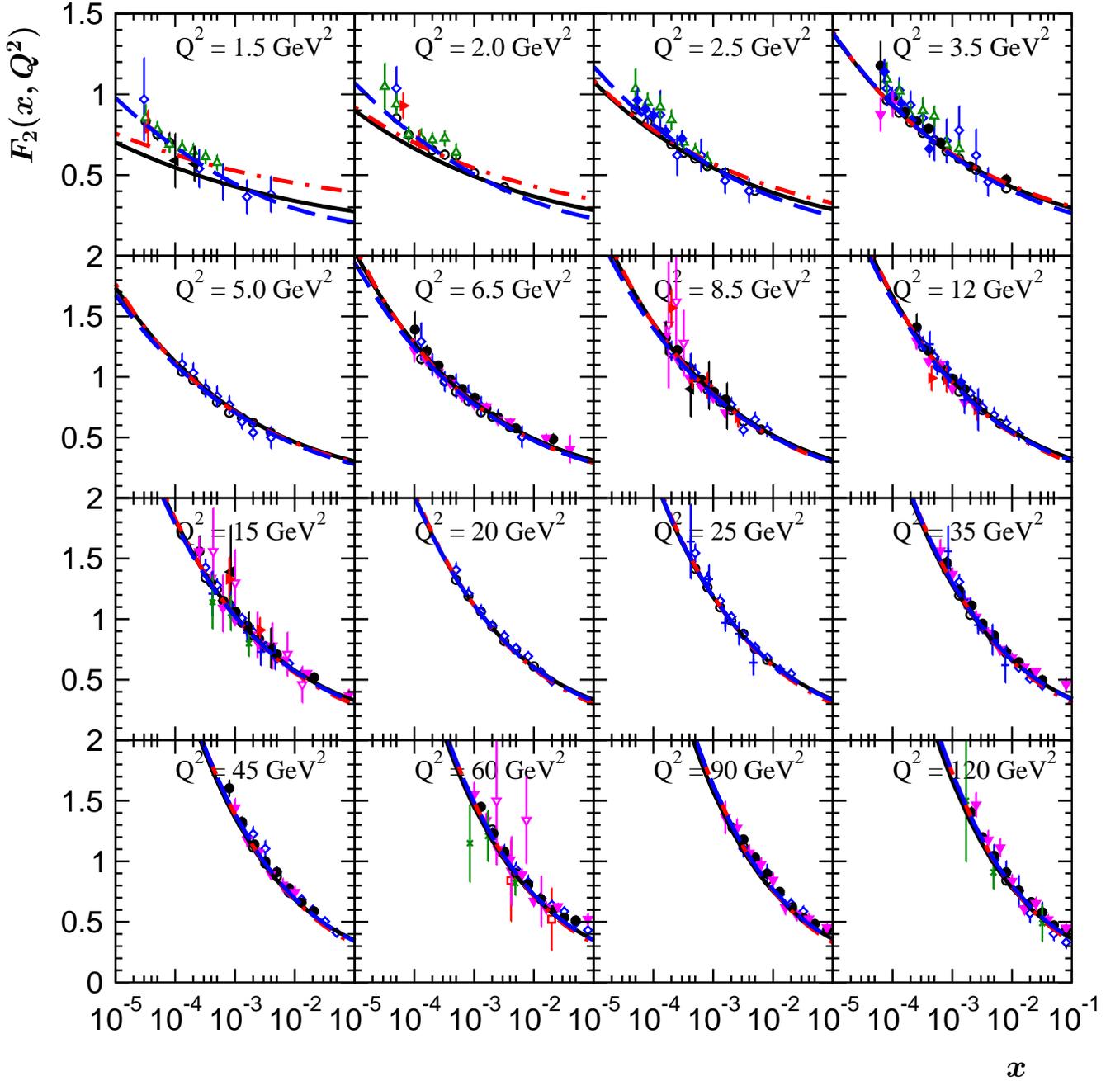}}
%
 \put(115,0){\bfseries\Large
  $\boldsymbol{x}$}
 \put(0,105){\bfseries\Large
  \begin{sideways}
   $\boldsymbol{F_2(x, Q^2)}$
  \end{sideways}}
\end{picture}
\end{center}
 \caption{\label{Fig:F2-t2HT}%
$F_2(x, Q^2)$ as a function of $x$ for different $Q^2$ bins.
The experimental points are
 the same as on Figure~\ref{Fig:F2-LOt2}.
The solid, black line represents the NLO fit alone with
$\chi^2/\text{n.d.f.} = 798/611 = 1.31$
[$A_G^{\tau2} = -.020, ~A_q^{\tau2} = .903, ~Q_0^2 = .495~\text{GeV}^2$].
The dash-dotted, red curve represents the BFKL-motivated estimations for higher
twist contribution to $F_2(x, Q^2)$ with the value of the parameter
$b = a^2/2$. The corresponding 
$\chi^2/\text{n.d.f.} = 629/609 = 1.03$
[$A_G^{\tau2} = .301, ~A_q^{\tau2} = .535, ~Q_0^2 = .631~\text{GeV}^2$ and
$A_G^{\tau4} = -.580~\text{GeV}^2, ~A_q^{\tau4} = 1.311~\text{GeV}^2$].
The dashed, blue curve is obtained from the fits at the NLO, when the
renormalon contributions of higher-twist terms have been incorporated.
The corresponding
$\chi^2/\text{n.d.f.} = 500/607 = 0.82$
[$A_G^{\tau2} = .041, ~A_q^{\tau2} = .824, ~Q_0^2 = .493~\text{GeV}^2$ and
${\mathrm a}^{\tau4}_G = -2.765~\text{GeV}^2,
~{\mathrm a}^{\tau4}_q = .676~\text{GeV}^2$,
${\mathrm a}^{\tau6}_G = .939~\text{GeV}^4,
~{\mathrm a}^{\tau6}_q = .252~\text{GeV}^4$].
}%
\end{figure}

\begin{figure}[t]
\begin{center}
  \setlength{\unitlength}{1.4mm}
\begin{picture}(125,125)   
 \put(5,5){
  \centering\epsfig{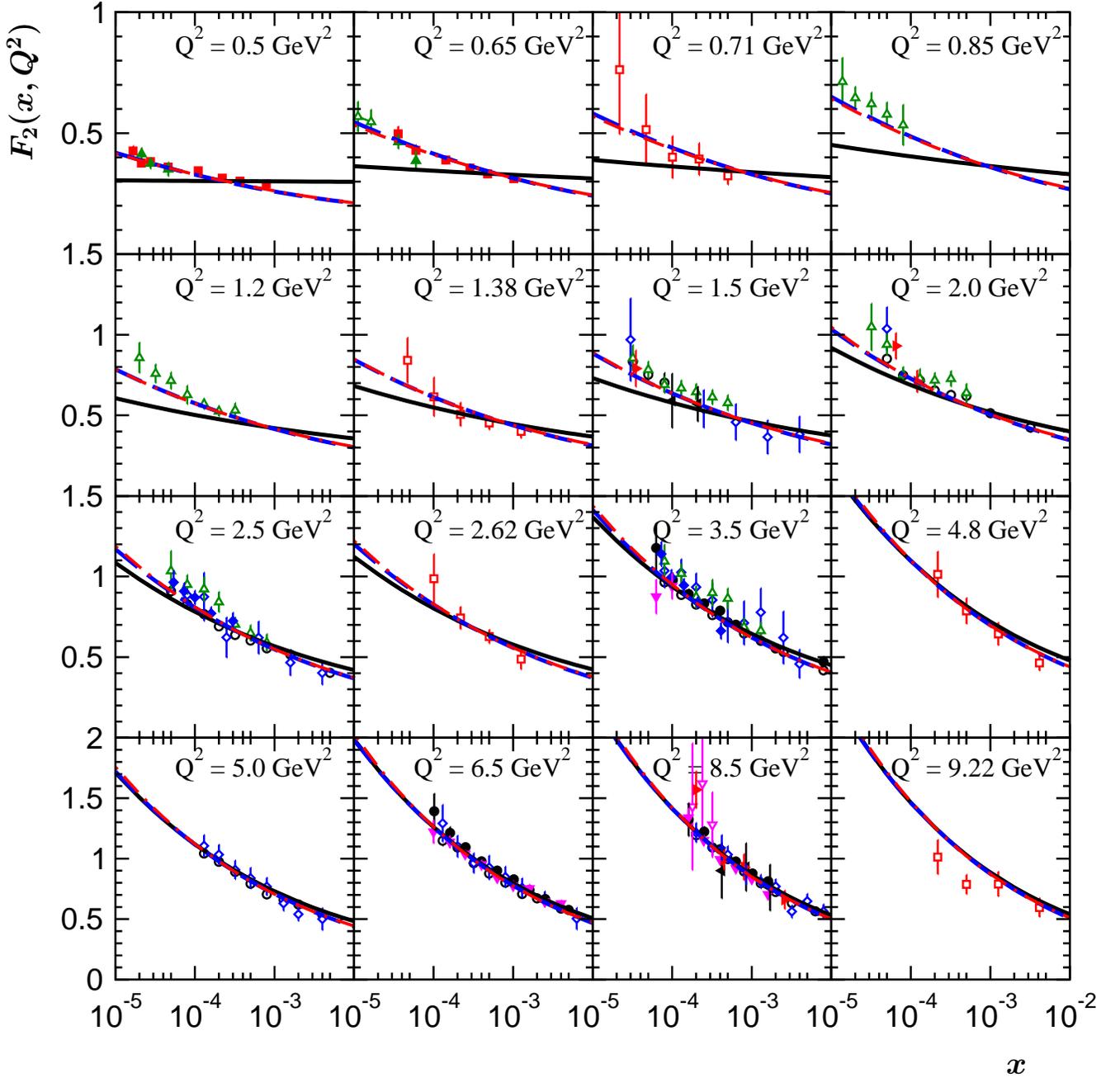}}
%
 \put(115,0){\bfseries\Large
  $\boldsymbol{x}$}
 \put(0,105){\bfseries\Large
  \begin{sideways}
   $\boldsymbol{F_2(x, Q^2)}$
  \end{sideways}}
\end{picture}
\end{center}
 \caption{\label{Fig:F2sm-t2HT}%
$F_2(x, Q^2)$ as a function of $x$ for different $Q^2$ bins.
The experimental points are
from H1\protect\cite{Adloff:1999,Adloff:1997,Aid:1996,Ahmed:1995,Abt:1993} (open points) and ZEUS\protect\cite{Chekanov:2001,Breitweg:2000,Breitweg:1999,Breitweg:1997,Derrick:1996:C72,Derrick:1996:C69,Derrick:1995,Derrick:1993} (solid points).
The solid, black line represents the NLO fit alone with
$\chi^2/\text{n.d.f.} = 798/611 = 1.31$
[$A_G^{\tau2} = -.020, ~A_q^{\tau2} = .903, ~Q_0^2 = .495~\text{GeV}^2$].
The dashed, blue curve is obtained from the fit at the NLO, when the
renormalon contributions of higher-twist terms have been incorporated.
The corresponding
$\chi^2/\text{n.d.f.} = 565/660 = 0.86$
[$A_G^{\tau2} = .279, ~A_q^{\tau2} = .640, ~Q_0^2 = .672~\text{GeV}^2$ and
${\mathrm a}^{\tau4}_G = -.143~\text{GeV}^2,
~{\mathrm a}^{\tau4}_q = .140~\text{GeV}^2$,
${\mathrm a}^{\tau6}_G = -.044~\text{GeV}^4,
~{\mathrm a}^{\tau6}_q = .043~\text{GeV}^4$].
The dash-dotted, red curve (hardly distinguished from the dashed one)
represents the fit at the LO together with the renormalon contributions
of higher-twist terms. The corresponding
$\chi^2/\text{n.d.f.} = 555/660 = 0.84$
[$A_G^{\tau2} = 1.211, ~A_q^{\tau2} = .539, ~Q_0^2 = .404~\text{GeV}^2$ and
${\mathrm a}^{\tau4}_G = -.002~\text{GeV}^2,
~{\mathrm a}^{\tau4}_q = .102~\text{GeV}^2$,
${\mathrm a}^{\tau6}_G = .001~\text{GeV}^4,
~{\mathrm a}^{\tau6}_q = .031~\text{GeV}^4$].
}%
\end{figure}

\begin{figure}[t]
\begin{center}
  \setlength{\unitlength}{1.4mm}
\begin{picture}(125,125)   
 \put(5,5){
  \centering\epsfig{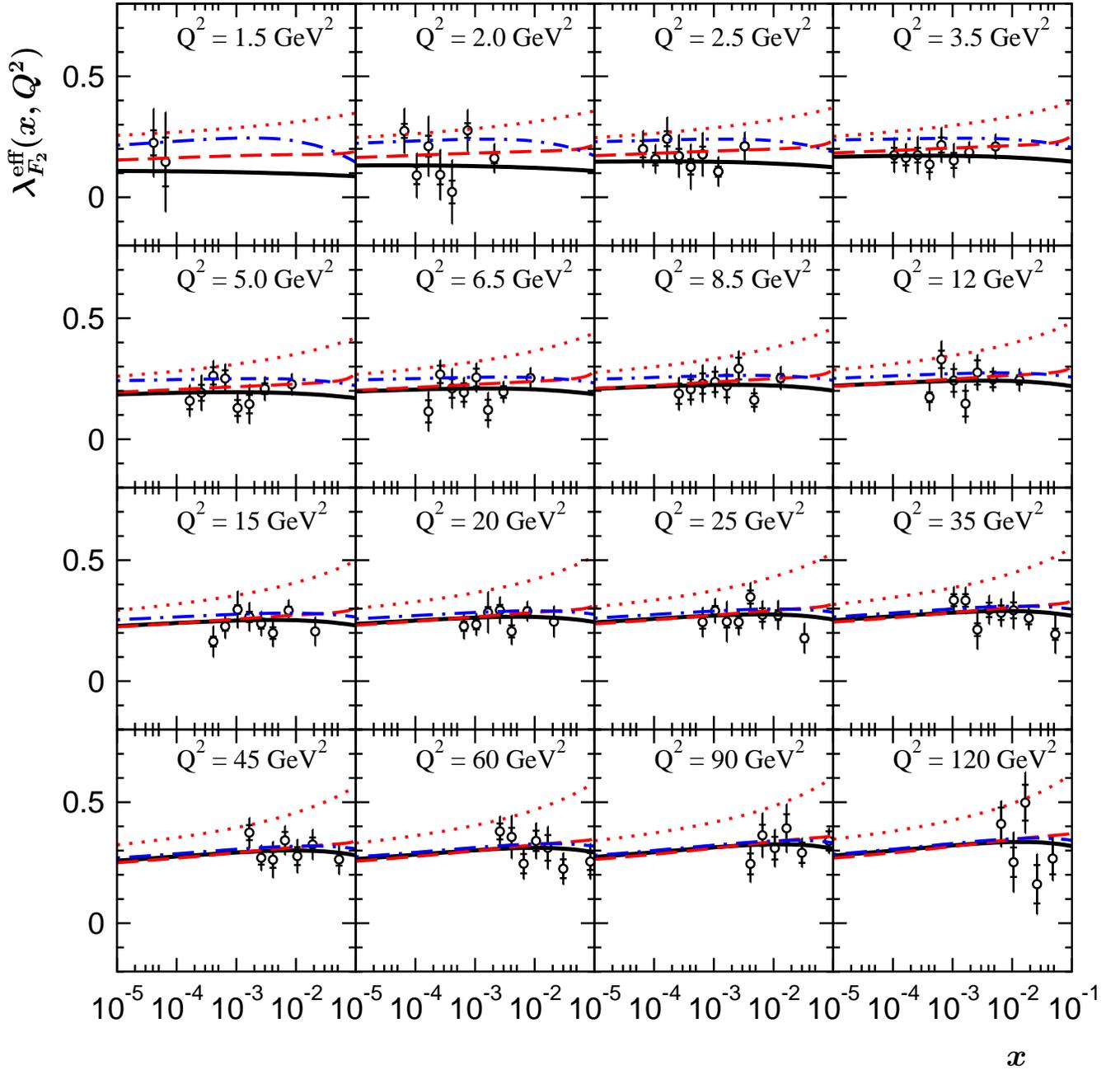}}
%
 \put(115,0){\bfseries\Large
  $\boldsymbol{x}$}
 \put(0,100){\bfseries\Large
  \begin{sideways}
   $\boldsymbol{\lambda_{F_2}^\text{eff}(x, Q^2)}$
  \end{sideways}}
\end{picture}
\end{center}
 \caption{\label{Fig:LH1-Rht}%
The derivative function (effective slope)
$\lambda_{F_2}^\text{eff} = \partial\ln{F_2(x, Q^2)}/\partial\ln{(1/x)}$
as a function of $x$ for different $Q^2$ bins. The experimental points and the
solid, black line (NLO fit with $\chi^2/\text{n.d.f.} = 798/611 = 1.31$%
) are the same as on Figure~\ref{Fig:LH1-LOt2}. All other curves are obtained
from the fits, when the renormalon contributions of higher-twist terms have
been incorporated.
The dashed, red one is the same as on the Figure~\ref{Fig:F2sm-t2HT} with the
corresponding
 $\chi^2/\text{n.d.f.} = 565/660 = 0.86$,
while the dash-dotted, blue line is the one from the Figure~\ref{Fig:F2-t2HT}
with
 $\chi^2/\text{n.d.f.} = 500/607 = 0.82$.
The dotted, red line corresponds to the asymptotic LO expression
$\lambda_{F_2, \text{as}}^{\text{eff}}(x, Q^2)$ in
Eq.~(\ref{SlopF2:tw4}), plotted at
 $\chi^2/\text{n.d.f.} = 573/667 = 0.87$
[$A_G^{\tau2} = 1.234, ~A_q^{\tau2} = .518, ~Q_0^2 = .407~\text{GeV}^2$ and
${\mathrm a}^{\tau4}_G =  .201~\text{GeV}^2,
~{\mathrm a}^{\tau4}_q = -.011~\text{GeV}^2$].
}%
\end{figure}

\begin{figure}[t]
\begin{center}
  \setlength{\unitlength}{1.4mm}
\begin{picture}(125,125)   
 \put(5,5){
  \centering\epsfig{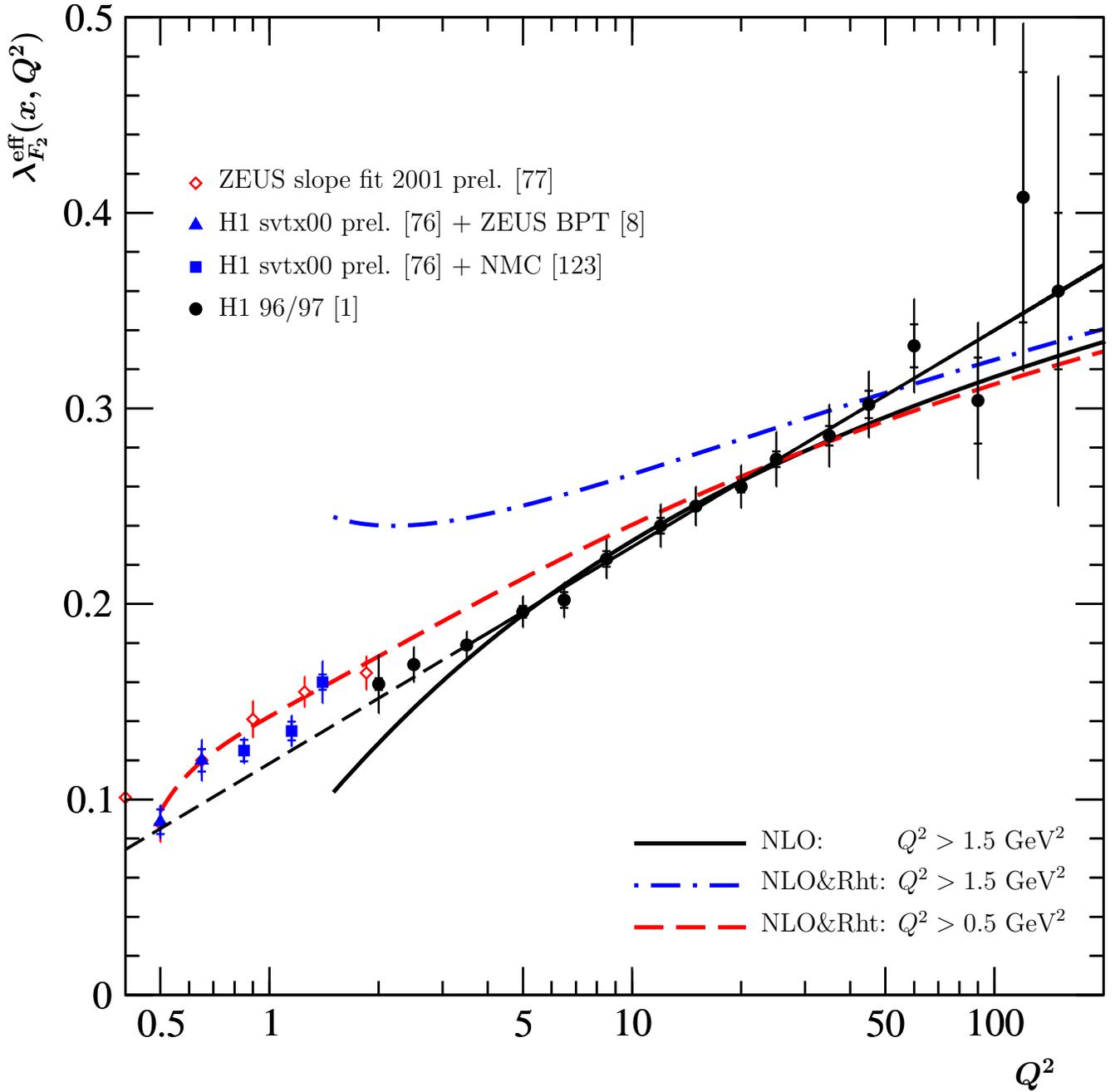}}
%
 \put(115,0){\bfseries\Large
  $\boldsymbol{Q^2}$}
 \put(0,103){\bfseries\Large
  \begin{sideways}
   $\boldsymbol{\lambda_{F_2}^\text{eff}(x, Q^2)}$
  \end{sideways}}
 \put(24,102.5){\large
  ZEUS slope fit 2001 prel. \protect\cite{Newman:2003}}
 \put(24, 97.6){\large
  H1 svtx00 prel. \protect\cite{Lastovicka:2002} $+$
  ZEUS BPT \protect\cite{Breitweg:2000}}
 \put(24, 92.7){\large
  H1 svtx00 prel. \protect\cite{Lastovicka:2002} $+$
  NMC \protect\cite{Arneodo:1997}}
 \put(24, 87.8){\large
  H1 96/97 \protect\cite{Adloff:2001}}
 \put(86,26.9){\large
 NLO: \hspace{0.9cm} $Q^2 > 1.5~\text{GeV}^2$}
 \put(86,22.3){\large
 NLO\&Rht: $Q^2 > 1.5~\text{GeV}^2$}
 \put(86,17.4){\large
 NLO\&Rht: $Q^2 > 0.5~\text{GeV}^2$}
\end{picture}
\end{center}
 \caption{\label{Fig:LQ2-Rht}%
The derivative function (effective slope)
$\lambda_{F_2}^\text{eff} = \partial\ln{F_2(x, Q^2)}/\partial\ln{(1/x)}$
as a function of $Q^2$. The experimental points are those H1 and ZEUS have
fitted their $x \le 0.01$ data to the form $F_2 = c(Q^2) x^{-\lambda(Q^2)}$:
black points -- H1 $F_2$ data \protect\cite{Adloff:2001};
blue squares -- H1 data \protect\cite{Lastovicka:2002} combined with NMC
data \protect\cite{Arneodo:1997};
blue triangles -- H1 data \protect\cite{Lastovicka:2002} combined with low
$Q^2$ ZEUS BPT data \protect\cite{Breitweg:2000};
open red diamonds -- preliminary ZEUS slope fit 2001
\protect\cite{Newman:2003}.
The inner error bars illustarte the statistical uncertainties, the full error
bars represent the statistical and systematic uncertanties added in quadrature.
The data are compared with a parametrization \protect\cite{Adloff:2001}
in which $\lambda(Q^2) = a\ln[Q^2/\Lambda^2]$ grows logarithmically with $Q^2$
$[a = .0481, ~\Lambda = 292~\text{MeV}]$,
using data for $Q^2 \ge 3.5~\text{GeV}^2$
(black solid straight line goes to approximated short-dashed one).
The solid, black line
 (NLO fit with $\chi^2/\text{n.d.f.} = 798/611 = 1.31$%
),
the long-dashed, red one
 (NLO\&Rht fit with $\chi^2/\text{n.d.f.} = 565/660 = 0.86$%
)
and the dash-dotted, blue one
 (NLO\&Rht fit with $\chi^2/\text{n.d.f.} = 500/607 = 0.82$%
)
are the same as on the previous Figure~\ref{Fig:LH1-Rht}. The value of $x$
was fixed to $10^{-3}$ for all curves.
}%
\end{figure}

\begin{figure}[t]
\begin{center}
  \setlength{\unitlength}{1.4mm}
\begin{picture}(125,125)   
 \put(5,5){
  \centering\epsfig{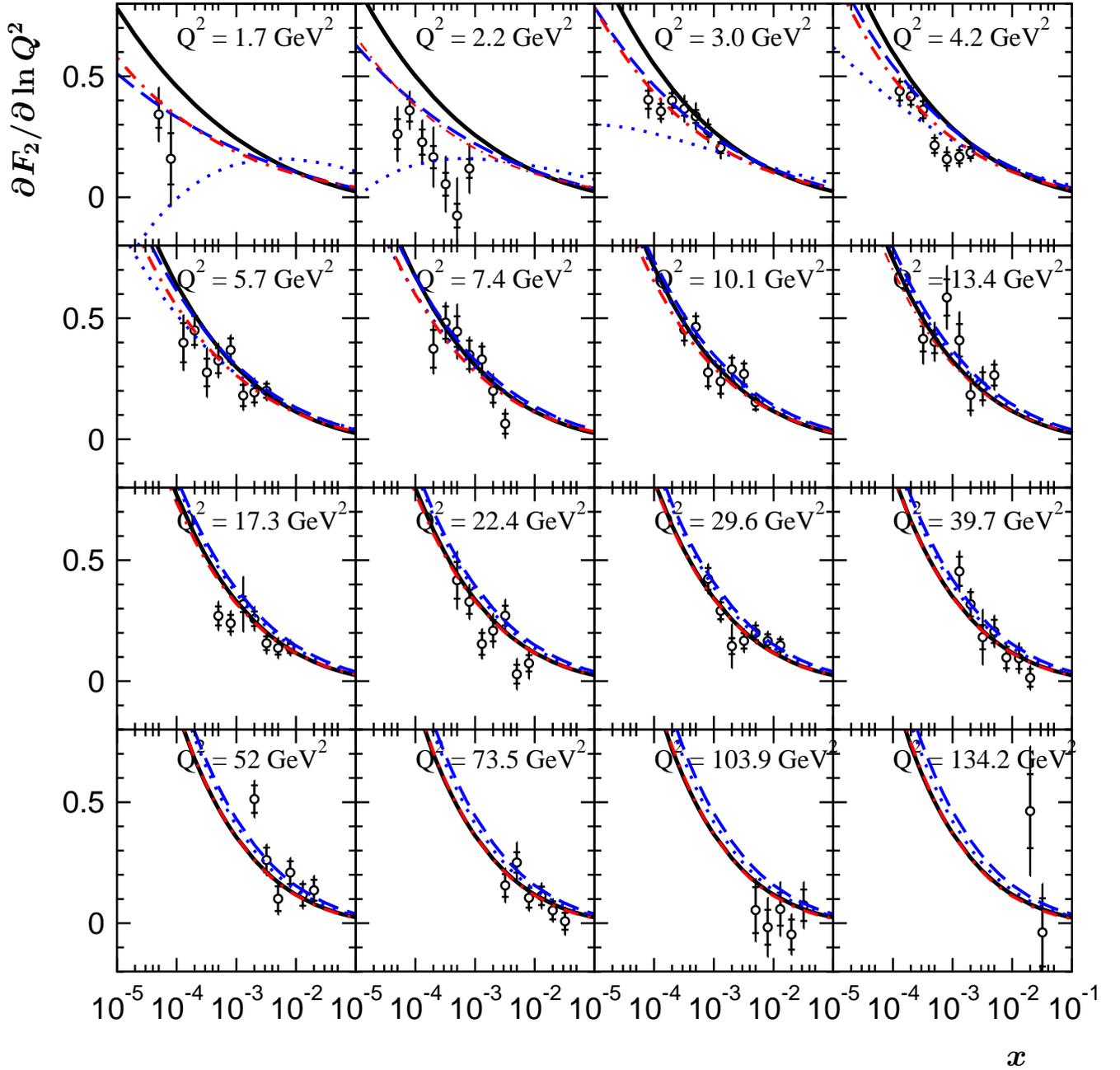}}
%
 \put(115,0){\bfseries\Large
  $\boldsymbol{x}$}
 \put(0,100){\bfseries\Large
  \begin{sideways}
   $\boldsymbol{\partial{F_2}/\partial\ln{Q^2}}$
  \end{sideways}}
\end{picture}
\end{center}
 \caption{\label{Fig:dF2-t2Rht}%
The derivative function $\partial{F_2(x, Q^2)}/\partial\ln{Q^2}$ taken at fixed
$Q^2$ and plotted as a function of $x$. The experimental points and the solid,
black line (NLO fit with
 $\chi^2/\text{n.d.f.} = 798/611 = 1.31$%
) are the same as on Figure~\ref{Fig:dF2-LOt2}.
The dashed and dotted, blue curves are obtained from the fit at the NLO, when
the renormalon contributions of higher-twist terms have been incorporated.
The dashed one is the same as on the Figure~\ref{Fig:F2sm-t2HT} with the
corresponding
$\chi^2/\text{n.d.f.} = 565/660 = 0.86$,
while the dotted line is the one from the Figure~\ref{Fig:F2-t2HT}
with
$\chi^2/\text{n.d.f.} = 500/607 = 0.82$.
The dash-dotted, red curve (hardly distinguished from the dashed one)
is the same as as on the Figure~\ref{Fig:F2sm-t2HT} and represents the fit
of data on structure function $F_2(x, Q^2)$ at the LO, the renormalon
contributions of higher-twist terms included. The corresponding
$\chi^2/\text{n.d.f.} = 555/660 = 0.84$.
}%
\end{figure}

The results for the slope $\lambda_{F_2}^\text{eff}(x, Q^2)$
are shown on Figs.~\ref{Fig:LH1-LOt2}, ~\ref{Fig:LH1-Rht} and
~\ref{Fig:LQ2-Rht} together with H1 and ZEUS experimental data
\cite{Adloff:2001,Lastovicka:2002,Breitweg:2000,Newman:2003}.

At Figs.~\ref{Fig:LH1-LOt2} and ~\ref{Fig:LH1-Rht}
we see very good agreement between theory and experiment
as with and without consideration of the HT
corrections.
Note that the asymptotic approximation does not work so well because at
large $Q^2$ values, i.e. at its range of applicability, there are 
experimental data only at quite large $x$ values: $x > 10^{-3}$.

Since the logarithmic $x$ derivative is compatible with independence of
$Q^2$, H1 and ZEUS have both fitted their data on the proton
structure function to the form $F_2 = c(Q^2) \, x^{-\lambda(Q^2)}$.
Fig.~\ref{Fig:LQ2-Rht} shows recent H1 and ZEUS fits
\cite{Adloff:2001,Lastovicka:2002,Breitweg:2000,Breitweg:1999}
for $\lambda(Q^2)$.
Some of them are preliminary only and extracted from Fig.~14 of the
recent review \cite{Newman:2003}.

The experimental data shows a rise of the slope
$\lambda_{F_2}^\text{eff}(x, Q^2)$ from the value $\sim 0.1$ at
$Q^2 \leq 1~\text{GeV}^2$ (so-called \emph{``soft pomeron range''})
to the value $\sim 0.3 \div 0.4$ at $Q^2 \geq 100~\text{GeV}^2$
(so-called \emph{``hard pomeron range''}) and $c \sim 0.18$ is consistent with
being constant.

In our opinion, the strong $Q^2$ dependence of the slope 
$\lambda_{F_2}^\text{eff}(x, Q^2)$ was observed firstly in Ref. 
\cite{Abramowicz:1991},
where fits of experimental data have been performed for the Regge-like PD
form. At high $Q^2$ side, the slope value is close to LO BFKL 
prediction ($\lambda_{F_2}^\text{eff}(x, Q^2)\sim 0.3 \div 0.4$), at smaller
$Q^2$ values $\lambda_{F_2}^\text{eff}(x, Q^2)\sim 0.2$, that is close
to model with Pomeron interactions \cite{Kaidalov:2001}
and to NLO BFKL predictions \cite{Kim:1998} based on
non-$\ms$-like renormalization schemes and
BLM resummation
 of large values of NLO corrections calculated
recently in \cite{Fadin:1998, Ciafaloni:1998}
(see also \cite{KotiLipa:2000, KotiLipa:2002}
). At low $Q^2$ 
the slope value coincides with Donnachie-Landshoff model, where
$\lambda_{F_2}^\text{eff}(x, Q^2)\sim 0.1$.

In a sence, the shape of the slope $\lambda_{F_2}^\text{eff}(x, Q^2)$
is in contrast with Regge asymptotics, where the corresponding slopes 
should be $Q^2$-independent. Note, however, that this $Q^2$-dependence can be 
described in phenomenological Regge-like models 
\cite{Capella:1994, Kaidalov:1998, Kaidalov:2000, Fiore:2000}.
There are also attempts (see \cite{Kotikov:1996:MPL,Kotikov:1996:YF})
 to recove the slope shape
in the Regge-like form of parton distributions considering the small $x$
asymptotics of DGLAP equation.
A quite natural explanation of the rise is given in the generalized DAS
approximation as it was shown in \cite{Kotikov:1999}. 

Quite recently the H1 96/97 data \cite{Adloff:2001}
(black circles on Fig.~\ref{Fig:LQ2-Rht}) has been analysed in
\cite{Kotikov:2003}, where good agreement has been found between
data and theoretical predictions based on generalized DAS approach.
For example, the rise can be described as $\ln \ln Q^2$, i.e. in pure
perturbative QCD. 
Incorparation of HT corrections gives a possibility to extend the
agreement to new preliminaty H1 and ZEUS data for quite low $Q^2$ values
(see dashed curve and the preliminary data near $Q^2 \sim 1$ GeV$^2$
on Fig.~\ref{Fig:LQ2-Rht}).

\subsection{Parton distributions}
 \label{Sec10:Fit:PD}

\begin{figure}[t]
\begin{center}
  \setlength{\unitlength}{1.4mm}
\begin{picture}(125,125)   
 \put(15,5){
  \centering\epsfig{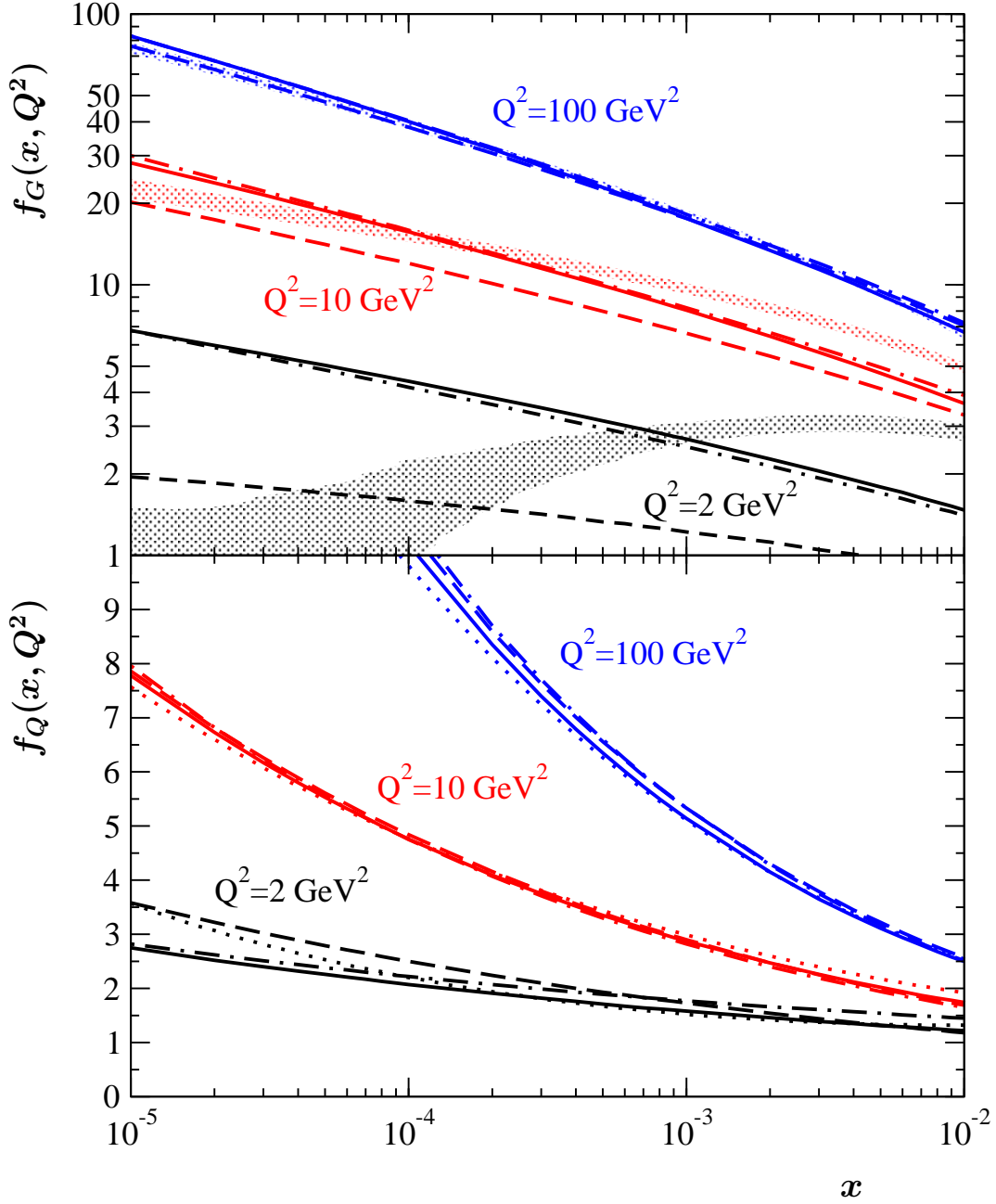}}
%
 \put(95,0){\bfseries\Large
  $\boldsymbol{x}$}
 \put(10,100){\bfseries\Large
  \begin{sideways}
   $\boldsymbol{f_G(x, Q^2)}$
  \end{sideways}}
 \put(10,45){\bfseries\Large
  \begin{sideways}
   $\boldsymbol{f_Q(x, Q^2)}$
  \end{sideways}}
\end{picture}
\end{center}
 \caption{\label{Fig:PDFs}%
The parton distributions $f_a(x, Q^2)$ as a function of $x$ for
 $Q^2 = 2, \, 10$ and $100~\text{GeV}^2$ compared to the NLO QCD predictions
of A02NLO \protect\cite{Alekhin:2003}, represented by black dots.
The solid lines represent the NLO fit alone with
$\chi^2/\text{n.d.f.} = 798/611 = 1.31$
[$A_G^{\tau2} = -.020, ~A_q^{\tau2} = .903, ~Q_0^2 = .495~\text{GeV}^2$].
The dash-dotted curves represent the BFKL-motivated estimation for the higher
twist contribution with the value of the parameter $b = a^2/2$.
The corresponding 
$\chi^2/\text{n.d.f.} = 629/609 = 1.03$
[$A_G^{\tau2} = .301, ~A_q^{\tau2} = 0.535, ~Q_0^2 = .631~\text{GeV}^2$ and
$A_G^{\tau4} = -.580~\text{GeV}^2, ~A_q^{\tau4} = 1.311~\text{GeV}^2$].
The dashed curves are obtained from the fits at the NLO, when the
renormalon contributions of higher-twist terms have been incorporated.
The corresponding
$\chi^2/\text{n.d.f.} = 565/660 = 0.86$
[$A_G^{\tau2} = .279, ~A_q^{\tau2} = .640, ~Q_0^2 = .672~\text{GeV}^2$ and
${\mathrm a}^{\tau4}_G = -.143~\text{GeV}^2,
~{\mathrm a}^{\tau4}_q = .140~\text{GeV}^2$,
${\mathrm a}^{\tau6}_G = -.044~\text{GeV}^4,
~{\mathrm a}^{\tau6}_q = .043~\text{GeV}^4$].
}%
\end{figure}

The results for the quark and gluon densities are shown on Fig.~\ref{Fig:PDFs}
together with 
the NLO QCD predictions
of A02NLO 
\cite{Alekhin:2003}, represented by dots.

As it was noted already in Sections ~\ref{Sec:7} and \ref{Sec:9},
there is very strong difference between the twist-two and total
parton distributions.
In the case of the twist-two parton densities $f_a^{\tau 2}(x,Q^2)$
the higher-twist corrections contribute to the Wilson coeffcient
functions, i.e. (in $\ms$-like factorization scheme used here) to the 
relation between the parton distributions $f_a^{\tau 2}(x,Q^2)$
and $F_2$.
Then, the higher-twist terms give additional power-like corrections
to the relation and, thus, change it.

Contrary to this, in the case of the total parton densities 
$f_a(x,Q^2)$ the coefficient functions are pure twist-two ones,
i.e. the relation between the parton distributions $f_a(x,Q^2)$ and
the structure function $F_2$ taken in the standard $\ms$-like way.
Thus, in the case the 
higher-twist corrections are responsable for the difference between the
twist-two parton distributions $f_a^{\tau 2}(x,Q^2)$ and the full ones
$f_a(x,Q^2)$.

For the quark density the difference between twist-two distributions
and total densities are not very strong.
In Fig.~\ref{Fig:PDFs} one can see
good agreement between quark distributions obtained in the 
different approximations. For the renormalon higher-twist corrections, 
our results very close to obtained by ZEUS Collaboration in 
\cite{Chekanov:2002}.

At high $Q^2$ values there is also good agreement between gluon distributions
obtained in the different approximations. For small $Q^2$ values, in the
renormalon model our total gluon density is strictly less then
twist-two one: for example, at $Q^2 = 2~\text{GeV}^2$ the ration 
$f_G(x, Q^2) / f_G^{\tau 2}(x, Q^2) < 1/3$.
Neverseless, there is a disagreement here between our results and the 
recent one from ZEUS Collaboration (see \cite{Chekanov:2002}) in the
range  of small $Q^2$ values: our total gluon density is higher
essentially the ZEUS one. A similar disagreement exist between 
our total gluon distribution and Alekhin one \cite{Alekhin:2003}
(see Fig. 9).
Thus, the deviation between our twist-two and total gluon distributions
is strong but less to have agreement with experimental data.

In our opinion, most part of the difference comes from neglection of
valent quark part $f_V(x,Q^2)$ in our article. 
The neglection is a quite standart tool at
small $x$ range (and quite large $Q^2$ values, where parton model is 
applicable), because $f_V(x,Q^2) \sim x^{\lambda_V}$ with $\lambda_V
\sim 0.3 \div 0.5$.

At low $Q^2$ values, however, the ignoring the valent and nonsinglet quark
distributions cannot be the correct approximation, because here  
the singlet parton distributions (at least the gluon density) start to 
fall when $x \to 0$. Moreover, at higher orders of perturbation theory
strong double-logarithmic terms contributte to the valent and nonsinglet 
quark distributions. The contributions can be evaluated in the framework
of BFKL-like approach and they can lead to essential decreasing of the
$\lambda_V$ value at low  $Q^2$ values (see 
\cite{Ermolaev:2000, Ermolaev:2001} and discussion therein).

Thus, in our model gluon density at small $Q^2$ values includes
effectively a contribution of the valent quark distributions and, thus,
is large essentially to compare with ZEUS and Alekhin predictions from 
\cite{Chekanov:2002} and \cite{Alekhin:2003}, respectively.

Note that the absence of the valent quarks can be partially responsable 
for some disagreement between theory and experiment for the derivation
$\partial F_2/\partial\ln Q^2$ (see Figs. 3 and 8 and discussions in
subsection D), which is depended strongly on gluon density.

We plan to return to the study the problem and to incorporate
the valent quark densities in our future investigations.

\section{Conclusions}
 \label{Concl}

In generalized DAS approximation we have incorporated HT
corrections for semi-analytical
solution of DGLAP equation obtained earlier in \cite{Kotikov:1999} 
 at LO and NLO levels in the leading twist
approximation for the flat initial condition.

The HT
corrections have been added in two models, the so-called
BFKL-like one and the renormalon one. In both models the HT
terms lead to improvement of the agreement with new precise experimental data
of H1 and ZEUS Collaborations. The elements of the renormalon model,
however, are essentially better defined and the model describes 
experimental data 
much better, especially at
very low $Q^2$ values ($Q^2 \geq 0.5~\text{GeV}^2$).

After verification of
all uncalculable parameters in our formulae from the fits
of $F_2$ data we apply our approach to compare with H1 data for 
the derivative $\partial F_2/\partial\ln Q^2$, with H1 and ZEUS data for 
the effective slope  $\lambda_{F_2}^\text{eff}(x, Q^2)$ data and 
with experimental predictions for the parton distributions.

We have found rather good agreement with data for the effective slope 
$\lambda_{F_2}^\text{eff}(x,  Q^2)$ and for the derivative $\partial
F_2/\partial\ln Q^2$ and also
with experimental predictions
for the quark distribution, but have some disagreement
with other results
for gluon densities at low $Q^2$ values
(see the subsection F
in the previous section
and discussions therein),
that needs an additional investigations.

As next steps we plan to add at low $Q^2$ values to our analysis some
phenomenological models of coupling constant. We hope to apply
the Shirkov-Solovtsov analitization \cite{Shirkov:1997,Solovtsov:1999}
 and a ``freezing'' procedure
(see, for example, Ref.~\cite{Badelek:1997} and discussions therein).

Moreover we plan also to add to our initial conditions (\ref{flat}) 
corrections $\sim \ln (1/x)$ and $\sim \ln^2 (1/x)$ obeying to Froassart 
restriction by analogy with consideration 
of these corrections in the
Regge-like 
small-$x$ asymptotics of parton distributions done earlier
in \cite{Jenkovszky:1993:PLB,JenKoti:1993}.

Addition of HT terms should be important also for high-energy cosmic rays,
where they can lead to quite important shadowing corrections for
cross-sections of neutrino-proton scattering studyed in DAS approach in
\cite{FioKoti:2003}. The subject will be considered in forthcoming article.

We are considering also to extend the application of the higher twist
corrections 
for the longitudinal srtucture function $F_L$. 
%
The consideration of $F_L$ should be very important essentially at low
$Q^2$ values, where $F_L$ should go to zero when $Q^2 \to 0$ 
\cite{Catani:1994,Catani:1996,KoLiZo:2002,KoLiPaZo:2001}
at low $x$ values based on $k_t$-factorization procedure 
\cite{Catani:1990:B242,Catani:1991:B366,Collins:1991,Levin:1991}).

In the QCD improved parton model the LO results for $F_L \sim \alpha_s(Q^2)$
and, thus, do not lead to zero values to the longitudinal srtucture function.
Moreover, the NLO corrections to $F_L$ are large and negative at low $x$
values (see 
\cite{vanNeerven:1991,Zijlstra:1991,Zijlstra:1992,Kazakov:1988,KazaKoti:1990,
Kazakov:1990,SanchezGuillen:1991,Kazakov:1992}
and, thus, give large negative contributions at low $Q^2$ range 
\cite{Keller:1991,Orr:1991,Berger:1993,Kotikov:1994:R=}.
Thus, they can lead to the
negative values for $F_L$ \cite{Kotikov:1994:R=,Martin:2001}
of perturbation theory and needs a resummation of large
corrections at low $Q^2$ values. Based on Grunberg approach 
\cite{Grunberg:1980, Grunberg:1984},
the resummation leads to recovering well-know Callan-Gross relation
$F_L = 0$ at asymptotics $x\to 0$ (see \cite{Kotikov:1994:R=}).

Thus, there are a quite conserval results for $F_L$ at low $x$ and $Q^2$
values.
The incorporation of the higher-twist corrections, which can be very
important namely in the case of the longitudinal srtucture function 
(see recent study \cite{Bartels:2000} and discussions therein), 
should give an 
additional important information about $F_L$ structure at low $x$ and $Q^2$
values.
%
%
Moreover, the measurement of $F_L$ should become possible in nearest future
(see discussions in Section~\ref{Sec:7}
of \cite{Cooper-Sarkar:1998}) with the proposed updates to the HERA machine,
which will yield very large integrated luminosity.
Note that some precise preliminary results for $F_L$ can be found
already in the recent review \cite{Newman:2003} 
and we plan to study of them in
nearest future.

\acknowledgments

%

We would like to thank J.~Bartels, S.~Alekhin
and especially V.~Chekelian and A.~Sidorov
for many valuable discussions and comments.



A.~Yu.~Illarionov would like to thanks the good peaple of the INFN for
providing him with food, shelter, money and oil while this work was
in progress.

A.~V.~Kotikov thanks the Alexander von Humboldt Foundation for the support 
of his stay at Hamburg and Karlsruhe Universities, where most part of
the  work has been done. He is also supported in part by the 
INTAS under Grant no.~366.

G.~Parente acknowledges the support of Galician research funds
(PGIDT00 PX20615PR) and Spanish CICYT (FPA2002-01161).

\section{Appendix A}
 \label{App:A}
\def\theequation{A\arabic{equation}}
\setcounter{equation}0

The twist-four and twist-six contributions in the framework of 
IR-renormalon model have been calculated in \cite{Dasgupta:1996}
(for nonsinglet case) and in \cite{Stein:1998} (for singlet one).
As we already noted in 
Section~\ref{Sec:2}, we neglect the nonsinglet component in our analysis. The
higher-twist corrections to singlet case contain sum of nonsinglet
and singlet higher-twist results. However, we have interest only to
$n \to 1$ asymptotics of the corrections, where nonsinglet part
of higher-twist corrections are neglected because exact Bjorken sum rule.

The singlet part of higher-twist corrections may be presented in the
following form
\begin{equation}
M^R_a(n, Q^2) \ = \ M_a(n, Q^2) \left[ 1 +
 \dfrac{{\mathrm a}_a^{\tau4}}{Q^2} \mu^{\tau4}_a
 \left( n,
 \ln\left(\dfrac{\left|{\mathrm a}_a^{\tau4}\right|}{Q^2}\right) \right)
+
 \dfrac{{\mathrm a}_a^{\tau6}}{Q^4} \mu^{\tau6}_a
 \left(n,
 \ln\left(\dfrac{\sqrt{\left|{\mathrm a}_a^{\tau6}\right|}}{Q^2}\right)
 \right) \right] \ .
\label{App:Mren}
\end{equation}

The quark contributions $\mu^{\tau4}_q( n, \ln({\mathrm A}/Q^2))$ and
$\mu^{\tau6}_q( n, \ln({\mathrm A}/Q^2))$ \cite{Stein:1998} may be
transformed to $n$-space
\begin{equation}
\mu^{\tau m}_a \left(n,
 \ln\left(\dfrac{\mathrm A}{Q^2}\right)\right) \, = \,
 \dfrac{8C_FT_Rf}{\beta_0^2} \left[ B^{\tau m}_a(n) +
 b^{\tau m}_a(n) \ \ln\left(\dfrac{\mathrm A}{Q^2}\right)
 \right] \ ,
\label{App:muq}
\end{equation}
%
%
where
\begin{align} 
B^{\tau4}_q(n) &=
   \dfrac{16}{15}\dfrac{1}{(n-1)^2} + \dfrac{22}{225}\dfrac{1}{n-1}
 + \dfrac{11}{3}\dfrac{1}{n+1} - \dfrac{25}{9}\dfrac{1}{n+2}
 - \dfrac{74}{75}\dfrac{1}{n+4}
\nonumber \\
&- \dfrac{12}{(n+1)^2} + \dfrac{6}{(n+2)^2}
 + \dfrac{4}{(n+1)^3} + \dfrac{12}{(n+2)^3} \ ,
\label{App:Bq:4} \\
b^{\tau4}_q(n) &=
 - \dfrac{8}{15}\dfrac{1}{n-1} + \dfrac{9}{n+1}
 - \dfrac{23}{3}\dfrac{1}{n+2} - \dfrac{4}{5}\dfrac{1}{n+4}
 - \dfrac{2}{(n+1)^2} - \dfrac{6}{(n+2)^2} \ ,
\label{App:bq:4} \\[0.2cm]
B^{\tau6}_q(n) &=
 - \dfrac{128}{105}\dfrac{1}{(n-1)^2} - \dfrac{572}{11025}\dfrac{1}{n-1}
 + \dfrac{52}{75}\dfrac{1}{n+1} + \dfrac{32}{9}\dfrac{1}{n+2}
 + \dfrac{16}{3}\dfrac{1}{n+3}
\nonumber \\
&- \dfrac{724}{75}\dfrac{1}{n+4} + \dfrac{452}{3675}\dfrac{1}{n+6}
 + \dfrac{16}{5}\dfrac{1}{(n+1)^2} - \dfrac{16}{(n+3)^2} \ .
\label{App:Bq:6} \\
b^{\tau6}_q(n) &=
   \dfrac{64}{105}\dfrac{1}{n-1} - \dfrac{8}{5}\dfrac{1}{n+1}
 - \dfrac{8}{3}\dfrac{1}{n+2} + \dfrac{8}{n+3}
 - \dfrac{24}{5}\dfrac{1}{n+4} + \dfrac{16}{35}\dfrac{1}{n+6} \ ,
\label{App:bq:6}
\end{align}

The gluon contributions $\mu^{\tau4}_G(n,\ln({\mathrm A}/Q^2))$ and
$\mu^{\tau6}_G(n, \ln({\mathrm A}/Q^2))$ may be estimated \cite{Stein:1998}
as
\begin{equation}
\mu^{\tau m}_G(n, \ln({\mathrm A}/Q^2)) \ = \
 \mu^{\tau m}_q(n, \ln({\mathrm A}/Q^2)) \; / \; \gamma^{(0)}_{Gq}(n) \ ,
\label{App:muG}
\end{equation}
where \cite{Buras:1980}
$$
\gamma^{(0)}_{Gq}(n) \ = \
  -4 C_F \, \dfrac{2 + n + n^2}{(n - 1) n (n + 1)}
$$
is the leading contribution to the gluon-quark anomalous dimension.

We have the interest to the asymptotics $n\to 1$, where the above
values may be represented as
\begin{align}
B^{\tau4}_q(n) &= - \dfrac{4}{15}
 \left(\dfrac{1}{\delta^2} + \dfrac{11}{120}\dfrac{1}{\delta} -
 \dfrac{2291}{3600} \right) + \mathcal{O}(\delta) \ , 
\quad
b^{\tau4}_q(n) = \dfrac{2}{15}
 \left(\dfrac{1}{\delta} - \dfrac{139}{120}\right) +
 \mathcal{O}(\delta) \ ,
\nonumber \\
B^{\tau6}_q(n) &= \dfrac{32}{105}
 \left( \dfrac{1}{\delta^2} + \dfrac{143}{3360}\dfrac{1}{\delta} -
 \dfrac{870637}{1411200} \right) + \mathcal{O}(\delta) \ , \quad
b^{\tau6}_q(n) = - \dfrac{16}{105}
 \left(\dfrac{1}{\delta} - \dfrac{3217}{3360}\right) +
 \mathcal{O}(\delta) \ ;
\label{App:Bbq:exp} \\[0.2cm]
B^{\tau4}_G(n) &= - \dfrac{2}{15C_F}
 \left(\dfrac{1}{\delta} + \dfrac{101}{120} \right) +
 \mathcal{O}(\delta) \ , \quad
b^{\tau4}_G(n) = \dfrac{1}{15C_F} +
 \mathcal{O}(\delta) \ ,
\nonumber \\
B^{\tau6}_G(n) &= \dfrac{16}{105C_F}
 \left(\dfrac{1}{\delta} + \dfrac{2663}{3360} \right) +
 \mathcal{O}(\delta) \ , \quad
b^{\tau6}_G(n) = - \dfrac{8}{105C_F} +
 \mathcal{O}(\delta) \ ,
\label{App:BbG:exp}
\end{align}
with $\delta = n - 1$.

\section{Appendix B}
 \label{App:B}
\def\theequation{B\arabic{equation}}
\setcounter{equation}0

  We present here the detailed analysis
\footnote{
Contrary to Ref.~\cite{Kotikov:1994:YF,Kotikov:1994:PRD} we use here the
variable $z=x/x_0$
}
of the method of replacing the convolution of two functions by a simple
product at small $~x$. We restrict ourselves to the accuracy
$\mathcal O(z)$. Some earlier presentations can be found in
\cite{Kotikov:1994:YF,Kotikov:1994:PRD} (here the accuracy
$\mathcal O(z^2)$ has been considered)
 and in \cite{Kotikov:1999}.

Let us to consider the set of PD with different forms:
\begin{itemize}
\item[(I)]
 Regge-like form $f_R(z) = z^{-\delta} \widetilde f(z)$, 
\item[(II)]
 Logarithmic-like form $f_L(z) = z^{-\delta} \ln(1/z) \widetilde f(z)$, 
\item[(III)]
 Bessel-like form
 $f_I(z) = z^{-\delta} \, {\hat{d} \ln(1/z)}^{k/2}
 \widetilde{I}_k\left(2\sqrt{\hat{d} \ln(1/z)}\right) \widetilde f(z)$
 with definition (\ref{ser:I_nu}) of the $\widetilde{I}_k$ function, 
\end{itemize}
where $\widetilde f(z)$ and its derivative
$\widetilde f^\prime(z) \equiv d \widetilde{f}(z)/dz$ are smooth at $z = 0$ and
both are equal to zero at $z = 1$:
$$
\widetilde{f}(1) \ = \ \widetilde{f}^\prime(1) \ = \ 0 \ .
$$

\noindent
\textsf{(1)} At the beginning we consider the basic integral
 with integer integer nonnegative $n$ values:
$$
 J^{(1)}_{\delta,i}(n,z) = z^n \otimes f_i(z) \equiv
  \int_{z}^{1} \dfrac{\mathrm{d} y}{y} \ y^n
  f_i\left(\dfrac{z}{y}\right)  \ , \ \  i = R, L, I \ .
$$

\noindent
\textit{(a)} \textit{Regge-like case}.
 Expanding $\widetilde f(z)$ near $\widetilde f(0)$, we have
\begin{align}
J^{(1)}_{\delta,R}(n,z) &=
 z^{-\delta} \int_{z}^{1}\mathrm{d} y \ y^{n+\delta-1} \left[
 \widetilde f(0) + \dfrac{z}{y} \widetilde f^{(1)}(0) + \ldots +
 \dfrac{1}{k!} \left(\dfrac{z}{y}\right)^k \widetilde f^{(k)}(0) + \ldots
 \right]
\nonumber  \\
&= z^{-\delta} \left[ \dfrac{1}{n+\delta} \ \widetilde f(0) +
 \mathcal{O}(z) \right]
\nonumber \\
&- z^n \left[ \dfrac{1}{n+\delta} \ \widetilde f(0) +
 \dfrac{1}{n+\delta-1} \ \widetilde f^{(1)}(0) + \ldots +
 \dfrac{1}{k!} \ \dfrac{1}{n+\delta-k} \ \widetilde f^{(k)}(0) + \ldots
 \right] \ .
\label{AppB:a}
\end{align}
Using the power-like large $x$ asymptotics
\begin{equation}
 f(z)  \ \sim \ (1 - z)^{\nu} \text{ when }  z \to 1 \ ,
\label{AppB:asympt}
\end{equation}
the second term on the r.h.s. of Eq.~(\ref{AppB:a}) can be summed:
\begin{equation}
J^{(1)}_{\delta,R}(n,z) = z^{-\delta} \left[\dfrac{1}{n+\delta} \ 
\widetilde f(0) +  \mathcal{O}(z) \right] \ + \
 z^n \ \dfrac{\Gamma(-(n+\delta)) \Gamma(1+\nu)}{\Gamma(1+\nu-n-\delta)} \
\widetilde f(0) \ .
\label{AppB:Rsum}
\end{equation}
Consider particular cases $n \geq 1$ and $n = 0$ separately:

\noindent
\textit{(a1)} If $n \geq 1$ then the second term in the r.h.s. of
(\ref{AppB:Rsum}) is negligible and we have
\begin{equation}
J^{(1)}_{\delta, R}(n, z) =
 z^{-\delta} \dfrac{1}{n+\delta} \ \widetilde f(0) +
 \mathcal{O}(z^{1- \delta}) \ = \
 \dfrac{1}{n+\delta} \ \widetilde f_R(z) +
 \mathcal{O}(z^{1- \delta}) \ .
\label{AppB:a1}
\end{equation}

\noindent
\textit{(a2)} If $n = 0$,  the r.h.s. of (\ref{AppB:Rsum})
can be rewritten as follows
\begin{align}
J^{(1)}_{\delta, R}(0,z) &= z^{-\delta}
 \left[ \dfrac{1}{\delta} \widetilde f(0) + \mathcal{O}(z) \right] \ + \
 \dfrac{\Gamma(-\delta) \Gamma(1+\nu)}{\Gamma(1+\nu-\delta)} \
 \widetilde f(0)
\nonumber \\
 &= \delta_R^{-1}(z) f_R(z) + \mathcal{O}(z^{1-\delta}) \ ,
\label{AppB:a2}
\end{align}
where
\begin{equation}
\dfrac{1}{\delta_R(z)}  \ = \ \dfrac{1}{\delta}
 \left[1- \dfrac{\Gamma(1-\delta) \Gamma(1+\nu)}{\Gamma(1+\nu-\delta)}
 z^{\delta} \right],
\label{AppB:dR}
\end{equation}
i.e. there is the correlation between small $x$ and large $x$ asymptotics
of parton distributions (see 
\cite{Kotikov:1991di,Kotikov:1997,JenKoti:1992}.
Note that the value 
$\delta_R^{-1}(z)$ is finite
at the limit $\delta \to 0$:
\begin{equation}
\lim_{\delta \to 0}\dfrac{1}{\delta_R(z)} \ = \ 
 \ln\left(\dfrac{1}{z}\right) - \left[\Psi(1+\nu) - \Psi(1)\right]
 \ \equiv \ \ln\left(\dfrac{1}{z}\right) - p(\nu) \ ,
\label{AppB:dR:lim0}
\end{equation}
where the Riemannian $\Psi$-function is the logarithmic derivation of the
$\Gamma$-function.

Remember that the large $x$ asymptotics are different in quark and gluon
cases, the values $\nu_q \approx 3$ and $\nu_G \approx 4$ are coming from
quark counting rules,
what lead to $p(\nu_q) \approx 11/6$ and $p(\nu_G) \approx 25/12$.

\noindent
\textit{(b)} \textit{Logarithmic-like case}.
Using the simple relation
$$
z^{-\delta} \ln(1/z) = d(z^{-\delta})/d\delta
$$
we immediately obtain:

\noindent
\textit{(b1)} $n \geq 1$ case:
\begin{align}
J^{(1)}_{\delta,L}(n,z) &= z^{-\delta} \ln(1/z) \left[
 \dfrac{1}{n+\delta} \left(1- \dfrac{1}{(n+\delta) \ln(1/z)}\right)
 \widetilde f(0) + \mathcal{O}(z) \right]
\nonumber \\
&= \dfrac{1}{n+\delta}
 \left(1- \dfrac{1}{(n+\delta) \ln(1/z)} \right) f_L(z) +
 \mathcal{O}(z^{1-\delta})
\nonumber \\
&= \dfrac{1}{n+\delta} \ f_L(z) +
 \mathcal{O}\left(\dfrac{1}{\ln(1/z)}\right) \ .
\label{AppB:b1}
\end{align}

\noindent
\textit{(b2)} $n = 0$ case:
\begin{align}
J^{(1)}_{\delta,L}(0,z) &= z^{-\delta} \ln(1/z) \left[
 \dfrac{1}{\delta} \left(1- \dfrac{1}{\delta \ln(1/z)}\right)
 \widetilde f(z) + \mathcal{O}(z) \right]
\nonumber \\
&+ \dfrac{\Gamma(-\delta) \Gamma(1+\nu)}{\Gamma(1+\nu-\delta)} \
 \widetilde f(0) \left[\Psi(1+\nu -\delta) - \Psi(-\delta) \right]
\nonumber \\
&= \delta_L^{-1}(z) \ f_L(z) + \mathcal{O}(z^{1- \delta}) \ ,
\label{AppB:b2}
\end{align}
where
\begin{align}
\dfrac{1}{\delta_L(z)} &\equiv
 \dfrac{z^{\delta}}{\ln(1/z)} \dfrac{d}{d\delta}
 \left(\dfrac{z^{-\delta}}{\delta_R(z)}\right) =
 \dfrac{1}{\delta_R(z)} + \dfrac{1}{\ln(1/z)}\dfrac{d}{d\delta}
 \left(\dfrac{1}{\delta_R(z)}\right)
\nonumber \\
&=  \dfrac{1}{\delta}
 \left[1 - \dfrac{1}{\ln(1/z)} \left(\dfrac{1}{\delta_R(z)} +
 \dfrac{\Gamma(1-\delta) \Gamma(1+\nu)}{\Gamma(1+\nu-\delta)}
 z^{\delta}
 \left[\Psi(1+\nu-\delta) - \Psi(1-\delta)\right] \right)\right] \ .
\label{AppB:dL}
\end{align}
The value $\delta_L^{-1}(z)$ is also finite at the limit $\delta \to 0$:
\begin{align}
\lim_{\delta \to 0} \dfrac{1}{\delta_L(z)}
 &= \dfrac{1}{2} 
 \ln\left(\dfrac{1}{z}\right) - \dfrac{1}{2\ln(1/z)} \left(
 \left[\Psi(1+\nu) - \Psi(1)\right]^2 -
 \left[\Psi^\prime(1+\nu) - \Psi^\prime(1)\right] \right)
\nonumber \\
 &= \dfrac{1}{2} 
 \ln\left(\dfrac{1}{z}\right) - \dfrac{1}{2\ln(1/z)}
 \left( p(\nu)^2 - p^\prime(\nu) \right) \ ,
\label{AppB:dL:lim0}
\end{align}
where the $\Psi^\prime$-function is the derivation of the
$\Psi$-function,
$p^\prime(\nu_q) \approx -49/36$ and $p^\prime(\nu_G) \approx -205/144$
are coming from quark counting rules.

\noindent
\textit{(c)} \textit{Bessel-like case}.
Representing Bessel function in the form
\begin{equation}
 z^{-\delta} \, {\hat{d} \ln(1/z)}^{k/2}
\widetilde{I}_k\left(2\sqrt{\hat{d} \ln(1/z)}\right) \ = \
 \sum^{\infty}_{n=0} \dfrac{1}{n!\Gamma(n+k+1)} 
 \left(\hat{d} \dfrac{d}{d\delta}\right)^{n+k} z^{-\delta} \ = \
\, {\hat{d} \left(\dfrac{d}{d\delta}\right)}^{k/2}
 \widetilde{I}_k\left(2\sqrt{\hat{d} \left(\dfrac{d}{d\delta}\right)}\right)
 z^{-\delta}
\label{AppB:c}
\end{equation}
and repeating the above analysis, we have

\noindent
\textit{(c1)} in the $n \geq 1$ case:
\begin{equation}
 J^{(1)}_{\delta,I}(n,z) \ = \
  \dfrac{1}{n+\delta} \ f_I(z) +
  \mathcal{O}\left(\sqrt{\dfrac{\hat{d}}{\ln(1/z)}}\right) \ ,
\label{AppB:c1}
\end{equation}

\noindent
\textit{(c2)} in the $n = 0$ case:
\begin{equation}
J^{(1)}_{\delta,I}(0,z) \ = \
 \dfrac{1}{\delta_I(z)} \ f_I(z) + \mathcal{O}(z^{1- \delta}) \ ,
\label{AppB:c2}
\end{equation}
where
\begin{equation}
\dfrac{1}{\delta_I(z)} =
 \dfrac{z^{\delta} \, {\hat{d}
 \left(\dfrac{d}{d\delta}\right)}^{k/2}}{\, 
{\hat{d} \ln(1/z)}^{k/2}
\widetilde{I}_k\left(2\sqrt{\hat{d} \ln(1/z)}\right)}
 \widetilde{I}_k\left(2\sqrt{\hat{d} \left(\dfrac{d}{d\delta}\right)}\right)
 \dfrac{z^{-\delta}}{\delta_R(z)} \ .
\label{AppB:dI}
\end{equation}

The value $\delta_I^{-1}(z)$ is also finite at the limit $\delta \to 0$:
\begin{equation}
\lim_{\delta \to 0} \dfrac{1}{\delta_I(z)} \ = \
 \sqrt{\dfrac{\ln(1/z)}{\hat{d}}}
 \dfrac{\widetilde{I}_{k+1}\left(2\sqrt{\hat{d} \ln(1/z)}\right)}{
        \widetilde{I}_k\left(2\sqrt{\hat{d} \ln(1/z)}\right)} \ \approx \
 \sqrt{\dfrac{\ln(1/z)}{\hat{d}}} - \dfrac{2k+1}{4 \hat{d}} + 
 \mathcal{O}\left(\sqrt{\dfrac{\hat{d}}{\ln(1/z)}}\right) \ ,
\label{AppB:dI:lim0}
\end{equation}
where the r.h.s. of (\ref{AppB:dI:lim0}) is obtained from the expansion
of the modified Bessel functions at $z \to 0$.

Note that we can represented the Eqs. (\ref{AppB:a2}), (\ref{AppB:b2}) and 
(\ref{AppB:c2}) formally as follows
\begin{equation}
\delta^{-1} \  f_B(z)  \ = \
\dfrac{1}{\delta_B(z)}  \  f_B(z) ~~~~ (B=R,L,I),
\label{AppB:delta}
\end{equation}
which
has been used in Sections III and V.\\

\noindent
\textsf{(2)} Since the HT coefficient functions $B^{\tau4,6}_q(n)$
contain the terms $\sim 1/(n-1)^2$
(see Eqs. (\ref{App:Bq:4}) and (\ref{App:Bq:6})),
we should consider also the second basic integral
 with 
integer nonnegative $n$ values:
$$
 J^{(2)}_{\delta,i}(n,z) = z^n \ln(1/z) \otimes f_i(z) \equiv
  \int_{z}^{1} \dfrac{\mathrm{d} y}{y} \ y^n \ln(1/y)
  f_i\left(\dfrac{z}{y}\right)  \ , \ \  i = R, L, I \ .
$$

It is easy to demonstrate that $$J^{(2)}_{\delta,i}(n,z) =
\dfrac{d}{da} \, J^{(1)}_{\delta,i}(n-a,z)|_{a=0},$$
that
symplyfies essentially the consideration of $J^{(2)}_{\delta,i}(n,z)$.\\

\noindent
\textit{(a)} \textit{Regge-like case}.
Repeating the analysis of the subsection (1a), we obtain easy that
\begin{eqnarray}
J^{(2)}_{\delta,R}(n,z) &=& z^{-\delta} \left[\dfrac{1}{(n+\delta)^2} \ 
\widetilde f(0) +  \mathcal{O}(z) \right] \nonumber \\
&+&
 z^n \ \dfrac{\Gamma(-(n+\delta)) \Gamma(1+\nu)}{\Gamma(1+\nu-n-\delta)} \
\left[ \ln \dfrac{1}{z} + \Psi(-(n+\delta)) - \Psi(1+\nu-n-\delta) \right]
\widetilde f(0) \ .
\label{AppB:R2sum}
\end{eqnarray}
Consider particular cases $n \geq 1$ and $n = 0$ separately:

\noindent
\textit{(a1)} If $n \geq 1$ then the second term in the r.h.s. of
(\ref{AppB:Rsum}) is negligible and we have
\begin{equation}
J^{(2)}_{\delta, R}(n, z) =
 z^{-\delta} \dfrac{1}{(n+\delta)^2} \ \widetilde f(0) +
 \mathcal{O}(z^{1- \delta}) \ = \
 \dfrac{1}{(n+\delta)^2} \ \widetilde f_R(z) +
 \mathcal{O}(z^{1- \delta}) \ .
\label{AppB:2a1}
\end{equation}

\noindent
\textit{(a2)} If $n = 0$,  the r.h.s. of (\ref{AppB:R2sum})
can be rewritten as follows
\begin{align}
J^{(2)}_{\delta, R}(0,z) &= z^{-\delta}
 \left[ \dfrac{1}{\delta^2} \widetilde f(0) + \mathcal{O}(z) \right] \ + \
 \dfrac{\Gamma(-\delta) \Gamma(1+\nu)}{\Gamma(1+\nu-\delta)} 
\left[ \ln \dfrac{1}{z} + \Psi(-\delta) - \Psi(1+\nu-\delta) \right]
\
 \widetilde f(0)
\nonumber \\
 &= \delta_R^{-2}(z) f_R(z) + \mathcal{O}(z^{1-\delta}) \ ,
\label{AppB:2a2}
\end{align}
where
\begin{equation}
\dfrac{1}{\delta^2_R(z)}  \ = \ - \dfrac{d}{d\delta} \, \dfrac{1}{\delta}
 \left[1- \dfrac{\Gamma(1-\delta) \Gamma(1+\nu)}{\Gamma(1+\nu-\delta)}
 z^{\delta} \right] \ \equiv \ - \dfrac{d}{d\delta} \, \dfrac{1}{\delta_R(z)}.
\label{AppB:d2R}
\end{equation}

Note that the value $\delta_R^{-2}(z)$ is finite
at the limit $\delta \to 0$:
\begin{equation}
\lim_{\delta \to 0}\dfrac{1}{\delta^2_R(z)} \ = \ \dfrac{1}{2}
 \left[ {\left(\lim_{\delta \to 0}\dfrac{1}{\delta_R(z)}\right)}^2
 - p^\prime(\nu) \right] \ ,
\label{AppB:d2R:lim0}
\end{equation}
where the value of $\lim_{\delta \to 0}\,(1/\delta_R(z))$ is given
in (\ref{AppB:dR:lim0}). 

\noindent
\textit{(b)} \textit{Logarithmic-like case}.
Following to the subsection (1b), we obtain:

\noindent
\textit{(b1)} $n \geq 1$ case:
\begin{align}
J^{(2)}_{\delta,L}(n,z) &=  \dfrac{1}{(n+\delta)^2} \ f_L(z) +
 \mathcal{O}\left(\dfrac{1}{\ln(1/z)}\right) \ .
\label{AppB:b21}
\end{align}

\noindent
\textit{(b2)} $n = 0$ case:
\begin{equation}
J^{(1)}_{\delta,L}(0,z) ~=~ \delta_L^{-2}(z) \ f_L(z) + 
\mathcal{O}(z^{1- \delta}) \ ,
\label{AppB:b22}
\end{equation}
where
\begin{equation}
\dfrac{1}{\delta^2_L(z)} ~=~ - \dfrac{d}{d\delta} \dfrac{1}{\delta_L(z)}
\label{AppB:d2L}
\end{equation}
and the value of $1/\delta_L(z)$ is given in (\ref{AppB:dL}).

The value $\delta_L^{-1}(z)$ is also finite at the limit $\delta \to 0$:
\begin{equation}
\lim_{\delta \to 0} \dfrac{1}{\delta_L(z)}
 ~=~  \dfrac{1}{6} {\left[ 
 \ln\left(\dfrac{1}{z}\right)\right]}^2 - \dfrac{1}{2}
 \left( p(\nu)^2 - p^\prime(\nu) \right) 
- \dfrac{1}{3\ln(1/z)}
 \left( p(\nu)^3 - 3p^\prime(\nu)p(\nu) + p^{\prime\prime}(\nu) \right)
\ ,
\label{AppB:d2L:lim0}
\end{equation}
where the $\Psi^{\prime\prime}$-function is the second derivation of the
$\Psi$-function,
$p^{\prime\prime}(\nu_q) \approx 251/108$ and $p^{\prime\prime}(\nu_G) 
\approx 2035/865$
are coming from quark counting rules.

\noindent
\textit{(c)} \textit{Bessel-like case}.
Following to the subsection (1c), we obtain:

\noindent
\textit{(c1)} in the $n \geq 1$ case:
\begin{equation}
 J^{(1)}_{\delta,I}(n,z) \ = \
  \dfrac{1}{(n+\delta)^2} \ f_I(z) +
  \mathcal{O}\left(\sqrt{\dfrac{\hat{d}}{\ln(1/z)}}\right) \ ,
\label{AppB:c21}
\end{equation}

\noindent
\textit{(c2)} in the $n = 0$ case:
\begin{equation}
J^{(1)}_{\delta,I}(0,z) \ = \
 \dfrac{1}{\delta^2_I(z)} \ f_I(z) + \mathcal{O}(z^{1- \delta}) \ ,
\label{AppB:c22}
\end{equation}
where
\begin{equation}
\dfrac{1}{\delta^2_I(z)} ~=~ - \dfrac{d}{d\delta} \dfrac{1}{\delta_I(z)}
\label{AppB:d2I}
\end{equation}
and the value of $\dfrac{1}{\delta_I(z)}$ is given in (\ref{AppB:dI}).

The value $\delta_I^{-2}(z)$ is also finite at the limit $\delta \to 0$:
\begin{equation}
\lim_{\delta \to 0} \dfrac{1}{\delta^2_I(z)} \ = \
\dfrac{\ln(1/z)}{\hat{d}} 
\dfrac{\widetilde{I}_{k+2}\left(2\sqrt{\hat{d} \ln(1/z)}\right)}{
        \widetilde{I}_k\left(2\sqrt{\hat{d} \ln(1/z)}\right)} \ \approx \
{\left( \lim_{\delta \to 0} \dfrac{1}{\delta_I(z)} - \dfrac{1}{4\hat{d}}
\right)}^2 + \dfrac{3(k+1)}{8 {\hat{d}}^2}
\label{AppB:d2I:lim0}
\end{equation}
Note that the r.h.s. of (\ref{AppB:dI:lim0}) is obtained from the expansion
of the modified Bessel functions at $z \to 0$.

Note that we can represented the Eqs. (\ref{AppB:2a2}), (\ref{AppB:b22}) and 
(\ref{AppB:c22}) formally as follows
\begin{equation}
\delta^{-2} \  f_B(z)  \ = \
\dfrac{1}{\delta^2_B(z)}  \  f_B(z) ~~~~ (B=R,L,I),
\label{AppB:delta2}
\end{equation}
which
has been used in Section V.\\

\noindent
\textsf{(3)} Consider the Mellin integral
$$
 I_{\delta}(z) \ = \ \widetilde K(z) \otimes f(z) \equiv
 \int_{z}^{1} \dfrac{\mathrm{d}y}{y} \ \hat{K}(y) \
 f\left(\dfrac{z}{y}\right)
$$
and define the moments of the kernel $\widetilde K(y)$ in the following form
$$
 K_n \ = \ \int_{0}^{1}\mathrm{d}y \ y^{n-2} \ \widetilde K(y) \ .
$$
In analogy with part \textsf{(1)} we have for the Regge-like case:
\begin{align}
 I_{\delta,R}(z) &=
  z^{-\delta} \int_{z}^{1}\mathrm{d}y \ y^{\delta-1} \ \widetilde K(y) \
  \left[\widetilde f(0) + \dfrac{z}{y} \ \widetilde f^{(1)}(0) + \ldots  +
  \dfrac{1}{k!} \left(\dfrac{z}{y}\right)^k \widetilde f^{(k)}(0) + \ldots
  \right]
\nonumber  \\
 &= z^{-\delta} \left[ K_{1+\delta} \ \widetilde f(0) + \mathcal{O}(z) \right]
\nonumber \\
 &- \left[ N_{1+\delta}(x) \widetilde f(0) +
  N_{\delta}(z) \widetilde f^{(1)}(0) + \ldots +
  \dfrac{1}{k!} \ N_{1+\delta-k}(z) \widetilde f^{(k)}(0) + \ldots
 \right] \ ,
\label{AppB:5}
\end{align}
where
$$
 N_{\eta}(z) \ = \
 \int_{0}^{1}\mathrm{d}y \ y^{\eta-2} \ \widetilde K(zy) \ .
$$

The case $K_{1+\delta} = 1/(n+\delta)$ corresponds to
$\widetilde{K}(y) = y^n$ and has been already considered in part
\textsf{(1)}. In the more general cases (for example,
$K_{1+\delta} = \Psi(1+\delta) + \gamma$) we can represent the ``moment''
$K_{1+\delta}$ as a series of the sort $\sum_{m=1} 1/(n+\delta+m)$.

So, for the initial integral at small $x$ we get the simple equation:
\begin{equation}
I_{\delta,R}(z) \ = \ z^{-\delta} \ K_{R,1+\delta} \ \widetilde f(z) +
 \mathcal{O}(z^{1- \delta})  \ = \
 K_{1+\delta} \ f_R(z) + \mathcal{O}(z^{1- \delta}) \ ,
\label{AppB:R-rule}
\end{equation}
where the coefficient $K_{R,1+\delta}$ coincides with the one 
$K_{1+\delta}$ in the case if $K_{n}$ does not contain the term $1/(n-1)$.
The coefficient $K_{R,1+\delta}$ contain the term $\delta_R^{-1}(z)$ if
the term $1/(n-1)$ contributed to $K_{n}$.
So, the function $K_{R,1+\delta}$ can be represented in the form:
\begin{equation}
 K_{R,1+\delta} \ = \  K_{1+\delta_R(z)} \ .
\label{AppB:K_R}
\end{equation}

Repeating the analysis of the subparts \textit{(b)} and \textit{(c)},
one easily obtains
\begin{align}
 I_{\delta,L}(n,z) &= K_{L,1+\delta} \ f_L(z) + 
  \mathcal{O}\left(\dfrac{1}{\ln(1/z)}\right)
\label{AppB:L-rule}\\
 I_{\delta,I}(n,z) &= K_{I,1+\delta} \ f_I(z) +
  \mathcal{O}\left(\sqrt{\dfrac{\hat{d}}{\ln(1/z)}}\right) \ , 
\label{AppB:I-rule}
\end{align}
where
\begin{align}
 K_{L,1+\delta} &= K_{1+\delta_L(z)} \ ,
\label{AppB:K_L}\\
 K_{I,1+\delta} &= K_{1+\delta_I(z)} \ .
\label{AppB:K_I}
\end{align}

Thus, in the non-singular case (i.e. in the case when $K_{n}$ does not
contain the term $1/(n-1)$) the results of tans formation of the Mellin
convolution to usual products depend only on the $\delta$ value but not
on the concrete shape of parton distribution. The presence of the term
$1/(n-1)$ in  $K_{n}$ leads to the results depending on numerical value
of $\delta$. If $\delta$ is large (more precisely, if
$z^{-\delta} \gg \text{const}$), the presence of the term $1/(n-1)$ in
$K_{n}$ leads to the term $1/\delta$ in the functions $K_{i,1+\delta}$
$(i=R,L,I)$ (because the term $z^{\delta}$ is negligible in expressions
for $1/\delta_i$) and the results do not also depend on the
concrete shape of parton distribution. If $\delta$ is small (i.e., if 
$z^{- \delta} \approx 1 + \delta \ln(1/z)$, that depends on concrete
$z$ values, of course), then the subasymptotic of parton distribution
starts to play and the function $K_{i,1+\delta}$ $(i=R,L,I)$ contains
the term $1/\delta_i$, which is determined by the both: asymptotics
and subasymptotics of parton distributions.


\newpage


\begin{table}
\caption{\label{Tab:H1+ZEUS:96/97}\sffamily
The result of the LO and NLO fits to H1 (1996/97) \protect\cite{Adloff:1999}
and ZEUS (1996/97) \protect\cite{Chekanov:2001} data for different low
$Q^2$ cuts.  In the fits $f$ is fixed to 4 flavors.}
\centering
\vspace{0.3cm}
\begin{tabular}{|l||c|c|c||r|} \hline \hline
~$Q^2~[\text{GeV}^2] ~ \ge $&
 $A_G^{\tau2}$ & $A_q^{\tau2}$ & $Q_0^2~[\text{GeV}^2]$ &
 $\chi^2 / n.o.p.$~ \\
\hline\hline
 LO(H1 96/97 \protect\cite{Adloff:1999}) &&&& \\
 ~1.5 & 0.797$\pm$.022 & 0.791$\pm$.026 & 0.304$\pm$.005 & 181/101 \\ 
 ~2.0 & 0.819$\pm$.022 & 0.781$\pm$.026 & 0.309$\pm$.005 & 139/98  \\ 
 ~2.5 & 0.869$\pm$.024 & 0.754$\pm$.027 & 0.319$\pm$.005 & 88/90   \\ 
 ~3.5 & 0.920$\pm$.028 & 0.733$\pm$.029 & 0.332$\pm$.006 & 61/81   \\
\hline
 LO(ZEUS 96/97 \protect\cite{Chekanov:2001}) &&&& \\
 ~2.7 & 0.918$\pm$.031 & 0.754$\pm$.040 & 0.317$\pm$.005 & 80/116 \\ 
 ~3.5 & 0.893$\pm$.034 & 0.780$\pm$.042 & 0.315$\pm$.006 & 76/111 \\
\hline\hline
 NLO(H1 96/97 \protect\cite{Adloff:1999}) &&&& \\
 ~1.5 &$-$.013$\pm$.015 & 0.893$\pm$.028 & 0.494$\pm$.009 & 201/101 \\ 
 ~2.0 &  0.003$\pm$.015 & 0.882$\pm$.028 & 0.505$\pm$.009 & 153/98  \\ 
 ~2.5 &  0.042$\pm$.017 & 0.850$\pm$.029 & 0.526$\pm$.010 & 95/90  \\ 
 ~3.5 &  0.082$\pm$.020 & 0.824$\pm$.032 & 0.554$\pm$.012 & 63/81   \\
\hline
 NLO(ZEUS 96/97 \protect\cite{Chekanov:2001}) &&&& \\
 ~2.7 &  0.061$\pm$.023 & 0.844$\pm$.044 & 0.523$\pm$.011 & 82/116  \\ 
 ~3.5 &  0.044$\pm$.025 & 0.871$\pm$.046 & 0.520$\pm$.012 & 78/111  \\
\hline\hline
 NLO(H1\protect\cite{Adloff:1999} $+$ ZEUS\protect\cite{Chekanov:2001})
 &&&& \\
 ~1.5 $(r_\text{Z} = .963)$
       & 0.010$\pm$.013 & 0.873$\pm$.024 & 0.506$\pm$.007 & 286/217
 [204/101, 82/116] \\ 
 ~2.0 $(r_\text{Z} = .964)$
       & 0.021$\pm$.013 & 0.864$\pm$.024 & 0.512$\pm$.007 & 233/214
 [154/98,  79/116] \\ 
 ~2.5 $(r_\text{Z} = .963)$
       & 0.046$\pm$.013 & 0.839$\pm$.024 & 0.524$\pm$.008 & 171/206
 [95/90,  76/116] \\ 
 ~3.5 $(r_\text{Z} = .962)$
       & 0.063$\pm$.015 & 0.829$\pm$.026 & 0.537$\pm$.008 & 140/192
 [66/81,  74/111] \\
\hline\hline
\end{tabular}
\end{table}

\begin{table}
\caption{\label{Tab:H1+ZEUS}\sffamily
The result of the LO and NLO fits to H1
\protect\cite{Adloff:1999,Adloff:1997,Aid:1996,Ahmed:1995,Abt:1993}
and ZEUS
\protect\cite{Chekanov:2001,Breitweg:2000,Breitweg:1999,Breitweg:1997,Derrick:1996:C72,Derrick:1996:C69,Derrick:1995,Derrick:1993}
data for different low $Q^2$ cuts and different $f$.}
\centering
\vspace{0.3cm}
\begin{tabular}{|l||c|c|c||r|} \hline \hline
~$Q^2~[\text{GeV}^2] ~ \ge $&
 $A_G^{\tau2}$ & $A_q^{\tau2}$ & $Q_0^2~[\text{GeV}^2]$ &
 $\chi^2 / n.o.p.$~~ \\
 \hline\hline
 LO $(f = 3)$
 &&&& \\
 ~0.5 ($r_\text{H1} = .933$, $r_\text{Z} = .955$) &
 1.216$\pm$.015 & 1.153$\pm$.015 & 0.306$\pm$.003 &
 1163/667 [488/292, 675/375] \\
 ~1.0 ($r_\text{H1} = .939$, $r_\text{Z} = .966$) &
 1.424$\pm$.023 & 0.977$\pm$.023 & 0.313$\pm$.003 &
 854/631 [389/279, 465/352]  \\
 ~1.5 ($r_\text{H1} = .946$, $r_\text{Z} = .969$) &
 1.472$\pm$.024 & 0.950$\pm$.023 & 0.317$\pm$.003 &
 775/614 [348/267, 427/347]  \\
 ~2.0 ($r_\text{H1} = .953$, $r_\text{Z} = .971$) &
 1.527$\pm$.025 & 0.923$\pm$.023 & 0.323$\pm$.003 &
 673/591 [273/252, 400/339]  \\
 ~2.5 ($r_\text{H1} = .958$, $r_\text{Z} = .971$) &
 1.589$\pm$.026 & 0.890$\pm$.024 & 0.330$\pm$.003 &
 580/573 [193/236, 387/337]  \\
 ~3.5 ($r_\text{H1} = .963$, $r_\text{Z} = .971$) &
 1.655$\pm$.030 & 0.866$\pm$.026 & 0.339$\pm$.004 &
 501/532 [142/210, 359/322]  \\
 \hline\hline
 LO $(f = 4)$
 &&&& \\
 ~0.5 ($r_\text{H1} = .934$, $r_\text{Z} = .957$) &
 0.641$\pm$.010 & 0.937$\pm$.012 & 0.295$\pm$.003 &
 1090/667 [455/292, 635/375] \\
 ~1.0 ($r_\text{H1} = .940$, $r_\text{Z} = .966$) &
 0.755$\pm$.015 & 0.821$\pm$.019 & 0.301$\pm$.003 &
 826/631 [373/279, 453/352]  \\
 ~1.5 ($r_\text{H1} = .947$, $r_\text{Z} = .969$) &
 0.784$\pm$.016 & 0.801$\pm$.019 & 0.304$\pm$.003 &
 754/614 [335/267, 419/347]  \\
 ~2.0 ($r_\text{H1} = .953$, $r_\text{Z} = .971$) &
 0.817$\pm$.017 & 0.780$\pm$.019 & 0.310$\pm$.003 &
 659/591 [264/252, 395/339]  \\
 ~2.5 ($r_\text{H1} = .958$, $r_\text{Z} = .971$) &
 0.855$\pm$.017 & 0.754$\pm$.020 & 0.316$\pm$.003 &
 570/573 [188/236, 382/337]  \\
 ~3.5 ($r_\text{H1} = .963$, $r_\text{Z} = .971$) &
 0.892$\pm$.020 & 0.737$\pm$.021 & 0.325$\pm$.004 &
 495/532 [140/210, 355/322]  \\
\hline \hline
 NLO $(f = 3)$
 &&&& \\
~0.5 ($r_\text{H1} = .929$, $r_\text{Z} = .951$) &
 $-$.094$\pm$.009 & 1.358$\pm$.015 & 0.515$\pm$.006 &
 1406/667 [599/292, 807/375] \\
 ~1.0 ($r_\text{H1} = .936$, $r_\text{Z} = .965$) &
   0.072$\pm$.014 & 1.114$\pm$.024 & 0.526$\pm$.006 &
 966/631 [455/279, 511/352]  \\
 ~1.5 ($r_\text{H1} = .944$, $r_\text{Z} = .968$) &
   0.109$\pm$.015 & 1.078$\pm$.025 & 0.535$\pm$.006 &
 863/614 [403/267, 460/347]  \\
 ~2.0 ($r_\text{H1} = .952$, $r_\text{Z} = .971$) &
   0.151$\pm$.016 & 1.045$\pm$.025 & 0.548$\pm$.006 &
 735/591 [311/252, 424/339]  \\
 ~2.5 ($r_\text{H1} = .958$, $r_\text{Z} = .970$) &
   0.198$\pm$.016 & 1.006$\pm$.025 & 0.564$\pm$.006 &
 620/573 [213/236, 407/337]  \\
 ~3.5 ($r_\text{H1} = .963$, $r_\text{Z} = .971$) &
   0.254$\pm$.019 & 0.972$\pm$.027 & 0.587$\pm$.007 &
 523/532 [151/210, 372/322]  \\
\hline \hline
 NLO $(f = 4)$
 &&&& \\
~0.5 ($r_\text{H1} = .932$, $r_\text{Z} = .955$) &
 $-$.142$\pm$.006 & 1.087$\pm$.012 & 0.478$\pm$.006 &
 1229/667 [514/292, 715/375] \\
 ~1.0 ($r_\text{H1} = .938$, $r_\text{Z} = .966$) &
 $-$.042$\pm$.011 & 0.929$\pm$.021 & 0.487$\pm$.006 &
 884/631 [407/279, 477/352]  \\
 ~1.5 ($r_\text{H1} = .946$, $r_\text{Z} = .969$) &
 $-$.020$\pm$.011 & 0.903$\pm$.021 & 0.495$\pm$.006 &
 798/614 [363/267, 435/347]  \\
 ~2.0 ($r_\text{H1} = .953$, $r_\text{Z} = .971$) &
   0.006$\pm$.012 & 0.877$\pm$.021 & 0.506$\pm$.006 &
 688/591 [282/252, 406/339]  \\
 ~2.5 ($r_\text{H1} = .958$, $r_\text{Z} = .971$) &
   0.035$\pm$.012 & 0.847$\pm$.022 & 0.520$\pm$.006 &
 589/573 [197/236, 392/337]  \\
 ~3.5 ($r_\text{H1} = .963$, $r_\text{Z} = .972$) &
   0.065$\pm$.014 & 0.826$\pm$.023 & 0.539$\pm$.007 &
 505/532 [143/210, 362/322]  \\
\hline \hline
\end{tabular}
\end{table}

\begin{table}
\caption{\label{Tab:H196/97:HT}\sffamily
The result of the LO and NLO fits to H1 (1996/97) \protect\cite{Adloff:1999}
data. Power corrections included for different values of the parameter $b$
and in the infrared renormalon case.
}
\centering
\vspace{0.3cm}
\begin{tabular}{|l||c|c||c|c||c|r|} \hline \hline
 H1 96/97 \protect\cite{Adloff:1999} &
 $A_G^{\tau2}$ & $A_q^{\tau2}$ &
 $A^{\tau4}_G$ $({\mathrm a}^{\tau4}_G)$ &
 $A^{\tau4}_q$ $({\mathrm a}^{\tau4}_q)$ &
 $Q_0^2~[\text{GeV}^2]$ & $\chi^2 / n.o.p.$ \\
 &&& ~~~~~ $({\mathrm a}^{\tau6}_G)$ &
     ~~~~~ $({\mathrm a}^{\tau6}_q)$ && \\
\hline\hline
~ LO $(f = 4)$ &&&&&& \\
~no $h\tau$ & 0.797$\pm$.022 & 0.791$\pm$.026 &
  --- & --- & 0.304$\pm$.005 & 181/101 \\
~$b = 0$    & 1.214$\pm$.060 & 0.426$\pm$.054 &
  --- & 0.969$\pm$.127 &  0.360$\pm$.009 &  124/101 \\
~$b = 1$    & 1.263$\pm$.070 & 0.436$\pm$.051 &
 $-$.496$\pm$.062 & 1.127$\pm$.142 & 0.388$\pm$.022 & 119/101 \\
~$b = a^2/2$& 1.321$\pm$.072 & 0.446$\pm$.049 &
 $-$.523$\pm$.065 & 1.205$\pm$.148 & 0.417$\pm$.023 & 106/101 \\
\hline
~$R\tau4$ & 1.155$\pm$.060 & 0.582$\pm$.032 &
 $-$.310$\pm$.171 & 0.230$\pm$.078 & 0.381$\pm$.020 & 56/101 \\
 &&& (0.000 fix) & (0.000 fix) && \\
~renorm.& 1.037$\pm$.121 & 0.668$\pm$.073 &
 $-$.011$\pm$.259 &$-$.007$\pm$.122 & 0.356$\pm$.035 & 54/101 \\
 &&& $-$.486$\pm$.841 & 0.084$\pm$.325 && \\
\hline\hline
~ NLO $(f = 4)$ &&&&&& \\
~no $h\tau$ & $-$.013$\pm$.015 & 0.893$\pm$.028 &
  --- & --- & 0.494$\pm$.009 & 201/101 \\ 
~$b = 0$    & $-$.024$\pm$.017 & 0.882$\pm$.029 &
  --- & $-$.001$\pm$.000 &  0.473$\pm$.017 &  199/101 \\
~$b = 1$    & 0.316$\pm$.047 & 0.474$\pm$.056 &
 $-$.542$\pm$.065 & 1.219$\pm$.147 & 0.600$\pm$.030 & 133/101 \\ 
~$b = a^2/2$ & 0.336$\pm$.045 & 0.492$\pm$.053 &
 $-$.603$\pm$.067 & 1.362$\pm$.152 & 0.635$\pm$.030 & 127/101 \\
\hline
~$R\tau4$ & 0.144$\pm$.078 & 0.764$\pm$.056 &
 $-$.692$\pm$.275 & 0.155$\pm$.021 & 0.576$\pm$.060 & 55/101 \\
 &&& (0.000 fix) & (0.000 fix) && \\
~renorm.& 0.102$\pm$.086 & 0.800$\pm$.066 &
 $-$1.327$\pm$1.218 & 0.310$\pm$.281 & 0.548$\pm$.067 & 54/101 \\
 &&& 0.412$\pm$.834 & 0.063$\pm$.144 && \\
\hline \hline
\end{tabular}
\end{table}

\begin{table}
\caption{\label{Tab:ZEUS96/97:HT}\sffamily
The result of the LO and NLO fits to ZEUS (1996/97)
\protect\cite{Chekanov:2001} data.
Power corrections included for different values of the parameter $b$
and in the infrared renormalon case.}
\centering
\vspace{0.3cm}
\begin{tabular}{|l||c|c||c|c||c|r|} \hline \hline
 ZEUS 96/97 \protect\cite{Chekanov:2001} &
 $A_G^{\tau2}$ & $A_q^{\tau2}$ &
 $A^{\tau4}_G$ $({\mathrm a}^{\tau4}_G)$ &
 $A^{\tau4}_q$ $({\mathrm a}^{\tau4}_q)$ &
 $Q_0^2~[\text{GeV}^2]$ & $\chi^2 / n.o.p.$ \\
 &&& ~~~~~ $({\mathrm a}^{\tau6}_G)$ &
     ~~~~~ $({\mathrm a}^{\tau6}_q)$ && \\
\hline\hline
~ LO $(f = 4)$ &&&&&& \\
~no $h\tau$ & 0.918$\pm$.031 & 0.754$\pm$.040 &
  --- & --- & 0.317$\pm$.005 & 80/116 \\
~$b = 0$    & 0.891$\pm$.067 & 0.780$\pm$.070 &
  --- &$-$.093$\pm$.203 & 0.314$\pm$.009 &  80/116 \\
~$b = 1$    & 0.910$\pm$.074 & 0.780$\pm$.068 &
 0.046$\pm$.101 &$-$.101$\pm$.229 & 0.324$\pm$.023 & 79/116 \\ 
~$b = a^2/2$  & 0.920$\pm$.069 & 0.786$\pm$.066 &
 0.083$\pm$.117 &$-$.179$\pm$.263 & 0.330$\pm$.019 & 78/116 \\
\hline
~$R\tau4$ & 0.980$\pm$.063 & 0.739$\pm$.050 &
 0.344$\pm$.329 &$-$.137$\pm$.135 & 0.343$\pm$.021 & 78/116 \\
 &&& (0.000 fix) & (0.000 fix) && \\
~renorm.& 0.859$\pm$.087 & 0.757$\pm$.074 &
 $-$2.439$\pm$1.207 & 1.014$\pm$.559 & 0.281$\pm$.024 & 68/116 \\
 &&& $-$10.66$\pm$3.60 & 4.99$\pm$1.78 && \\
\hline\hline
~ NLO $(f = 4)$ &&&&&& \\
~no $h\tau$ &  0.061$\pm$.023 & 0.844$\pm$.044 &
  --- & --- & 0.523$\pm$.011 & 82/116 \\
~$b = 0$    &  0.067$\pm$.030 & 0.849$\pm$.046 &
  --- &$-$.001$\pm$.002 &  0.533$\pm$.034 & 81/116 \\
~$b = 1$    & 0.062$\pm$.015 & 0.859$\pm$.026 &
 0.020$\pm$.002 &$-$.044$\pm$.005 & 0.534$\pm$.017 & 81/116 \\
~$b = a^2/2$  & 0.071$\pm$.055 & 0.866$\pm$.073 &
 0.046$\pm$.122 & $-$.101$\pm$.275 & 0.549$\pm$.037 & 80/116 \\
\hline
~$R\tau4$ & 0.083$\pm$.081 & 0.823$\pm$.078 &
 $-$.046$\pm$.313 & 0.016$\pm$.041 & 0.533$\pm$.054 & 81/116 \\
 &&& (0.000 fix) & (0.000 fix) && \\
~renorm.& $-$.329$\pm$.068 & 1.242$\pm$.094 &
 $-$1.599$\pm$.643 & $-$.177$\pm$.173 & 0.312$\pm$.027 & 64/116 \\
 &&& $-$16.008$\pm$2.451 & 2.253$\pm$0.492 && \\
\hline \hline
\end{tabular}
\end{table}

\begin{table}
\caption{\label{Tab:H1+ZEUS:HT}\sffamily
The result of the LO and NLO fits to H1
\protect\cite{Adloff:1999,Adloff:1997,Aid:1996,Ahmed:1995,Abt:1993}
and ZEUS
\protect\cite{Chekanov:2001,Breitweg:2000,Breitweg:1999,Breitweg:1997,Derrick:1996:C72,Derrick:1996:C69,Derrick:1995,Derrick:1993}
data at $Q^2 \ge 1.5~\text{GeV}^2$.
Power corrections included for different values of the parameter $b$
and in the infrared renormalon case.
}
\centering
\vspace{0.3cm}
\begin{tabular}{|l||c|c||c|c||c|l|} \hline \hline
 H1\protect\cite{Adloff:1999,Adloff:1997,Aid:1996,Ahmed:1995,Abt:1993}
 $+$ ZEUS\protect\cite{Chekanov:2001,Breitweg:2000,Breitweg:1999,Breitweg:1997,Derrick:1996:C72,Derrick:1996:C69,Derrick:1995,Derrick:1993} &
 $A_G^{\tau2}$ & $A_q^{\tau2}$ &
 $A^{\tau4}_G$ $({\mathrm a}^{\tau4}_G)$ &
 $A^{\tau4}_q$ $({\mathrm a}^{\tau4}_q)$ &
 $Q_0^2~[\text{GeV}^2]$ & $\chi^2 / n.o.p.$ \\
~~  $Q^2 \ge 1.5~\text{GeV}^2$ &
 && ~~~~~ $({\mathrm a}^{\tau6}_G)$ &
    ~~~~~ $({\mathrm a}^{\tau6}_q)$ && \\
\hline\hline
~ LO $(f = 3)$ &&&&&& \\
~no $h\tau$ ($r_\text{H1} = .946$, $r_\text{Z} = .969$) &
 1.472$\pm$.024 & 0.950$\pm$.023 &   --- & --- & 0.317$\pm$.003 &
 775/614 [348/267, 427/347] \\
~$b = 0$ ($r_\text{H1} = .953$, $r_\text{Z} = .970$) &
 2.083$\pm$.056 & 0.513$\pm$.042 &
  --- & 1.275$\pm$.103 &  0.372$\pm$.006 &
 628/614 [246/267, 382/347] \\
~$b = 1$ ($r_\text{H1} = .954$, $r_\text{Z} = .971$) &
 2.164$\pm$.068 & 0.528$\pm$.040 &
 $-$.623$\pm$.049 & 1.422$\pm$.113 & 0.401$\pm$.014 &
 616/614 [240/267, 376/347] \\
~$b = a^2/2$ ($r_\text{H1} = .954$, $r_\text{Z} = .972$) &
 2.224$\pm$.067 & 0.546$\pm$.039 &
 $-$.617$\pm$.051 & 1.431$\pm$.116 & 0.421$\pm$.013 &
 591/614 [224/267, 367/347] \\
\hline
~$R\tau4$ & 2.012$\pm$.062 & 0.687$\pm$.029 &
 $-$.279$\pm$.135 & 0.326$\pm$.088 & 0.390$\pm$.012 & 503/614 \\
 ~~ ($r_\text{H1} = .959$, $r_\text{Z} = .973$) &&&
 (0.000 fix) & (0.000 fix) && ~ [151/267, 352/347] \\
~renorm. & 1.826$\pm$.100 & 0.784$\pm$.050 &
 $-$.064$\pm$.185 & 0.006$\pm$.092 & 0.360$\pm$.019 & 498/614 \\
 ~~ ($r_\text{H1} = .959$, $r_\text{Z} = .972$) &&&
 $-$1.245$\pm$.718 & .525$\pm$.383 && ~ [149/267, 349/347] \\
\hline\hline
~ LO $(f = 4)$ &&&&&& \\
~no $h\tau$ ($r_\text{H1} = .947$, $r_\text{Z} = .969$) &
 0.784$\pm$.016 & 0.801$\pm$.019 &   --- & --- & 0.304$\pm$.003 &
 754/614 [335/267, 419/347] \\
~$b = 0$ ($r_\text{H1} = .953$, $r_\text{Z} = .970$) &
 1.157$\pm$.036 & 0.461$\pm$.035 &
  --- & 0.950$\pm$.082 &  0.353$\pm$.005 &
 625/614 [244/267, 381/347] \\
~$b = 1$ ($r_\text{H1} = .954$, $r_\text{Z} = .971$) &
 1.202$\pm$.042 & 0.477$\pm$.033 &
 $-$.478$\pm$.040 & 1.088$\pm$.091 & 0.383$\pm$.014 &
 612/614 [237/267, 375/347] \\ 
~$b = a^2/2$ ($r_\text{H1} = .954$, $r_\text{Z} = .972$) &
 1.232$\pm$.041 & 0.494$\pm$.032 &
 $-$.481$\pm$.042 & 1.107$\pm$.096 & 0.402$\pm$.013 &
 586/614 [220/267, 366/347] \\
\hline
~$R\tau4$ & 1.125$\pm$.037 & 0.582$\pm$.024 &
 $-$.172$\pm$.100 & 0.162$\pm$.047 & 0.376$\pm$.012 & 505/614 \\
 ~~ ($r_\text{H1} = .959$, $r_\text{Z} = .973$) &&&
 (0.000 fix) & (0.000 fix) && ~ [153/267, 352/347] \\
~renorm.& 0.990$\pm$.060 & 0.679$\pm$.043 &
 $-$.009$\pm$.161 & $-$.019$\pm$.092 & 0.345$\pm$.017 & 497/614 \\
 ~~ ($r_\text{H1} = .959$, $r_\text{Z} = .972$) &&&
 $-$.980$\pm$.497 & 0.276$\pm$.196 && ~ [149/267, 348/347] \\
\hline\hline
~ NLO $(f = 3)$ &&&&&& \\
~no $h\tau$ ($r_\text{H1} = .944$, $r_\text{Z} = .968$) &
 0.109$\pm$.015 & 1.078$\pm$.025 & --- & --- & 0.535$\pm$.006 &
 863/614 [403/267, 460/347] \\ 
~$b = 0$ ($r_\text{H1} = .945$, $r_\text{Z} = .968$) &
 0.101$\pm$.018 & 1.073$\pm$.025 &
  --- &$-$.001$\pm$.001 &  0.527$\pm$.011 &
 862/614 [401/267, 461/347] \\
~$b = 1$ ($r_\text{H1} = .953$, $r_\text{Z} = .970$) &
 0.600$\pm$.043 & 0.563$\pm$.043 &
 $-$.735$\pm$.052 & 1.655$\pm$.117 & 0.661$\pm$.020 &
 669/614 [273/267, 396/347] \\
~$b = a^2/2$ ($r_\text{H1} = .953$, $r_\text{Z} = .970$) &
 0.630$\pm$.042 & 0.585$\pm$.041 &
 $-$.783$\pm$.053 & 1.771$\pm$.120 & 0.691$\pm$.019 &
 654/614 [265/267, 389/347] \\
\hline
~$R\tau4$ & 0.410$\pm$.067 & 0.864$\pm$.041 &
 $-$.660$\pm$.185 & 0.265$\pm$.024 & 0.643$\pm$.034 & 505/614 \\
 ~~ ($r_\text{H1} = .960$, $r_\text{Z} = .973$) &&&
 (0.000 fix) & (0.000 fix) && ~ [151/267, 354/347] \\
~renorm.& 0.188$\pm$.080 & 0.985$\pm$.045 &
 $-$3.081$\pm$1.019 & .957$\pm$.315 & 0.530$\pm$.041 & 498/614 \\
 ~~ ($r_\text{H1} = .959$, $r_\text{Z} = .972$) &&&
  .036$\pm$.422 & .437$\pm$.164 && ~ [149/267, 349/347] \\
\hline\hline
~ NLO $(f = 4)$ &&&&&& \\
~no $h\tau$ ($r_\text{H1} = .946$, $r_\text{Z} = .969$) &
 $-$.020$\pm$.011 & 0.903$\pm$.021 & --- & --- & 0.495$\pm$.006 &
 798/614 [363/267, 435/347] \\
~$b = 0$ ($r_\text{H1} = .946$, $r_\text{Z} = .969$) &
 $-$.024$\pm$.013 & 0.899$\pm$.022 &
  --- & 0.00$\pm$.0004 &  0.488$\pm$.011 &
 798/614 [362/267, 436/347] \\
~$b = 1$ ($r_\text{H1} = .953$, $r_\text{Z} = .970$) &
 0.288$\pm$.029 & 0.515$\pm$.037 &
 $-$.535$\pm$.042 & 1.205$\pm$.095 & 0.602$\pm$.019 &
 645/614 [256/267, 389/347] \\
~$b = a^2/2$ ($r_\text{H1} = .954$, $r_\text{Z} = .971$) &
 0.301$\pm$.028 & 0.535$\pm$.035 &
 $-$.580$\pm$.044 & 1.311$\pm$.100 & 0.631$\pm$.019 &
 629/614 [248/267, 381/347] \\
\hline
~$R\tau4$ & 0.156$\pm$.041 & 0.734$\pm$.035 &
 $-$.522$\pm$.153 & 0.141$\pm$.014 & 0.579$\pm$.031 & 506/614 \\
 ~~ ($r_\text{H1} = .960$, $r_\text{Z} = .973$) &&&
 (0.000 fix) & (0.000 fix) && ~ [151/267, 355/347] \\
~renorm.& 0.041$\pm$.045 & 0.824$\pm$.034 &
 $-$2.765$\pm$.968 & .676$\pm$.240 & 0.493$\pm$.037 & 500/614 \\
 ~~ ($r_\text{H1} = .959$, $r_\text{Z} = .973$) &&&
 .939$\pm$.718 & .252$\pm$.099 && ~ [151/267, 349/347] \\
\hline \hline
\end{tabular}
\end{table}

\begin{table}
\caption{\label{Tab:Rht}\sffamily
The result of the LO and NLO fits to H1
\protect\cite{Adloff:1999,Adloff:1997,Aid:1996,Ahmed:1995,Abt:1993}
and ZEUS
\protect\cite{Chekanov:2001,Breitweg:2000,Breitweg:1999,Breitweg:1997,Derrick:1996:C72,Derrick:1996:C69,Derrick:1995,Derrick:1993}
 at $Q^2 \ge 0.5~\text{GeV}^2$.
Power corrections included in the infrared renormalon case.
}
\centering
\vspace{0.3cm}
\begin{tabular}{|l||c|c||c|c||c|l|} \hline \hline
 H1\protect\cite{Adloff:1999,Adloff:1997,Aid:1996,Ahmed:1995,Abt:1993}
 $+$ ZEUS\protect\cite{Chekanov:2001,Breitweg:2000,Breitweg:1999,Breitweg:1997,Derrick:1996:C72,Derrick:1996:C69,Derrick:1995,Derrick:1993} &
 $A_G^{\tau2}$ & $A_q^{\tau2}$ &
 $ {\mathrm a}^{\tau4}_G $ &
 $ {\mathrm a}^{\tau4}_q $ &
 $Q_0^2~[\text{GeV}^2]$ & $\chi^2 / n.o.p.$ \\
 ~~ $Q^2 \ge 0.5~\text{GeV}^2$ &
 && ${\mathrm a}^{\tau6}_G$ &
    ${\mathrm a}^{\tau6}_q$ && \\
\hline\hline
 LO $R\tau4$ $(f = 3)$ & 2.212$\pm$.050 & 0.602$\pm$.027 &
 0.238$\pm$.019 &$-$.014$\pm$.005 & 0.428$\pm$.008 & 569/667 \\
 ~ ($r_\text{H1} = .953$, $r_\text{Z} = .975$) &&&
 (0.000 fix) & (0.000 fix) && ~ [192/292, 377/375] \\
 LO $(f = 3)$ & 2.161$\pm$.055 & 0.633$\pm$.029 &
 $-$.002$\pm$.024 & 0.165$\pm$.024 & 0.421$\pm$.010 & 553/667 \\
 ~ ($r_\text{H1} = .955$, $r_\text{Z} = .974$) &&&
 $-$.017$\pm$.017 & 0.053$\pm$.010 && ~ [181/292, 372/375] \\
\hline
 LO $R\tau4$ $(f = 4)$ & 1.234$\pm$.031 & 0.518$\pm$.023 &
 0.201$\pm$.016 &$-$.011$\pm$.003 & 0.407$\pm$.008 & 573/667 \\
 ~ ($r_\text{H1} = .953$, $r_\text{Z} = .975$) &&&
 (0.000 fix) & (0.000 fix) && ~ [193/292, 380/375] \\
 LO $(f = 4)$ & 1.211$\pm$.033 & 0.539$\pm$.023 &
 $-$.002$\pm$.020 & 0.102$\pm$.015 & 0.404$\pm$.009 & 555/667 \\
 ~ ($r_\text{H1} = .955$, $r_\text{Z} = .974$) &&&
 0.001$\pm$.010 & 0.031$\pm$.005 && ~ [182/292, 373/375] \\
\hline\hline
 NLO $R\tau4$ $(f = 3)$ & 1.014$\pm$.057 & 0.521$\pm$.034 &
 0.632$\pm$.048 & 0.188$\pm$.026 & 0.956$\pm$.031 & 621/667 \\
 ~ ($r_\text{H1} = .951$, $r_\text{Z} = .975$) &&&
 (0.000 fix) & (0.000 fix) && ~ [220/292, 401/375] \\
 NLO $(f = 3)$ & 0.617$\pm$.058 & 0.742$\pm$.038 &
 $-$.129$\pm$.102 & 0.224$\pm$.022 & 0.746$\pm$.030 & 562/667 \\
 ~ ($r_\text{H1} = .956$, $r_\text{Z} = .975$) &&&
 $-$.203$\pm$.053 & 0.061$\pm$.010 && ~ [182/292, 380/375] \\
\hline
 NLO $R\tau4$ $(f = 4)$ & 0.485$\pm$.033 & 0.476$\pm$.029 &
 0.556$\pm$.042 & 0.071$\pm$.008 & 0.826$\pm$.026 & 617/667 \\
 ~ ($r_\text{H1} = .950$, $r_\text{Z} = .975$) &&&
 (0.000 fix) & (0.000 fix) && ~ [222/292, 395/375] \\
 NLO $(f = 4)$ & 0.279$\pm$.038 & 0.640$\pm$.034 &
 $-$.143$\pm$.100 & 0.140$\pm$.015 & 0.672$\pm$.029 & 565/667 \\
 ~ ($r_\text{H1} = .955$, $r_\text{Z} = .974$) &&&
 $-$.044$\pm$.050 & 0.043$\pm$.007 && ~ [184/292, 381/375] \\
\hline \hline
\end{tabular}
\end{table}

\newpage


\end{document}